\documentclass[journal]{IEEEtran}     
\usepackage{comment}
\usepackage{booktabs} % For formal tables

\usepackage{hyperref}
\usepackage{diagrams}
\usepackage{color}
\usepackage{xcolor}
\usepackage{cite}
\usepackage{amssymb,latexsym,amsfonts,amsmath}
\renewcommand{\theequation}{\arabic{equation}}
\usepackage{graphicx}
\usepackage{subcaption}
\usepackage{stfloats}

\usepackage{enumitem}
\setlist[enumerate,1]{label=$\bullet$,leftmargin=5.5mm}

\def\diag{\textup{diag}}

\newtheorem{theorem}{Theorem}

\newtheorem{problem}[theorem]{Problem}
\newtheorem{proposition}[theorem]{Proposition}
\newtheorem{corollary}[theorem]{Corollary}

\newtheorem{definition}[theorem]{Definition}
\newtheorem{example}[theorem]{Example}

\newtheorem{remark}[theorem]{Remark}
\newtheorem{assumption}[theorem]{Assumption}

\newcommand{\n}{\mathcal{N}}

\newcommand{\C}{\mathcal{C}}

\newcommand{\N}{\mathbb{N}}%
\newcommand{\R}{\mathbb{R}}%
\def\bl#1{{\textcolor{black}{{\bf}#1}}}
\def\blue#1{{\textcolor{black}{{\bf}#1}}}

\begin{document}
 \title{Controller Synthesis of Collaborative Signal Temporal Logic Tasks for Multi-Agent Systems via Assume-Guarantee Contracts}

\author{Siyuan Liu, Adnane Saoud,~\IEEEmembership{Member,~IEEE}, and Dimos V. Dimarogonas,~\IEEEmembership{Fellow,~IEEE} 
	\thanks{The work is supported in part by 
the ERC LEAFHOUND Project, the Swedish Research Council (VR), the 
Horizon Europe EIC project SymAware (101070802), Digital Futures,
and the Wallenberg AI, Autonomous Systems and Software Program (WASP) funded by the Knut and Alice Wallenberg (KAW) Foundation.}% <-this % stops a space
	\thanks{S. Liu and D. Dimarogonas are with the Division of Decision and Control Systems,  KTH Royal Institute of Technology, Stockholm, Sweden. Email: {\tt\small \{siyliu,dimos\}@kth.se}}%
	\thanks{A. Saoud is with the College of Computing, University Mohammed VI Polytechnic, Benguerir, Morocco. Email: {\tt\small adnane.saoud@um6p.ma}}}

\maketitle

%%%%%%%%%%%%%%%%%%%%%%%%%%%%%%%%%%%%%%%%%%%%%%%%%%%%%%%%%%%%%%%%%%%%%%%%%%%%%%%%
\begin{abstract}
This paper considers the problem of controller synthesis of a fragment of signal temporal logic (STL) specifications for large-scale multi-agent systems, where the agents are dynamically coupled and subject to collaborative tasks.
A compositional framework based on continuous-time assume-guarantee contracts is developed to break the complex and large synthesis problem into subproblems of manageable sizes. 
We first show how to formulate the collaborative STL tasks as assume-guarantee contracts by leveraging the idea of prescribed performance control. 
The concept of contracts is used to establish our compositionality result, which allows us to guarantee the satisfaction of a global contract by the multi-agent system when all agents satisfy their local contracts. 
Then, a closed-form continuous-time feedback controller is designed to enforce local contracts over the agents in a distributed manner, which further guarantees global task satisfaction based on the compositionality result.
The effectiveness of our results is demonstrated by two numerical examples.
\end{abstract}
%\begin{IEEEkeywords}
%Assume-Guarantee Contracts;
%\end{IEEEkeywords}

\begin{IEEEkeywords}
Multi-agent systems, signal temporal logics, formal methods, assume-guarantee contracts, distributed control, prescribed performance control.
\end{IEEEkeywords}

\section{Introduction}
Over the last few decades, multi-agent systems have received increasing attention with a wide range of applications in various areas, including multi-robot systems, social networks, 
autonomous driving, and smart grids.
%With the rapid technological advancements in smart cities, large-scale and interconnected multi-agent systems are becoming  prevalent, subjecting to high-level specifications that are difficult to handle using classical control design approaches. 
Revolving around the control of multi-agent systems, extensive research works have been established in the past two decades that deal with tasks such as formation control \cite{tanner2003stable}, consensus \cite{ren2005consensus,olfati2007consensus}, and coverage \cite{cortes2004coverage}; see \cite{mesbahi2010graph} for an overview. 

In recent years, there has been a growing trend in the planning and control of single- and multi-agent systems under more complex high-level specifications expressed as temporal logics.  
Temporal logic, including linear temporal logics (LTL) and signal temporal logics (STL), \bl{resembles natural languages and offers a powerful framework for rigorously reasoning about the temporal behaviors of systems. Thus, they  
can be used to express complex high-level specifications that extend beyond standard control objectives \cite{belta2017formal}}. 
One potential solution that appeared more recently is the \emph{correct-by-construction synthesis scheme} \cite{tabuada2009verification,belta2017formal}, which is built upon formal methods-based approaches  to formally verify or synthesize certifiable controllers against rich specifications given by temporal logic formulae. 
Planning and control of temporal logic formulae are addressed in \cite{kloetzer2008fully,fainekos2009temporal} for single agent and in \cite{loizou2004automatic,guo2015multi,liu2017distributed} for multi-agent systems.  
However, 
when encountering large-scale systems, most of the above-mentioned approaches suffer severely from the \emph{curse of dimensionality} due to their high computational complexity, which limits their applications to systems of moderate size. 
To tackle this complexity issue, different compositional approaches were developed for the analysis and control of large-scale and interconnected systems. 
The most common types of compositional approaches are based on input-output properties (e.g., small-gain or dissipativity properties) \cite{jiang1994small,7857702,kim2017small,jagtap2020compositional,liu2021symbolic} and assume-guarantee contracts \cite{kim2017small,saoud2021assume,sharf2021assume,7857702,saoud2020contract, ghasemi2020compositional,barttac}, respectively. 
Specifically, the notion of assume-guarantee contracts (AGCs) prescribes properties that a component must guarantee under assumptions on the behavior of its environment (or its neighboring agents)\cite{benveniste2018contracts}. 

%The compositional framework proposed in the present paper will fall into the second category by leveraging assume-guarantee contracts (AGCs). 

The main goal of this paper is to develop a compositional framework for the controller synthesis of  signal temporal logic (STL) formulae on large-scale multi-agent systems.
The synthesis of STL properties for control systems has attracted a lot of attention in the last few years.
%which can express more complex tasks with real-time and real-valued constraints \cite{maler2004monitoring}.
\bl{Compared to LTL, which is a propositional temporal logic that deals only with discrete-time signals, STL \cite{maler2004monitoring} is a predicate logic interpreted over continuous-time signals that allows to formulate more expressive tasks with real-time and real-valued constraints. Moreover, STL naturally entails space robustness \cite{donze2010robust} 
which enables one to assess the robustness of satisfaction.}
Despite the strong expressivity of STL formulae, 
the synthesis of control systems under STL specifications is known to be challenging due to its nonlinear, nonconvex, noncausal, and nonsmooth semantics. 
In \cite{raman2014model}, 
the problem of synthesizing STL tasks on discrete-time systems is handled using model predictive control (MPC) where space robustness is encoded as mixed-integer linear programs.
The results in \cite{larsCDC} established a connection between prescribed performance control (PPC) and the robust semantics of STL specifications, based on which continuous-time feedback control laws are derived for multi-agent systems \cite{lindemann2019feedback}.
%by providing a least violating solution for conflicting STL specifications \cite{lindemann2019feedback}.
Barrier function-based approaches are proposed for the collaborative control of STL tasks for multi-agent systems~\cite{lindemann2020barrier}. 
\blue{The recent results in \cite{ye2024decentralized} developed efficient decentralized and distributed control frameworks to enforce a prescribed-time stability property for multi-agent systems, which allows dynamical interactions with limited communications among agents.} 

%The results in \cite{lindemann2018control} proposed for the first time a synthesis approach for STL tasks by leveraging a notion of time-varying control barrier functions. 
%Other typical optimization-based approaches to this synthesis problem are summarized in \cite[Section 5]{belta2019formal}. 
%Motivated by the emerging challenges in designing large-scale systems under complex specifications, in this paper, we propose a compositional approach for the synthesis of STL tasks of continuous-time interconnected systems using AGC.

%We consider the interconnection of finitely many agents, whose dynamics are given by nonlinear control-affine systems. 

In this paper, we consider continuous-time and interconnected multi-agent systems under a fragment of STL specifications. In this setting, each agent is subject to a local STL task that may depend on the behavior of other agents, whereas the interconnection of agents is induced by the dynamical couplings between each other.
In order to provide a compositional framework to synthesize distributed controllers for a multi-agent system, we first formulate the desired STL formulae as prescribed performance control problems, and then introduce assume-guarantee contracts for the agents by leveraging the derived prescribed performance functions for the STL tasks.  
Two concepts of contract satisfaction, i.e., \emph{weak satisfaction} and \emph{uniform strong satisfaction} (cf. Definition \ref{asg}) are introduced to establish our compositionality results using assume-guarantee reasoning, i.e., if all agents satisfy their local contracts, then the global contract is satisfied by the multi-agent system. 
In particular, we show that \emph{weak satisfaction} of the local contracts is sufficient to deal with multi-agent systems with acyclic interconnection topologies, while \emph{uniform strong satisfaction} is needed to reason about general interconnection topologies containing cycles.
Based on the compositional reasoning, we then present a controller synthesis approach in a distributed manner. 
In particular, continuous-time closed-form feedback controllers are derived for the agents that ensure the satisfaction of local contracts, thus leading to the satisfaction of global contract by the multi-agent system.
%The proposed decentralized framework allows us to design local controllers to enforce local contracts independently, while guaranteeing the satisfaction of the global STL for the interconnected system.  
To the best of our knowledge, this paper is the first to handle STL specifications on multi-agent systems using assume-guarantee contracts.
Thanks to the derived closed-form control strategy and the distributed framework, our approach requires very low computational complexity compared to existing results in the literature, which mostly rely on discretizations in state space or time \cite{tabuada2009verification,belta2017formal}.

{\bf{Related work:}} 
While AGCs have been extensively used in the computer science community~\cite{benveniste2018contracts,nuzzo2015compositional}, new frameworks of AGCs for dynamical systems with continuous state-variables have been proposed recently in~\cite{saoud2021assume,saoud2020contract} for continuous-time systems, and ~\cite{kim2017small},~\cite[Chapter 2]{saoud2019compositional} for discrete-time systems. In this paper, we follow the same behavioral framework of AGCs for continuous-time systems proposed in~\cite{saoud2021assume}. In the following, we provide a comparison with the approach proposed in~\cite{saoud2020contract,saoud2021assume}. A detailed comparison between the framework in~\cite{saoud2021assume}, the one in~\cite{kim2017small} and existing approaches from the computer science community~\cite{benveniste2018contracts,nuzzo2015compositional} can be found in~\cite[Section 1]{saoud2021assume}.

The contribution of the paper is threefold:
\begin{enumerate}[fullwidth,itemindent=0em]
    \item At the level of compositionality rules: The authors in~\cite{saoud2021assume} rely on a notion of \emph{strong contract satisfaction} to provide a compositionality result (i.e., how to go from the satisfaction of local contracts at the component's level to the satisfaction of the global specification for the interconnected system) under the condition of the set of guarantees (of the contracts) being closed. In this paper, we are dealing with STL specifications, which are encoded as AGCs made of open sets of assumptions and guarantees. The non-closedness of the set of guarantees makes the concept of contract satisfaction proposed in~\cite{saoud2021assume} not sufficient to establish a compositionality result. For this reason, in this paper, we introduce the concept of \emph{uniform strong contract satisfaction} and show how the proposed concept makes it possible to go from the local satisfaction of the contracts at the component's level to the satisfaction of the global STL specification at the interconnected system's level.
    \item At the level of the considered control objectives: When the objective is to synthesize controllers to enforce the satisfaction of AGCs for continuous-time systems, to the best of our knowledge, existing approaches in the literature 
    %makes it possible to 
    can only deal with the particular class of invariance AGCs\footnote{where the set of assumptions and guarantees of the contract are described by invariants.} in \cite{saoud2020contract}, where the authors used symbolic control techniques to synthesize controllers. In this paper, we present a new approach to synthesize controllers for a more general class of AGCs, where the set of assumptions and guarantees are described by STL formulae, by leveraging tools in the spirit of prescribed performance-based control.
    \item \bl{At the level of distributed control of STL tasks}: Compared to the results in \cite{lindemann2019feedback} which use 
    similar PPC-based strategies for the control of STL tasks for multi-agent systems,  
 %   Compared to the results in \cite{lindemann2019feedback} which use similar PPC-based strategies for the control of STL tasks for multi-agent systems,  
%However, the feedback control law presented in \cite{lindemann2019feedback} requires that all the neighboring agents share identical STL tasks. Instead, 
our proposed control law allows for distinct collaborative STL tasks over the agents, 
%thus being more applicable in real-world scenarios. 
whereas \cite{lindemann2019feedback} is restricted to identical STL tasks shared by all neighboring agents. 
A preliminary investigation of our results appeared in \cite{liu2022compositional}. Our results here improve and extend those in \cite{liu2022compositional} in the following directions. 
\bl{First, the compositional approach developed in \cite{liu2022compositional} is tailored to interconnected systems without communication or collaboration among the components, and thus can only deal with non-collaborative STL tasks. 
Here, we deal with general multi-agent systems with three types of interactions among agents, including dynamical couplings, communication/sensing, and task dependencies, which facilitate the handling of collaborative STL tasks. 
Second, different from the result in \cite{liu2022compositional}, we present two different compositionality results which show that weak satisfaction of contracts is sufficient to deal with acyclic interconnections (in terms of dynamical couplings) while strong satisfaction is needed to reason about general interconnections containing cycles.} 
%Finally, a distributed controller design approach is developed here that can handle collaborative STL tasks, while \cite{liu2022compositional} can only deal with non-collaborative ones.
\end{enumerate}

%local detection-and-repair scheme is proposed in possibly conflicting STL specifications  

%This arXiv paper is an extended version of the paper submitted to IEEE Control Systems Letters.

% 1. CPS are complex integrations, complex specifications including STL are hard to control. Large-scale CPS. 
% Abstraction-based techniques were proposed. Limitation: curse of dimensionality.

% 2. Assume guarantee contracts is a promising tool to handle this issue. Literature of AGC. All the papers till Adnane's Auto. 

% 3. In this paper, we propose for the first time a framework to handle the compositional synthesis of STL tasks using assume-guarantee contracts. 
% Introduce STL (the powerfulness of STL compared to other LTL tasks) and related literature for the synthesis of STL.	

% 4. Methodology of this paper: 
% continuous-time AGC and closed-form continuous feedback control laws. abstraction free

%\subsection{Motivations}
%
%
%\subsection{Our Contributions}
%
%\subsection{Related Works}
%
%\subsection{Organization}
\vspace{-0.3cm}
\section{Preliminaries and Problem Formulation}\label{sec:pre}

{\bf{Notation:}} 
%We denote the set of real, positive real, nonnegative real, and positive integer numbers by $\mathbb{R}$, $\mathbb{R}^+$, $\mathbb{R}_{\geq 0} $, and $\mathbb{N}$, respectively. 
We denote by $\R$ and $\N$ the set of real and non-negative integers, respectively. These symbols are annotated with subscripts to restrict them in the usual way, e.g., $\R_{>0}$ denotes the positive real numbers.
We denote by $\mathbb{R}^n$ an $n$-dimensional Euclidean space and by $\mathbb{R}^{n\times m}$ a space of real matrices with $n$ rows and $m$ columns. We denote by  $I_n$ the identity matrix of size $n$, and by $\mathbf{1}_n = [1,\dots,1]^\mathsf{T}$ the vector of all ones of size $n$. We denote by $\diag(a_1,\dots,a_n)$ the diagonal matrix with diagonal elements being $a_1,\dots,a_n$.
%Given an ordered pair $(x,y)$, we denote by $\overline{\textbf{Proj}}(x,y)$ the projection of the pair onto the first coordinates, i.e., $\overline{\textbf{Proj}}(x,y)=x$. 	Similarly, given a Cartesian product $X \times Y$ of sets $X$ and $Y$,  $\overline{\textbf{Proj}}(X \times Y)$ denotes the projection of the  Cartesian product onto the first coordinates, i.e., $\overline{\textbf{Proj}}(X \times Y)=X$.	
Consider sets $S_1, S_2, \ldots, S_n$.
We denote by $\prod_{i=1}^n S_i$ the Cartesian product of $S_1, S_2, \ldots, S_n$. For a set $K$, the cardinality of $K$ is denoted $|K|$.
For each $j \in\{1,2, \ldots, n\}$, the $j$th projection on $S=\prod_{i=1}^n S_i$ is the mapping $\operatorname{proj}^j: S \rightarrow S_j$ defined by: $\operatorname{proj}^j\left(s_1, s_2, \ldots, s_j, \ldots, s_n\right)=s_j$ for all $\left(s_1, s_2, \ldots, s_n\right) \in S$. Moreover, we further define $\overline{\textbf{Proj}}^j(S)=S_j$ for all $j \in\{1,2, \ldots, n\}$.

% Consider sets $S_1, S_2, \ldots, S_j, \ldots, S_n$.
% We denote by $\prod_{i=1}^n S_i$ the Cartesian product of $S_1, S_2, \ldots, S_n$.
% For each $j \in\{1,2, \ldots, n\}$, the $j$th projection on $S=\prod_{i=1}^n S_i$ is the mapping $\overline{\textbf{Proj}}^j: S \rightarrow S_j$ defined by: $\overline{\textbf{Proj}}^j\left(s_1, s_2, \ldots, s_j, \ldots, s_n\right)=s_j$ for all $\left(s_1, s_2, \ldots, s_n\right) \in S$.

\vspace{-0.25cm}

\subsection{Signal Temporal Logic (STL)} \label{STLsec}

Signal temporal logic (STL) is a predicate logic based on continuous-time signals, which consists of predicates $\mu$ that are obtained by evaluating a continuously differentiable predicate function $\mathcal{P}:\mathbb{R}^{n}\rightarrow \mathbb{R}$. Specifically,   
\blue{$\mathcal{P}$ is the predicate function associated with $\mu$ which assigns the respective true or false boolean value of $\mu$ as:}
$\mu:=\begin{cases} \top & \text{if}\ \mathcal{P}(x) \geq 0\\ \perp & \text{if}\ \mathcal{P}(x) < 0, \end{cases}$ for $x \in \mathbb{R}^{n}$, %For instance, consider the predicate $\mu:=(x\geqslant1)$, which can be expressed by $h(x): = x-1$. 
\bl{where $\top$ and $\perp$ denotes true and false, respectively.
We consider in this paper an STL fragment that is defined as} \bl{
\begin{align}\label{syntax1}
\psi &::= \top \mid   \mu \mid \neg \mu \mid \psi_1 \wedge \psi_2, \\ \label{syntax2}
\phi &::= G_{[a,b]}\psi \mid F_{[a,b]} \psi  \mid F_{[\underline{a},\underline{b}]}G_{[\bar{a}, \bar{b}]}\psi,
%\mid F_{[a,b]}G_{[a,b]}\psi
\end{align}}where $\mu$ is the predicate, $\psi$ in \eqref{syntax2} and $\psi_1, \psi_2$ in \eqref{syntax1} are formulae of class $\psi$ given in \eqref{syntax1}. 
The operators $\neg, \wedge, G_{[a,b]}, F_{[a,b]}$ denote the negation, conjunction, always, and eventually operators, respectively, with $a, b \in \mathbb{R}_{\geq 0}$ and $a \leq b$.
\blue{The meaning of these operators will be specified later by the definition of STL semantics. 
}
We refer to $\psi$ given in \eqref{syntax1} as non-temporal formulae, i.e., boolean formulae, while $\phi$ is referred to as temporal formulae due to the use of always- and eventually-operators.
\bl{We use $(\mathbf{x}, t) \models\phi$ to denote that the state trajectory $\mathbf{x}: \mathbb{R}_{\geq 0}  \rightarrow X \subseteq \mathbb{R}^{n}$ satisfies $\phi$  at time $t$, where $\models$ denotes the satisfaction relation.}
The trajectory $\mathbf{x}: \mathbb{R}_{\geq 0}  \rightarrow X \subseteq \mathbb{R}^{n}$ satisfying formula $\phi$ is denoted by $(\mathbf{x}, 0) \models\phi$.
\bl{The STL semantics \cite[Definition 1]{maler2004monitoring} of the fragment in \eqref{syntax1}-\eqref{syntax2} can be recursively given by: }
\begin{align*}
(\mathbf{x}, t)\models\mu &\iff \mathcal{P}(\mathbf{x}(t))\geq 0\\
(\mathbf{x}, t)\models\neg\phi &\iff \neg((\mathbf{x}, t)\models\phi)\\
 (\mathbf{x}, t)\models\phi_{1}\wedge\phi_{2} &\iff  (\mathbf{x}, t)\models \phi_{1}\wedge(\mathbf{x}, t)\models\phi_{2} \\
  (\mathbf{x}, t)\models F_{[a, b]}\phi &\iff  \exists \bar t \in [t+a, t+b]  \text{ s.t. } (\mathbf{x}, \bar t)\models \phi 
\end{align*}
The always-operator can be derived as $G_{[a, b]}\phi=\neg F_{[a, b]}{\neg\phi}$.

\bl{Next, we recall that STL is naturally equipped with a quantitative semantics, called robust semantics \cite[Def. 3]{donze2010robust} (also referred to as space robustness).
This quantitative semantics can be interpreted as ``how much (robustly) a signal $\mathbf{x}$ satisfies or violates a STL formula $\phi$". Formally, space robustness for the STL operators considered in this work is defined as:}
\begin{align}\notag
    \rho^{\mu}(\mathbf{x}, t)& := \mathcal{P}(\mathbf{x}(t))\\ \notag
\rho^{\neg\phi}(\mathbf{x}, t)&:=-\rho^{\phi}(\mathbf{x}, t)\\\notag
\rho^{\phi_{1}\wedge\phi_{2}}(\mathbf{x},  t)&: =\min(\rho^{\phi_{1}}(\mathbf{x},t), \rho^{\phi_{2}}(\mathbf{x}, t))\\\notag
\rho^{F_{[a, b]}\phi}(\mathbf{x}, t)&: = \max_{t_{1}\in[t+a, t+b]}\rho^{\phi}(\mathbf{x}, t_{1})\\ \notag
\rho^{G_{[a, b]}\phi}(\mathbf{x}, t)&:   = \min_{t_{1}\in[t+a, t+b]}\rho^{\phi}(\mathbf{x}, t_{1}) \\ \notag
\rho^{F_{[\underline{a},\underline{b}]}G_{[\bar{a}, \bar{b}]}\phi}(\mathbf{x}, t)&:   = \max_{t_{1}\in[t+\underline{a}, t+\underline{b}]}\min_{t_{2}\in[t_{1}+\bar{a}, t_{1}+\bar{b}]}\rho^{\phi}(\mathbf{x}, t_{2})
\end{align}
where \bl{$\rho^{\phi}(\mathbf{x}, t)$ are real-valued functions mapping from $X \times \mathbb{R}_{\geq 0}$ to $\mathbb{R}$ and are referred to as \emph{robustness functions}.} 
\blue{Note that every STL formula $\phi$ is equipped with a robustness function $\rho^{\phi}(\mathbf{x}, t)$ for which it holds that: $(\mathbf{x}, t)\models\phi$ if $\rho^{\phi}(\mathbf{x}, t) \!> \!{0}$~\cite[Proposition 16]{fainekos2009robustness}.}
The robustness functions will be employed in the controller design process to enforce STL satisfaction. 
Similarly to \cite{aksaray2016q}, the non-smooth conjunction can be approximated by smooth functions as $\rho^{\phi_{1}\wedge\phi_{2}}(\mathbf{x}, t)\approx- \frac{1}{\eta} \ln(\exp(-\eta\rho^{\phi_{1}}(\mathbf{x}, t))+\exp(-\eta\rho^{\phi_{2}}(\mathbf{x}, t)))$.
Note that $- \frac{1}{\eta} \ln(\exp(-\eta\rho^{\phi_{1}}(\mathbf{x}, t))+\exp(-\eta\rho^{\phi_{2}}(\mathbf{x}, t))) \leq \min(\rho^{\phi_{1}}(\mathbf{x},t), \rho^{\phi_{2}}(\mathbf{x}, t))$ holds for any choice of $\eta >0$, and the equality holds as $\eta \rightarrow \infty$. Thus, we have $(\mathbf{x}, t) \models \phi_1 \wedge \phi_2$ as long as $- \frac{1}{\eta} \ln(\exp(-\eta\rho^{\phi_{1}}(\mathbf{x}, t))+\exp(-\eta\rho^{\phi_{2}}(\mathbf{x}, t)))\!> \!0$.

Note that the considered STL fragment allows us to encode concave temporal tasks, which is a necessary assumption used later for the design of closed-form, continuous feedback controllers (cf. Assumption \ref{assmprho1}). \blue{It should be mentioned that by leveraging the results in e.g., \cite{lindemann2020efficient}, it is possible to  expand our results to full STL semantics. }   

\vspace{-0.25cm}
\subsection{{Multi-agent systems and interconnection topologies}} \label{subsec:mas}

Consider a team of $N \in \N$ agents $\Sigma_i$, $i \in I = \{1,\dots,N\}$. 
%study the interconnection of finitely many continuous-time control agents. 
%Here, let us first introduce some notations from graph theory for the definition of interconnected systems. 
%Consider a network consisting of $N \in \N$ control agents $\Sigma_i$, $i \in I = \{1,\dots,N\}$. 
%We denote the interconnection topology of the network by the directed graph $\G = (I, \E)$, where $I = \{1,\dots,N\}$ is the set of vertices with each vertex $i \in I$ labeled with agent $\Sigma_i$, and $\E \subseteq I \times I$ is the set of ordered pairs $(i,j)$, $i,j \in I$, which represents a binary connectivity relation in the directed graph, 
%Given a multi-agent system $\Sigma$ with $N \in \N$ agents $\Sigma_i$, 
Each agent $\Sigma_i$ 
%is a continuous-time control system, 
%the definition of which is formally given as follows.
%\begin{definition}  
%	\label{agent} (Continuous-time control system)
%An agent $\Sigma_i$ 
is a tuple $\Sigma_i = (X_i,U_i,W_i,f_i,g_i,h_i)$, where 
\begin{enumerate} 
\item $X_i = \mathbb{R}^{n_i}$, $U_i =\mathbb{R}^{m_i}$ and $W_i = \mathbb{R}^{p_i}$ are the state, external input, and internal input spaces, respectively;
%\item $f : X \times U \times W \rightarrow X$ is a continuous map satisfying the following Lipschitz assumption: for every $x \in X$, there exists a neighborhood $\mathcal{D}$ of $x$ such that $\Vert f(x_1,u_1,w_1) - f(x_2,u_2,w_2)\Vert \leq L \Vert x_1 - x_2\Vert$  for all $x_1,x_2 \in \mathcal{D}$, all $u \in U$, and all $w_1, w_2 \in W$; (locally Lipschitz)
\item $f_i:\mathbb{R}^{n_i} \rightarrow \mathbb{R}^{n_i}$ is the flow drift, $g_i:\mathbb{R}^{n_i} \rightarrow \mathbb{R}^{n_i\times m_i}$ is the external input matrix, and $h_i: \mathbb{R}^{p_i} \rightarrow  \mathbb{R}^{n_i}$ is the internal input map. 
\end{enumerate} 
A trajectory of $\Sigma_i$ is \blue{a} \bl{uniformly continuous} map $(\mathbf{x}_i,\mathbf{w}_i)\!:\!\mathbb{R}_{\geq 0} \!\rightarrow\! X_i \!\times\!W_i$ such that 
%there exists an external input trajectory $\mathbf{u}_i\!:\! \mathbb{R}_{\geq 0} \!\rightarrow\! U_i$ such that 
for all $t \!\geq\! 0$ 
\begin{equation}
\label{eqn:subsys}
    \mathbf{\dot x}_i(t)  = f_i(\mathbf{x}_i(t))+ g_i(\mathbf{x}_i(t))\mathbf{u}_i(t)+h_i(\mathbf{w}_i(t)),
\end{equation}
where $\mathbf{u}_i\!:\! \mathbb{R}_{\geq 0} \!\rightarrow\! U_i$ is the external input trajectory, and $\mathbf{w}_i :\!\mathbb{R}_{\geq 0} \rightarrow W_i$ is the internal input trajectory. 
%\end{definition}
Note that $u_i \in U_i$ are ``external" inputs served as interfaces for controllers, and $w_i  \in  W_i$ are termed as ``internal" inputs describing the physical interaction between agents which may be unknown to agent $\Sigma_i$. 
\bl{
Note that the considered nonlinear control affine systems as in \eqref{eqn:subsys} have been extensively studied in nonlinear control theory \cite{sastry2013nonlinear,isidori2013nonlinear}. 
Control affine systems with drift terms are very general, since they can model a wide range of real-world physical systems, including various types of vehicle dynamics and control systems, e.g., dynamical systems subject to underactuation or nonholonomic motion constraints.
}

\begin{assumption}\label{assmp:lipschitz}
Consider agent $\Sigma_i$ as in \eqref{eqn:subsys}.
The functions $f_i : \mathbb{R}^{n_i} \rightarrow  \mathbb{R}^{n_i}$, $g_i  : \mathbb{R}^{n_i} \rightarrow  \mathbb{R}^{n_i \times m_i}$, and $h_i : \mathbb{R}^{p_i} \rightarrow  \mathbb{R}^{n_i}$ are locally Lipschitz continuous, and $g_i(\mathbf{x}_i)g_i(\mathbf{x}_i)^\top$ is positive definite for all $\mathbf{x}_i \in \mathbb{R}^{n_i}$.
\end{assumption}
\begin{remark}
\bl{We assume that the functions that appeared in the system dynamics \eqref{eqn:subsys} are locally Lipschitz continuous, which will be used later in Theorem \ref{theorem} to ensure the boundedness of $f_i(\mathbf{x}_i), g_i(\mathbf{x}_i), h_i(\mathbf{x}_i)$ in bounded domains.
Note that any continuously differentiable function is locally Lipschitz. 
Remark that $g_i(\mathbf{x}_i)g_i(\mathbf{x}_i)^\top$ is positive definite if and only if $g_i(\mathbf{x}_i)$ has full row rank, which implies that $m \geq n$.}
\blue{Note that in the case where $m = n$, it implies that the system is fully actuated. 
This assumption captures for instance the dynamics of omnidirectional robots or room temperature regulation as in Sec. \ref{sec: case}, and other practical control applications including robotic manipulators with one actuator per degree of freedom.}
\end{remark}

A multi-agent system consisting of $N \in \N$ agents $\Sigma_i$ is formally defined as follows.

\begin{definition}\label{intersys} (Multi-agent system)
Given $N \in \N$  agents $\Sigma_i$, $i \in \{1,\ldots,N\}$, as described in \eqref{eqn:subsys}. A multi-agent system denoted by $\Sigma=\mathcal{I}(\Sigma_1,\dots,\Sigma_N)$ %and equipped with a directed graph $\G = (I, \E)$, 
is a tuple $\Sigma = (X,U,f,g)$ where 
\begin{enumerate}
\item $X = \prod_{i\in I} X_i$ and $U = \prod_{i\in I} U_i$ are the state and external input spaces, respectively;
%\item $f(x,u) \Let [f_1(x_1,u_1,w_1);\dots;f_N(x_N,u_N,w_N)]$ where $x = [x_1;\dots;x_N]$, and $u =[u_1;\dots;u_N]$, with internal inputs constrained by $w_i = [x_{j_1},\dots,x_{j_p}]$, where $\n_i = \{j_1,\dots,j_p\}$, for all $i \in I$;
\item $f:\mathbb{R}^{n} \rightarrow \mathbb{R}^{n}$ is the flow drift and $g:\mathbb{R}^{n} \rightarrow \mathbb{R}^{n\times m}$ is the external input matrix defined as : 
$f(\mathbf{x}(t)) =  [f_1(\mathbf{x}_1(t))+h_1({\mathbf{w}_1(t)});
      \dots; f_N(\mathbf{x}_N(t))+h_N({\mathbf{w}_N(t)})]$, $
    g(\mathbf{x}(t)) =  \diag(g_1(\mathbf{x}_1(t)),$  $\dots, g_N(\mathbf{x}_N(t))),$ 
where $n=\sum_{i\in I}n_i$, $m=\sum_{i\in I}m_i$, $\mathbf{x} = [\mathbf{x}_1;\dots;\mathbf{x}_N]$,  $\mathbf{w}_i(t) =[\mathbf{x}_{j_1}(t);\dots;\mathbf{x}_{j_{|\n^a_i|}}(t)]$, for all $i \in I$, where $\n^a_i$ denotes the set of agents providing internal inputs to  $\Sigma_i$.
\end{enumerate}
A trajectory of $\Sigma$ is then \blue{a}  \bl{uniformly continuous} map
$\mathbf{x}\!:\!\mathbb{R}_{\geq 0} \!\rightarrow\! X$
such that for all $t \geq 0$, 
$\mathbf{\dot x}(t)  = f(\mathbf{x}(t))+ g(\mathbf{x}(t))\mathbf{u}(t)$,
where $\mathbf{u} = [\mathbf{u}_1;\dots;\mathbf{u}_N]$ is the external input trajectory.
\end{definition}

\vspace{-0.25cm}

\subsection{\bl{Interaction topologies among agents}}

\bl{Note that the interaction topology plays an important role in the analysis and synthesis of networked multi-agent systems \cite{mesbahi2010graph}.
In this paper, we consider a general setup in the synthesis of multi-agent systems by considering three types of interactions among agents, which are determined by the \emph{dynamical interconnection topology}, 
\emph{communication topology}, and \emph{task dependency topology}, respectively. Throughout the paper, we will use the graph-based representations of these network topologies, as described in the following.
}

\bl{\bf{Dynamical interconnection graph $\mathcal{G}^a=(I,E^a)$}:} 
In this paper, we consider the existence of physical interactions in terms of unknown dynamical couplings between agents, captured by $h_i(\mathbf{w}_i)$ as in \eqref{eqn:subsys}. 
%Therefore, for each agent $i \in I$, the neighboring agents inducing dynamical couplings with agent $i \in I$ are regarded as its \emph{adversarial} neighbors, denoted by $\n^a_i$. 
We denote by a directed graph $\mathcal{G}^a=(I,E^a)$ as the \emph{interconnection graph} capturing the dynamical couplings among the $N$ agents, where $I = \{1,\dots,N\}$, and $(j,i)\in E^a$, indicate that agent $j$ provides internal inputs to agent $i$ through dynamical couplings in \eqref{eqn:subsys}. 
Thus, we formally have $\n^a_i = \{ j \in I | (j,i) \in E^a\}$.
Given an interconnection graph $\mathcal{G}^a$, we define $I^{Init} = \{i \in I |\n^a_i = \emptyset\}$ as the set of agents who do not have any adversarial neighbors that impose unknown dynamical couplings on them. 
Note that in Definition \ref{intersys}, the interconnection structure indicates that the internal input  $\mathbf{w}_i$ of an agent $\Sigma_i$ is the stacked state $[\mathbf{x}_{j_1};\dots;\mathbf{x}_{j_{|\n^a_i|}}]$ of its adversarial neighbors $\Sigma_j$, $j \in \n^a_i$. 
%Note that the definition of a multi-agent system boils down to the tuple $\Sigma = (X,U,f,g)$ since it has null internal inputs.

\bl{\bf{Communication graph $\mathcal{G}^c=(I,E^c)$}:} 
We denote by an undirected graph $\mathcal{G}^c=(I,E^c)$ the \emph{communication graph} among the $N$ agents, where $I = \{1,\dots,N\}$, $(i,j)\in E^c$, $\forall i,j \in I$ are unordered pairs indicating communication links between  agents $i$ and $j$. \bl{The existence of a communication link between agents $i$ and $j$ reflects the fact that these two agents can exchange information directly through communication channels or active sensing, and thus the control input of agent $i$ can depend on the state of agent $j$ and vice versa.}

\bl{\bf{Task dependancy graph $\mathcal{G}^t = (I, E^t)$}:} 
We consider a multi-agent system  subject to a global STL specification $\bar \phi$, which is decomposed into local ones $\phi_i$, $i = 1, \dots, N$ \footnote{\bl{By ``global" and ``local" STL specifications, we mean the tasks defined for the entire network and individual agents, respectively.}}.
%Each agent $\Sigma_i$ is subject to a STL formula $\phi_i$ of class $\phi$ given as in \eqref{syntax2}. 
\bl{Note that the satisfaction of task $\phi_i$ do not only depend on the behavior of $\Sigma_i$, but may also depend on the behavior of other agents $I\setminus \{i\}$} \footnote{ \bl{By saying that the satisfaction of a task $\phi_i$ ``depends" on the behavior of an agent $\Sigma_j$, we mean that the predicate function $\mathcal{P}_i$ of $\phi_i$ (as defined in Sec.~\ref{STLsec} before \eqref{syntax1}) is a function of $\mathbf{x}_j$.}}. 
\bl{
%We say that $\phi_i$ depends on $\Sigma_j$,  $j \in I\setminus \{i\}$,  
If the satisfaction of $\phi_i$ depends on the behavior of more than one agent, $\phi_i$ is referred to as a \emph{collaborative task}.} If $\phi_i$ only depends on one agent $\Sigma_i$, $\phi_i$ is called a \emph{non-collaborative task}. 
\bl{To characterize the task dependencies in the multi-agent system,  we define the \emph{task dependency graph} as $\mathcal{G}^t = (I, E^t)$, which is a directed graph with} 
$I = \{1,\dots,N\}$,  and $(i,j)\in E^t$ if and only if the formula $\phi_i$ depends on $\Sigma_j$. 

\begin{assumption}\label{assump:communication}
For each $\phi_i$ depending on agent $\Sigma_j$, $j \in I$, agent $\Sigma_i$ can communicate with  $\Sigma_j$, i.e., we have $\mathcal{G}^t \subseteq \mathcal{G}^c$ with $(i, j) \!\in\! E^t \!\implies\! (i, j) \!\in \! E^c$.
\end{assumption}
Assumption \ref{assump:communication} implies that the task dependencies are compatible with the communication graph topology of the multi-agent system. 
\bl{If this assumption does not hold, one can leverage communication-aware task decomposition techniques, such as the one proposed in our recent work \cite{marchesini2024communication}, to decompose the global STL task such that the resulting task dependency graph is compatible with the given communication topology.}

%In this work, we are interested in the \emph{cycles} in the interconnection graph $\mathcal{G}^a=(I,E^a)$ of multi-agent systems. 
%Note that cycles in a graph are paths that run along the edges in a graph which end where they started without any other vertices being repeated. Formally, 
Note that a path is a sequence $v_1v_2\dots v_m$ of nodes in the graph, such that $(v_i, v_{i+1}) \in E$, $\forall i \in \{1,\dots, m-1\}$, i.e., every consecutive pair of nodes is an edge in the graph. A cycle is a path $v_1v_2\dots v_m$ where $v_m = v_1$ and the vertices $v_1, \dots,v_{m-1}$ are distinct. A directed acyclic graph (DAG) is a directed graph with no directed cycles \cite{gross2005graph}. 

The following assumption \blue{will be used only in Theorem \ref{theorem} for the design of the distributed control law}.
\begin{assumption}\label{assmp:task}
%Consider the task dependency graph $\mathcal{G}^t = (I, E^t)$ of the multi-agent system. 
%For each cluster of agents $\bar \Sigma_k = \mathcal{I}(\Sigma_{k_1},\dots,\Sigma_{k_{|I_k|}})$, where each agent $\Sigma_{k_i}$ is subject to STL task $\phi_{k_i}$, $i \in \{1, \ldots, |I_k|\}$.
The task dependency graph  $\mathcal{G}^t = (I, E^t)$ of the multi-agent system is a directed acyclic graph.
%, i.e., there is no directed cycles in $\mathcal{G}^t$.}
%For all $i , i' \in \{1, \ldots, |I_k|\}$ with $i > i'$, it holds that $(i, i') \notin E^t$.
%$\phi_i$ does not depend on $\Sigma_j$.
%$(i, i') \notin E^t$. 
\end{assumption}
\blue{Note that although Assumption \ref{assmp:task}
imposes the task dependencies to follow a tree structure, one can leverage existing task decomposition or assignment approaches to rewrite cyclic task graphs into acyclic ones under mild assumptions \cite{marchesini2024communication}.  
}

\vspace{-0.2cm}

\subsection{{Problem statement}} \label{sec:statement}

We now have all the ingredients to provide a formal statement of the problem considered in the paper:

\begin{problem}
\label{pro:1}
Consider a multi-agent system $\Sigma\!=\!(X,$ $U,f,g)$ as in Definition \ref{intersys}, consisting of $N$ agents $\Sigma_i$, $i \in \{1,\ldots,N\}$, and given a  global STL specification 
$\bar \phi$, where $\bar \phi=\land_{i=1}^N\phi_i$ and $\phi_i$ are  local  STL tasks of the form in \eqref{syntax2}. Synthesize local controllers $\mathbf{u}_i$ for agents $\Sigma_i$ such that $\Sigma$ satisfies the specification $\bar \phi$. 
%\PJ{I guess we are not providing solution for any STL...we can only deal with a fragment of STL, right?}
\end{problem}

In the remainder of the paper,  
we first introduce the concepts of assume-guarantee contracts and contract satisfaction in Section~\ref{subsec:agc}. Based on these notions of contracts, our main compositionality results are presented in Sections \ref{subsec:compo1} and \ref{subsec:compo2}, which are used to tackle acyclic and cyclic interconnection graphs, respectively.
Then, we show that STL tasks can be casted as prescribed performance functions in Section \ref{subsec: ppc}, and then formulated as assume-guarantee contracts in Section  \ref{subsec:stl_to_agc}. 
% Then, we present our main compositionality result based on a notion of assume-guarantee contracts as in Section \ref{subsec:agc}, which allows us to tackle the synthesis problem in a distributed fashion.
%We will further explain in Section \ref{subsec:stl_to_agc} on how to assign assume-guarantee contracts tailored to the funnel-based formulation of STL tasks.
A closed-form continuous-time control law will be derived in Section \ref{sec:localcontroller} to enforce local contracts over agents in a distributed manner. 

\vspace{-0.2cm}
\section{Assume-Guarantee Contracts and Compositional Reasoning}\label{sec: agc}
In this section, we present a compositional approach based on assume-guarantee contracts which enables us to reason about the properties of a multi-agent system based on the those of its components.
We will first introduce the new notions of assume-guarantee contracts and contract satisfations for multi-agent systems. Then, two compositionality results will be presented which are tailored to multi-agent systems with acyclic and cyclic interconnection topologies, respectively.

\bl{In the next subsection, we first split the agents into clusters induced by the task dependency topology among the agents. Clusters will be leveraged to define assume-guarantee contracts representing the desired collaborative STL tasks. 
%compositional result using assume-guarantee contracts. 
}

\vspace{-0.2cm}
\subsection{{Clusters induced by the task dependencies}}\label{subsec:task}

Consider a multi-agent system with task dependency graph $\mathcal{G}^t = (I, E^t)$. We say that $I' \subseteq I$ is a  \emph{maximal dependency cluster} \cite{guo2013reconfiguration} \bl{if $I'$ is a \emph{weakly connected component} \cite{mesbahi2010graph} in $\mathcal{G}^t$}, i.e., 
$\forall i, j \in I'$, $i$ and $j$ are connected\footnote{Here, $i$ and $j$ are said to be connected if there is an \emph{undirected path} $i, k_1, \ldots, k_m, j$ such that: $(i, k_1)\in E^t$  and/or $(k_1,i) \in E^t$ hold; $(k_1, k_2)\in E^t$  and/or $(k_1, k_2) \in E^t$ hold;$\ldots$; $(k_m, j)\in E^t$  and/or $(k_m, j) \in E^t$ hold.} $\mathcal{G}^t$ and $\nexists i \in I', i' \in I \setminus I'$ such that $i$ and $i'$ are connected.
Thus, a multi-agent system under tasks $\{\phi_1, \dots, \phi_N\}$ induces $K \leq N$ maximal dependency clusters $I_k$, $k \in \bar I = \{1,\dots,K\}$.
Note that these clusters are maximal, i.e., there are no task dependencies between different clusters.

\begin{example}\label{eg2}
\begin{figure}[!t]
	\centering
	\includegraphics[width=.15\textwidth]{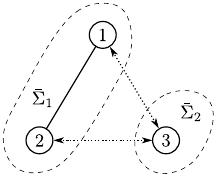}
 	\caption{A multi-agent system with two clusters $I_1 = \{1,2\}$ and $I_2 = \{3\}$.\bl{The solid lines indicate the communication links between agents, and the dotted lines with arrows indicate the dynamical couplings between clusters.}} \label{fig:toynet}  
  \vspace{-0.7cm}
\end{figure}
  Consider a multi-agent system of three agents $\Sigma_i$, $i \in \{1,2,3\}$, as shown in Fig. \ref{fig:toynet}. 
  Agent $\Sigma_1$ is subject to a collaborative STL task $\phi_1 := G_{[0,10]} (\Vert \mathbf{x}_1 - \mathbf{x}_2\Vert \leq 3)$, i.e., agents 1 and 2 should always stay close to each other by $3$ within time interval $[0,10]$; agent $\Sigma_2$ is subject to a non-collaborative STL task $\phi_2 := F_{[5,10]} (\Vert \mathbf{x}_2 - [50,20]^\mathsf{T}\Vert \leq 5)$, and agent $\Sigma_3$ is subject to a non-collaborative STL task $\phi_3 := G_{[0,10]} (\Vert \mathbf{x}_3 - [20,80]^\mathsf{T}\Vert \leq 10)$.   
  The task dependency graph of the multi-agent system thus induces two maximal dependency clusters $I_1 = \{1,2\}$ and $I_2 = \{3\}$.  
\end{example}

\bl{Let us denote a cluster by $\bar \Sigma_k=\mathcal{I}(\Sigma_{k_1},\dots,\Sigma_{k_{|I_k|}}) = (\bar X_k,\bar U_k,\bar W_k,\bar f_k,\bar g_k,\bar h_k)$,
which results from the interconnection of agents $\Sigma_i$, $i \in I_k= \{{k_1},\ldots,{k_{|I_k|}}\}$, $k \in \{1,\dots,K\}$, induced by the task dependency graph. 
The formal definition of maximal dependency clusters is provided in Definition \ref{def:cluster} in the appendix for completeness.  
A cluster's dynamics capture the collaborative behavior of all the agents involved within the same group whose tasks depend on each other. }
We define by $\n_i^c = I_k \setminus \{i\}$ 
the set of \emph{cooperative neighbors} of each agent $\Sigma_i$,  $i \in I_k$, and the \emph{cooperative
internal input} by the stacked state $\mathbf{z}_i(t) =[\mathbf{x}_{j_1}(t);\dots;\mathbf{x}_{j_{\bl{|I_k|-1}}}(t)]$, with $Z_i = \prod_{j\in \n_i^c} X_j$. 
\blue{For the sake of simplicity, we assume that $\n_i^a \cap \n_i^c = \emptyset$, which implies that there are no dynamical couplings among agents in the same cluster. However, the proposed results can be readily extended to handle the case that $\n_i^a \cap \n_i^c \neq \emptyset$ by deriving one more layer of compositional reasoning for each cluster; see detailed discussions at the end of Sec. \ref{subsec:agc}.} 
Note that, since an agent can only belong to one cluster, it can be shown easily that the interconnection of all the clusters forms the same multi-agent system as the interconnection of all the single agents as in Definition \ref{intersys}, i.e., $\mathcal{I}(\Sigma_1,\!\dots,\Sigma_N) \!=\! \mathcal{I}(\bar \Sigma_1,\!\dots,\bar \Sigma_K)$ where $\bar \Sigma_k \!=\! \mathcal{I}(\Sigma_{k_1},\!\dots,\Sigma_{k_{|I_k|}})$ are the clusters.

% Define here the mutual specification of a cluster of agents $I_k$ as:
% $$
% \bar \phi_k =\bigwedge_{i \in I_k} \phi_i, k \in \{1,\dots,K\}.
% $$

We further define for each cluster $\bar \Sigma_k$ the STL formula as $\bar \phi_k = \land_{i=1}^{I_k} \phi_i$.
If the satisfaction of $\bar \phi_k$ for each cluster $I_k$ is guaranteed, it holds by definition that the satisfaction of all the individual formulae $\phi_i, i \in I_k$ is guaranteed as well.
% the multi-agent system is subject to an STL formula $\phi$ which is decomposed into $K \geq 0$ STL formulas $\phi_1, \dots, \phi_K$ of the form in \eqref{syntax2}, i.e., $\phi=\land_{k=1}^K\phi_k$. 
%Each agent $\Sigma_i$ is subject to a local STL formula $\phi_i$ of the form in \eqref{syntax2}. 
As defined earlier, the satisfaction of $\bar\phi_k$, $k \in \{1,\dots,K\}$, depends on the set of agents $I_k \subseteq I$.  Note that we have by the definition of maximal dependency cluster that $I_1 \cup \dots \cup I_K = I$ and $I_{k} \cap I_{k'} = \emptyset$ for all $k, k' \in  \{1,\dots,K\}$ with $k \neq k'$. 
Induced by the task dependency graph, 
the global STL specification can be written as $\bar \phi =\land_{k=1}^K \bar\phi_k$. 
%Note that each set of agents $I_k$ is a maximal dependency cluster as defined above. 
Although there are no task dependencies between agents in different clusters $I_{k}$ and $I_{k'}$, these agents might be dynamically coupled as shown in each agent's dynamics \eqref{eqn:subsys} based on the interconnection graph $\mathcal{G}^a$.

As mentioned in Subsection \ref{subsec:mas}, we defined the \emph{interconnection graph}  $\mathcal{G}^a=(I,E^a)$ for the multi-agent system with each agent $\Sigma_i$ being a vertex in the graph. 
Fur future use, we further denote by a directed graph $\bar{\mathcal{G}}^a=(\bar I,\bar E^a)$ as the \emph{cluster interconnection graph} capturing the dynamical couplings among the $K$ clusters in the multi-agent system, where $\bar I = \{1,\dots, K\}$, and $(j,i)\in \bar E^a$ indicate that cluster $I_j$ provides internal inputs to cluster $I_i$, i.e., $\exists j_l \in I_j, i_{l'} \in I_i$ such that  $(j_l,i_{l'})\in E^a$. 

 \vspace{-0.2cm} 

%Before introducing the compositionality results, 
 
\subsection{{Assume-guarantee contracts for multi-agent systems}} \label{subsec:agc}

In this subsection, we introduce a notion of continuous-time assume-guarantee contracts to establish our compositional framework. A new concept of contract satisfaction is defined which is tailored to the PPC-based formulation of STL specifications as discussed in Subsection \ref{subsec: ppc}.

%Now, let us provide the definition of assume-guarantee contracts and a new concept of contract satisfaction for continuous-time systems. 
%
%\begin{figure}[b]
%	\centering
%	\begin{subfigure}[b]{0.45\textwidth}
	%		\centering
	%	\includegraphics[width=\textwidth]{from_wts1.pdf}	
	%		\caption{}
	%		%\vspace{-0.2cm}
	%		\label{fig:traj}
	%	\end{subfigure}
%\hfill
%	\begin{subfigure}[b]{0.45\textwidth}
	%		\centering
	%		\includegraphics[width=\textwidth]{contract_strong3.pdf}
	%		\caption{}
	%		%\vspace{-0.2cm}
	%		\label{fig:barrier}
	%	\end{subfigure}
%\caption{Weakly satisfaction $\Sigma \models \mathcal{C}$ v.s. strongly satisfaction $\Sigma \models_{us} \mathcal{C}$}
%\end{figure}

\begin{definition} (Assume-guarantee contracts) \label{asg}
	Consider an agent
	$\Sigma_i \!=\! (X_i,U_i,W_i,f_i,g_i,h_i)$, and the cluster $\bar \Sigma_k$ that $\Sigma_i$ belongs to as in Definition \ref{def:cluster}.  An assume-guarantee contract for $\Sigma_i$ is a tuple $\C_i \!=\! (A_i^a, G_i)$ where
	\begin{enumerate}
		\item $A_i^a: \mathbb{R}_{\geq 0}  \rightarrow  W_i = \prod_{j\in \n_i^a} X_j$ is a set of assumptions on the adversarial internal input trajectories, i.e., the state trajectories of its adversarial neighboring agents; % \prod_{j\in \n_i^a} X_j
%		\item $A_i^c: \mathbb{R}_{\geq 0}  \rightarrow  Z_i = \prod_{j\in \n_i^c}		X_j$ is a set of assumptions on the cooperative internal input trajectories, i.e., the state trajectories of its cooperative neighboring agents;	
		\item 
  %$G_i: \mathbb{R}_{\geq 0}  \rightarrow  \bar X_k = X_i \times \prod_{j\in I_k \setminus \{i\}} X_j$
  $G_i: \mathbb{R}_{\geq 0}  \rightarrow  \bar X_k = X_{k_1} \times \ldots  \times X_i \times \ldots \times \blue{X_{k_{\left|I_k\right|}}}$  
%  \AS{$G_i: \mathbb{R}_{\geq 0}  \rightarrow  P(\bar X_k,i)$, where the map $P(\bar X_k,i)$ is defined for $k \in \{1,\ldots,K\}$ and $i \in I_k$ as follows: $P(\bar X_k,i)= X_i \times X_{k_1} \times \ldots X_{i-1} \times X_{i+1}  \times \ldots \times X_{I_k}$}
  is a set of guarantees on the state trajectories and cooperative internal input trajectories.
	\end{enumerate}

We say that $\Sigma_i$ (\emph{weakly}) satisfies $\mathcal{C}_i$, denoted by $\Sigma_i \models \mathcal{C}_i$, if for any trajectory  $(\mathbf{x}_i,\mathbf{w}_i)\! : \! \mathbb{R}_{\geq 0}\! \rightarrow\! X_i  \!\times\! W_i$ of $\Sigma_i$, the following holds:
for all $t \in \mathbb{R}_{\geq 0}$, if $\mathbf{w}_i(s) \!\in\! A_i^a(s)$ for all $s \in [0,t]$, then we have $(\mathbf{x}_i(s), \mathbf{z}_i(s))\in G_i(s)$\footnote{From here on, we slightly abuse the notation by denoting $(\mathbf{x}_i(s), \mathbf{z}_i(s))$ as the stacked vector $[\mathbf{x}_{j_1}(s);\dots;\mathbf{x}_{j_{|I_k|}}(s)] \in \bar X_k $ for the sake of brevity.} for all $s \in [0,t]$. 
%\AS{then we have $\bar{\mathbf{x}}_k(s) \in G_i(s)$ for all $s \in [0,t]$}

% 2. We say that $\Sigma_i$ (\emph{weakly}) satisfies $\mathcal{C}_i$, denoted by $\Sigma_i \models \mathcal{C}_i$, if for any trajectory  $(\mathbf{x}_i,\mathbf{w}_i)\! : \! \mathbb{R}_{\geq 0}\! \rightarrow\! X_i  \!\times\! W_i$ of $\Sigma_i$, the following holds:
% there exists $\delta_i>0$ such that:

% for all $t \geq \delta_i$, $\mathbf{w}_i(s) \!\in\! A_i^a(s)$ for all $s \in [0,t]$, and $\mathbf{z}_i(s) \!\in\! A_i^c(s)$ for all $s \in [0,t-\delta_i]$, we have $(\mathbf{x}_i(s), \mathbf{z}_i(s))\in G_i(s)$ for all $s \in [0,t]$; 

% for all $t < \delta_i$, $\mathbf{w}_i(s) \!\in\! A_i^a(s)$ for all $s \in [0,t]$, and $\mathbf{z}_i(0) \!\in\! A_i^c(0)$, we have $(\mathbf{x}_i(s), \mathbf{z}_i(s))\in G_i(s)$ for all $s \in [0,t]$.

We say that $\Sigma_i$ \emph{uniformly strongly} satisfies $\mathcal{C}_i$, denoted by $\Sigma_i \models_{us} \mathcal{C}_i$, if for any trajectory $(\mathbf{x}_i,\mathbf{w}_i) \!:\! \mathbb{R}_{\geq 0}  \!\rightarrow\! X_i  \!\times\! W_i$ of $\Sigma_i$, the following holds: 
there exists $\delta_i>0$ such that for all $t \in \mathbb{R}_{\geq 0}$, if $\mathbf{w}_i(s) \!\in\! A_i^a(s)$ for all $s \in [0,t]$, then we have 
 %$\bar{\mathbf{x}}_k(s)\in G_i(s)$ 
 $(\mathbf{x}_i(s), \mathbf{z}_i(s))\in G_i(s)$ 
 for all $s \in [0,t+\delta_i]$.
 %\AS{then we have $\bar{\mathbf{x}}_k(s) \in G_i(s)$ for all $s \in [0,t+\delta_i]$.}
\end{definition}

\begin{figure}[!t]
	\centering
	
	\subcaptionbox{Weak satisfaction.\label{fig:weak}}
	{\includegraphics[width=.16\textwidth]{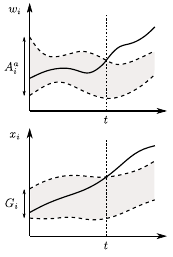}}
 \hspace{0.5cm}
	\subcaptionbox{Strong satisfaction.\label{fig:strong}}
	{\includegraphics[width=.16\textwidth]{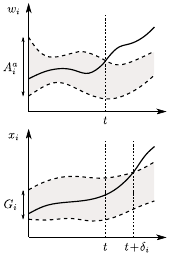}}
	\caption{An illustration of weak and strong satisfaction of a contract.  Top: trajectories of
the internal inputs $w_i$. Bottom: trajectories of the state $x_i$. \bl{Intuitively, the weak (or strong) satisfaction states that if the assumptions $A_i^a$ are satisfied up to time $t \in \mathbb{R}_{\geq 0}$, the system's state belongs to  $G_i$ at least until time $t$ ($t+\delta_i$ with $\delta_i >0$ in the case of strong satisfaction)}.
} 
\label{fig:contractsatisfaction}
 \vspace{-0.7cm}
\end{figure}

\blue{
Intuitively, an assume-guarantee contract in Definition \ref{asg} states that if the restriction on the adversarial internal input trajectories to the agent belongs to $A_i^a$, then the restriction of the state trajectories of the agent belongs to $G_i$.
}
Note that $\Sigma_i \models_{us} \mathcal{C}_i$ implies $\Sigma_i \models \mathcal{C}_i$. An illustration of weak and strong satisfaction of a contract is depicted in Fig.~\ref{fig:contractsatisfaction}.
\blue{We say that the assumption set $A_i^a$ (or guarantee set $G_i$) of a contract $\C_i \!=\! (A_i^a, G_i)$ is \emph{closed} if for all $t \in \mathbb{R}_{\geq 0}$, $A_i^a (t)$ (or $G_i (t)$) is closed, 
i.e., each $A_i^a (t)$ (or $G_i (t)$) contains all of its limit points; 
otherwise, it is \emph{open}.}
As it can be seen in the above definition, the set of guarantees $G_i$ for each agent $\Sigma_i$ is on both the state trajectories of $\Sigma_i$ and the cooperative internal input trajectories, i.e.,
the state trajectories of its cooperative neighbors $\Sigma_j$, $j \in \n_i^c$. For later use, let us define %$\overline{\textbf{Proj}}^i(G_{i}) : \mathbb{R}_{\geq 0}  \rightarrow  X_i$ such that 
$\overline{\textbf{Proj}}^i(G_{i}(t)):=\{x_i\in X_i | \exists x_{k_j} \in X_{k_j}, \forall {k_j} \in I_k, \text{s.t. } [x_{k_1};\ldots;x_i;\ldots;x_{k_|I_k|}]\in G_{i}(t)\}$.
%\AS{For $G \subseteq \bar X_k$ and $i \in I_k$ we define $\overline{\textbf{Proj}}_i(G):=\{x_i\in X_i | \exists x_{k_j} \in X_{k_j}, {k_j} \in I_k, \text{s.t. } [x_{k_1};\ldots;x_i;\ldots;x_{k_|I_k|}]\in G\}$}
%where for all $t \in \mathbb{R}_{\geq 0}$, is the $i_{th}$ projection of the set. $\overline{\textbf{Proj}}(Y):=\{y\in \mathbb{R}^n | \exists \hat{y}\in\mathbb{R}^n, \text{s.t. } [y;\hat{y}]\in Y\}$

It should be mentioned that multi-agent systems have no assumptions on internal inputs since they have trivial null internal input sets as in Definition \ref{intersys}. Hence, an AGC for a multi-agent system $\Sigma = \mathcal{I}(\Sigma_1,\dots,\Sigma_N)$ will be denoted by $\mathcal{C}=(\emptyset, G)$ with $G \subseteq X$.  
The concept of contract satisfaction by a multi-agent system $\Sigma$ is similar to those for the single agents by removing the conditions on internal inputs:
We say that $\Sigma$ (\emph{weakly}) satisfies $\mathcal{C}$, denoted by $\Sigma \models \mathcal{C}$, if for any trajectory  $\mathbf{x}\! : \! \mathbb{R}_{\geq 0}\! \rightarrow\! X $ of $\Sigma$, and for all $s \in \mathbb{R}_{\geq 0} $, we have $\mathbf{x}(s)\in G(s)$.
% Also note that  interconnected systems have trivial null assumptions on internal inputs. Hence, a contract for an interconnected system $\mathcal{I}(\Sigma_1,\dots,\Sigma_N)$ will be denoted by $\mathcal{C}=(\emptyset, G)$.  
% \textcolor{blue}{
	% Provide an example of an assume guarantee contract in terms of funnel as in (\ref{goalineq}) to motivate the next remark.}
% As it can been seen in Example \ref{eg: asg}, the assumptions and guarantees of a contract
% are the funnels for STL formulae, which can be considered as time-varying safe sets (\textcolor{red}{to discuss}) (see e.g.~\cite{8796018}). 
%The reason of our choice follows from the fact that a large fragment of STL can be encoded into time-varying safe sets. 
%Indeed, in the previous section, we have shown how the considered fragment of STL can be written in terms of funnels, which are considered as time-varying safe sets. 
%Remark that Assumption \ref{assump:communication} implies that for each agent $\Sigma_i$ under STL specification $\phi_k$, it holds that $I_k \setminus \{i\} \subseteq \n^c_i$. 
Similarly, we can define assume-guarantee contracts for clusters of agents in the multi-agent system.

\begin{definition}\label{def:asg_cluster}
Consider a cluster of agents $\bar \Sigma_k = (\bar X_k,\bar U_k,\bar W_k,\bar f_k,\bar g_k,\bar h_k)$
 as in Definition \ref{def:cluster}. An assume-guarantee contract for the cluster can be defined as $\bar \C_k \!=\! (\bar A_k^a, \bar G_k)$, where $\bar A_k^a: \mathbb{R}_{\geq 0}  \rightarrow \bl{\bar W_k} = \prod_{i\in I_k} W_i$ and $\bar G_k : \mathbb{R}_{\geq 0}  \rightarrow  \bar X_k$. 
\end{definition}
The concept of contract satisfaction by the clusters can be defined similarly as in Definition \ref{asg} and is provided in the appendix in Definition \ref{agc_cluster}.
The following proposition will be used to show the main compositionality result of this section. 

\begin{proposition}\label{proposition1}
Consider a cluster of agents $\bar \Sigma_k=\mathcal{I}(\Sigma_{k_1},\dots,\Sigma_{k_{|I_k|}})$,  where each agent is associated with a local assume-guarantee contract $C_i \!=\! (A_i^a, G_i)$, $i \in I_k$. Consider the assume-guarantee contract $\bar \C_k \!=\! (\bar A_k^a, \bar G_k)$ for the cluster $\bar \Sigma_k$, as in Definition \ref{def:asg_cluster} constructed from the local assume-guarantee contracts $\mathcal{C}_i$, $i\in I_k$ as follows: $\bar A_k^a = \prod_{i\in I_k}A_i^a$ and $\bar G_k = \cap_{i \in I_k} G_i$.
Then, we have $\bar \Sigma_k \models \bar \C_k$ if $\Sigma_i \models \C_i$ for all $i \in I_k$;  $\bar \Sigma_k \models_{us} \bar \C_k$ if $\Sigma_i \models_{us} \C_i$ for all $i \in I_k$. 
% the satisfaction of   assume-guarantee contract for the cluster $\C_k$ is guaranteed  if and only if the satisfaction of all the local assume-guarantee contracts for the agents  $\C_i$, $i \in \{{k_1},\ldots,{k_{|I_k|}}\}$ are guaranteed. 
\end{proposition}
\begin{IEEEproof}
The proof can be found in the appendix.
\end{IEEEproof}  
\blue{
Let us remark that Proposition \ref{proposition1} holds under the assumption that $\n_i^a \cap \n_i^c = \emptyset$. 
In the case that $\n_i^a \cap \n_i^c \neq \emptyset$, one can derive similar compositionality results for each cluster with acyclic or cyclic dynamical interconnections among agents in the same cluster by following the same reasoning as in Theorems \ref{thm:main_dag} and \ref{thm:main}.
}
In the following subsections, we present our main compositionality results %that allow us to reason about the properties of multi-agent systems based on contracts satisfied by the agents. 
by providing conditions under which one can go from the satisfaction of local contracts at the agent's level to the satisfaction of a global contract for the multi-agent system.

\vspace{-0.3cm}
\subsection{{Compositional reasoning for acyclic interconnections}}\label{subsec:compo1} 
In this subsection, we first present the compositionality result for multi-agent systems with \emph{acyclic interconnections} among the clusters, i.e., the cluster interconnection graph is a  directed acyclic graph (DAG).

\begin{theorem}
	\label{thm:main_dag}
Consider a multi-agent system  $\Sigma = \mathcal{I}(\Sigma_1,\dots,\Sigma_N)$ as in Definition \ref{intersys} consisting of $K$ clusters induced by the \blue{task dependency graph $\mathcal{G}^t$}.
	%equipped with a directed graph $\G = (I, \E)$. 
Assume that the cluster interconnection graph  $\bar{\mathcal{G}}^a=(\bar I,\bar E^a)$  is a directed acyclic graph. 	 
	Suppose each agent $\Sigma_i$ is associated with a contract $\C_i \!=\! (A_i^a, G_i)$.
Let $\C  = (\emptyset,G)= (\emptyset,\prod_{k\in \bar I } \bar G_{k})$ be the corresponding contract for $\Sigma$, with $\bar G_k = \cap_{i \in I_k} G_i$, $k \in \bar I$. Assume the following conditions hold:
	\begin{enumerate}[leftmargin = 1.5em] 		
		\item[i)] for all $i \in I$, $\Sigma_i \models \C_i$;
		\item[ii)] \textcolor{black}{for all $i \in I$, $\prod_{j\in \n_i^a} \overline{\textbf{Proj}}^j(G_{j}(t)) \!\!\subseteq \!\!A_{i}^a(t)$} for all $t \in \mathbb{R}_{\geq 0}$.   
	\end{enumerate}
	Then, $\Sigma \models \mathcal{C}$.
\end{theorem}
\begin{IEEEproof}
The proof can be found in the appendix.
\end{IEEEproof}

\begin{figure}[!t]
	\centering
	
	\subcaptionbox{Acyclic cluster interconnection graph.\label{fig:acy}}
	{\includegraphics[width=.18\textwidth]{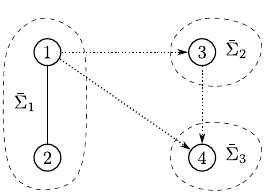}}
 \hspace{0.5cm}
	\subcaptionbox{Cyclic cluster interconnection graph.\label{fig:cyc}}
	{\includegraphics[width=.15\textwidth]{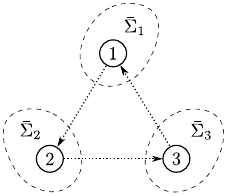}}
	\caption{An illustration of a multi-agent system with acyclic or cyclic cluster interconnection graphs. The solid lines indicate the communication links between agents, and the dotted lines with arrows indicate the dynamical couplings between clusters. } 
	\label{fig:intergraph}
\vspace{-0.7cm}
\end{figure}

The above theorem provides us a compositionality result for multi-agent systems with acyclic cluster interconnection graphs via assume-guarantee contracts.
However, we should mention that weak satisfaction is generally insufficient to reason about general networks with an interconnection graph containing cycles. Interested readers are referred to \cite[Example 6]{saoud2021assume} for a counter-example. 
A more general result is presented in the next subsection which allows us to deal with cyclic interconnections. 
An illustration of multi-agent systems with acyclic and cyclic cluster interconnection graphs is depicted in Fig.~\ref{fig:intergraph}.

\vspace{-0.25cm}
\subsection{{Compositional reasoning for cyclic interconnections}}\label{subsec:compo2}

%\textcolor{blue}{If we are not going to distinguish the case of weak and strong, it would be better to remove the interconnection topology to simplify.}

Next, we present a compositionality result on multi-agent systems with possibly \emph{cyclic interconnections} between clusters.

\begin{theorem}
\label{thm:main}
	Consider a multi-agent system  $\Sigma = \mathcal{I}(\Sigma_1,\dots,\Sigma_N)$ as in Definition \ref{intersys} consisting of $K$ clusters induced by the \blue{task dependency graph $\mathcal{G}^t$}.	 
	Suppose each agent $\Sigma_i$ is associated with a contract $\C_i \!=\! (A_i^a, G_i)$. Let $\C  = (\emptyset,G)= (\emptyset,\prod_{k\in \bar I } \bar G_{k})$ be the corresponding contract for $\Sigma$, with $\bar G_k = \cap_{i \in I_k} G_i$, $k \in \bar I$.
Assume the following conditions hold:
 \begin{enumerate}[leftmargin = 1.5em] 
 \item[i)] for all $i \in I$, for any trajectory $(\mathbf{x}_i,\mathbf{w}_i):\mathbb{R}_{\geq 0} \rightarrow X_i\times W_i$ of $\Sigma_i$, $\mathbf{x}_i(0) \in \overline{\textbf{Proj}}^i(G_i(0))$;
 \item[ii)] for all $i \in I$, $\Sigma_i \models_{us} \C_i$;
 \item[iii)] for all $i \in I$, $\prod_{j\in \n_i^a} \overline{\textbf{Proj}}^j(G_{j}(t))\! \subseteq \!\!A_{i}^a(t)$ for all $t \in \mathbb{R}_{\geq 0}$. % $\prod_{j\in \n_i^c} \overline{\textbf{Proj}}(G_{j}) \subseteq A_{i}^c$. 
% \item if for all $t \geq 0 $, $\mathbf{x}(t) \in G(t)$, then $\Sigma$ satisfies the signal temporal logic task $\phi$;
 \end{enumerate}
 Then, $\Sigma \models \mathcal{C}$.
\end{theorem}
\begin{IEEEproof}
The proof can be found in the appendix.
\end{IEEEproof}  

%The above theorem presents a compositionality result that works for multi-agent systems with both acyclic and cyclic cluster interconnection graphs due to the fact that $\Sigma_i \models_{us} \mathcal{C}_i$ implies $\Sigma_i \models \mathcal{C}_i$.

\blue{
Note that condition (ii) in Theorem~\ref{thm:main_dag} and condition (iii) in Theorem~\ref{thm:main} impose a compatibility condition between the local contracts of neighboring agents, which is essential for the compositional reasoning. Intuitively, this condition requires that for each agent $i$, the guarantee set of its adversarial neighboring agents $j\!\in \!\!\n_i^a$ is a subset of the assumption set of agent $i$. 
}

\begin{remark}
It is important to note that while in the definition of the strong contract satisfaction in~\cite{saoud2021assume} the parameter $\delta$ may depend on time, our definition of assume-guarantee contracts requires a uniform $\delta$ for all time. The reason for this choice is that the uniformity of $\delta$ is critical in our compositional reasoning, since we do not require the set of guarantees to be closed as in~\cite{saoud2021assume}. See~\cite[Example 9]{saoud2021assume} for an example, showing that the compositionality result does not hold using the concept of strong satisfaction when the set of guarantees of the contract is open. Indeed, as it will be shown in the next section, the set of guarantees of the considered contracts are open and one will fail to provide a compositionality result based on the classical (non-uniform) notion of strong satisfaction in~\cite{saoud2021assume}.
\end{remark}

 \vspace{-0.2cm}
\subsection{From weak to strong satisfaction of contacts}

Now, we provide an important result in Proposition \ref{weaktostrong} to be used to prove the main theorem in the next section, which shows how to go from \emph{weak} to \emph{uniform strong satisfaction} of AGCs by relaxing the assumptions. 
The following notion of $\varepsilon$-closeness of trajectories is needed to measure the distance between continuous-time trajectories.

\begin{definition}\label{closeness} ($\varepsilon$-closeness of trajectories) 
Let $Z \subseteq \mathbb{R}^n$.
Consider $\varepsilon > 0$ and two continuous-time trajectories $z_1 :  \mathbb{R}_{\geq 0}  \rightarrow Z $ and $z_2 :  \mathbb{R}_{\geq 0}  \rightarrow Z $. Trajectory $z_2$ is said to be $\varepsilon$-close to  $z_1$, \blue{if for all $t \in \mathbb{R}_{\geq 0} $, $\Vert z_1(t) -z_2(t) \Vert  \leq \varepsilon$}. We define the $\varepsilon$-expansion of $z_1$ by : $\mathcal{B}_{\varepsilon}(z_1) = \{ z' : \mathbb{R}_{\geq 0}  \rightarrow Z \mid z' \text{ is } \varepsilon  \text{-close to } z_1\}$. For set $A = \{ z: \mathbb{R}_{\geq 0} \rightarrow Z\}$, $\mathcal{B}_{\varepsilon}(A) = \cup_{z\in A} \mathcal{B}_{\varepsilon}(z)$.
\end{definition}
 
Now, we introduce the following proposition which will be used later to prove our main theorem.

\begin{proposition}\label{weaktostrong}
(From weak to uniformly strong satisfaction of AGCs)
Consider an agent $\Sigma_i=(X_i,U_i,W_i, f_i,g_i,h_i)$ associated with a local AGC $\C_i \!=\! (A_i^a, G_i)$. If trajectories of $\Sigma_i$ are uniformly continuous and if there exists an $\varepsilon>0$ such that $\Sigma_i \models \C_i^{\varepsilon}$ with $ \C_i^{\varepsilon} = (\mathcal{B}_{\varepsilon}(A_i^a), G_i)$, then $\Sigma_i \models_{us} \C_i$.
\end{proposition}
\begin{IEEEproof}
The proof can be found in the appendix. 
\end{IEEEproof}

In the next subsection, we show how to cast STL formulae into  
time-varying \bl{prescribed performance} functions which will be leveraged later to design continuous-time assume-guarantee contracts. 
%In this subsection, we show how to recast the problem of synthesizing the fragment of STL as in \eqref{syntax1}-\eqref{syntax2} into a funnel-based control framework. 

%prescribed performance control (PPC) framework. 
%In this way, it will allow us to formulate STL formulae as time-varying assume-guarantee contracts.
%This will also enable us to design a closed-form feedback control law based on a funnel-based control strategy to enforce local STL tasks on agents.
 
 \vspace{-0.2cm}
\subsection{Casting STL as \bl{prescribed performance} functions}\label{subsec: ppc}

As mentioned earlier, we consider that each agent in the multi-agent system is subject to an STL task $\phi_i$ which does not only depend on the behavior of agent $\Sigma_i$, but also on other agents $\Sigma_j$, $j \in I$. 
\bl{We denote by $I_{\phi_i} = \{i_{1},\ldots,i_{P_i}\}$ the set of agents that are involved in $\phi_i$, where $P_i$ indicates the total number of agents that are involved in $\phi_i$. 
We further define $\mathbf{x}_{\phi_i}(t) = [\mathbf{x}_{i_{1}}(t); \ldots; \mathbf{x}_{i_{\bl{P_i}}}(t)] \in \mathbb{R}^{n_{P_i}}$, where $n_{P_i} := n_{i_{1}}+\ldots+n_{i_{P_i}}$.}

In order to formulate the STL tasks as assume-guarantee contracts, we first show in this subsection how to cast STL formulae into \bl{prescribed performance} functions.   
Note that the idea of casting STL as \bl{prescribed performance} functions was originally proposed in \cite{larsCDC}.  

First, let us define a \bl{prescribed performance} function
$\gamma_i(t) =  (\gamma_i^0 - \gamma_i^{\infty})\exp(-l_it) + \gamma_i^{\infty}$,   
where $l_i,t\in\R_{\geq 0}$, $\gamma_i^0,\gamma_i^{\infty} \in \R_{> 0}$ with $\gamma_i^0\geq \gamma_i^{\infty} $.
Consider the robust semantics of STL introduced in Subsection \ref{STLsec}. For each agent with STL specification $\phi_i$ in \eqref{syntax2} with the corresponding non-temporal formula $\psi_i$ inside the $F, G, FG$ operators as in \eqref{syntax2}, we can achieve  $0 < r_i \leq \rho_i^{\phi_i}(\mathbf{x}_{\phi_i},0) \leq \rho_{i}^{max}$ by prescribing a temporal behavior to $\rho_i^{\psi_i}(\mathbf{x}_{\phi_i}(t))$ through a properly designed function $\gamma_i$ and parameter $\rho_{i}^{max}$, and the prescribed region %\vspace{-0.15cm}
\begin{align} \notag
&-\gamma_i(t) + \rho_{i}^{max}  <  \rho_i^{\psi_i}(\mathbf{x}_{\phi_i}(t))  <  \rho_{i}^{max} \\     \label{goalineq}
& \iff  - \gamma_i(t) < \rho_i^{\psi_i}(\mathbf{x}_{\phi_i}(t)) -  \rho_{i}^{max}  < 0.
\end{align}
Note that functions $\gamma_i : \mathbb{R}_{\geq 0}  \rightarrow \mathbb{R}_{> 0}$, $i\in \{1,\ldots,N\}$, are positive, continuously differentiable, bounded, and non-increasing \cite{bechlioulis2008robust}. 
% By properly choosing performance functions $\gamma_i$, which prescribes temporal behavior, and combining with $\rho_i^\psi(\mathbf{x}_i(t))$, 
The design of $\gamma_i$ and $\rho_{i}^{max}$ that leads to the satisfaction of $0 < r_i \leq \rho_i^{\phi_i}(\mathbf{x}_{\phi_i},0) \leq \rho_{i}^{max}$  through \eqref{goalineq} will be discussed in Remark \ref{remark:funneldesign} in Section \ref{Subsec:localcontroller}. %\PJ{do we need $r_i$?} 
%Furthermore, we make the following two assumptions on the functions $\rho_i^{\psi_i}(\mathbf{x}_i(t))$ for non-temporal formulae $\psi_i$ contained in \eqref{syntax1}-\eqref{syntax2}.
%\PJ{which assumptions you are talking about?}

To better illustrate the satisfaction of STL tasks using PPC-based strategy, we provide the next  example. 

\begin{example}
\begin{figure}[!t]
	\centering
	%\quad
	\subcaptionbox{Prescribed region $(-\gamma_1(t) + \rho_{1}^{max},  \rho_{1}^{max})$ (dashed lines) for $\phi_1:= F_{[0,8]}\psi_1$, s.t. $\rho_1^{\phi_1}(\mathbf{x},0) \geq r_1$ with $r_1 = 0.1$ (dotted line). \label{fig:funnel_f}}
	{\includegraphics[width=.4\textwidth]{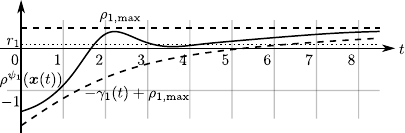}}
	\subcaptionbox{Prescribed region $(-\gamma_2(t) + \rho_{2}^{max}, \rho_{2}^{max})$  (dashed lines) for $\phi_2 := G_{[0,8]}\psi_2$, s.t. $ \rho_2^{\phi_2}(\mathbf{x},0) \geq r_2$ with $r_2 = 0.1$ (dotted line). \label{fig:funnel_g}}
	{\includegraphics[width=.4\textwidth]{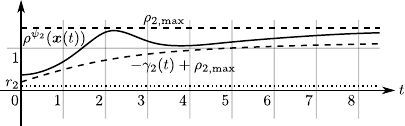}}
	%\quad
	\caption{Prescribed regions for STL formulae. } 
	%Performance bounds are indicated by dashed lines. Evolution of $\rho^{\psi_{i}}(\boldsymbol{x}_i(t))$ are depicted using solid lines.}
	\label{fig:funnel_example}
	\vspace{-0.7cm}
\end{figure}
Consider STL formulae $\phi_1 := F_{[0,8]}\psi_1$ and $\phi_2 := G_{[0,8]}\psi_2$ with $\psi_1 = \mu_1$ and $\psi_2 = \mu_2$, where $\mu_1$ and $\mu_2$ are associated with predicate functions $\mathcal{P}_1(\mathbf{x}) = \mathcal{P}_2(\mathbf{x}) = \mathbf{x}$. 
Figs.~\ref{fig:funnel_f} and \ref{fig:funnel_g} show the defined region in \eqref{goalineq} prescribing a desired temporal behavior to satisfy $\phi_1$ and $\phi_2$, respectively. 
Specifically, it can be seen that
$\rho_1^{\psi_1}(\mathbf{x}(t)) \in (-\gamma_1(t) + \rho_{1}^{max},  \rho_{1}^{max})$ and $\rho_2^{\psi_1}(\mathbf{x}(t)) \in (-\gamma_2(t) + \rho_{2}^{max}, \rho_{2}^{max})$ for all $t \in \mathbb{R}_{\geq 0}$ as in Fig.~\ref{fig:funnel_example}. This shows that \eqref{goalineq} is satisfied for all $t \in \mathbb{R}_{\geq 0}$. 
Then, the connection between atomic formulae $\rho_i^{\psi_{i}}(\mathbf{x}(t))$ and temporal formulae $\rho_i^{\phi_i}(\mathbf{x},0)$, is made by the choice of  $\gamma_1$, $\gamma_2$, $\rho_{1}^{max}$, and $\rho_{2}^{max}$. For example,  the lower bound $-\gamma_1(t) + \rho_{1}^{max}$ in Fig.~\ref{fig:funnel_f} ensures that $\rho_1^{\psi_1}(\mathbf{x}(t))  \geq r_1 = 0.1$ for all $t \geq 6$, which guarantees that the STL task $\phi_1$ is robustly satisfied by  $\rho_1^{\phi_1}(\mathbf{x},0) \geq r_1$.  
% $-\gamma_1(t) + \rho_{1}^{max} < \rho_1^{\psi_1}(\mathbf{x}(t)) < \rho_{1}^{max} $ and $-\gamma_2(t) + \rho_{2}^{max} < \rho_2^{\psi_1}(\mathbf{x}(t)) < \rho_{2}^{max}$.
\end{example}

% \begin{example}
% Consider STL formulae $\phi_1 := F_{[0,8]}\psi_1$ and $\phi_2 := G_{[0,8]}\psi_2$ with $\psi_1 = \mu_1$ and $\psi_2 = \mu_2$, where $\mu_1$ and $\mu_2$ are associated with predicate functions $\mathcal{P}_1(\mathbf{x}) = \mathcal{P}_2(\mathbf{x}) = \mathbf{x}$. 
% Figs.~\ref{fig:funnel_f} and \ref{fig:funnel_g} show the funnel in \eqref{goalineq} prescribing a desired temporal behavior to satisfy $\phi_1$ and $\phi_2$, respectively. 
% Specifically, it can be seen that
% $\rho_1^{\psi_1}(\mathbf{x}(t)) \in (-\gamma_1(t) + \rho_{1}^{max},  \rho_{1}^{max})$ and $\rho_2^{\psi_1}(\mathbf{x}(t)) \in (-\gamma_2(t) + \rho_{2}^{max}, \rho_{2}^{max})$ for all $t \in \mathbb{R}_{\geq 0}$ as in Fig.~\ref{fig:funnel_example}. This shows that \eqref{goalineq} is satisfied for all $t \in \mathbb{R}_{\geq 0}$. 
% Then, the connection between atomic formulae $\rho_i^{\psi_{i}}(\mathbf{x}(t))$ and temporal formulae $\rho_i^{\phi_i}(\mathbf{x},0)$, is made by the choice of  $\gamma_1$, $\gamma_2$, $\rho_{1}^{max}$, and $\rho_{2}^{max}$. For example,  the lower funnel $-\gamma_1(t) + \rho_{1}^{max}$ in Fig.~\ref{fig:funnel_f} ensures that $\rho_1^{\psi_1}(\mathbf{x}(t))  \geq r_1 = 0.1$ for all $t \geq 6$, which guarantees that the STL task $\phi_1$ is robustly satisfied by  $\rho_1^{\phi_1}(\mathbf{x},0) \geq r_1$.  
% % $-\gamma_1(t) + \rho_{1}^{max} < \rho_1^{\psi_1}(\mathbf{x}(t)) < \rho_{1}^{max} $ and $-\gamma_2(t) + \rho_{2}^{max} < \rho_2^{\psi_1}(\mathbf{x}(t)) < \rho_{2}^{max}$.
% \end{example}

%\textcolor{blue}{Some examples would be helpful}

In the sequel, STL tasks will be formulated as contracts by leveraging the above-presented PPC-based framework. We will then design local controllers enforcing the local contracts over the agents (cf. Section \ref{sec:localcontroller}, Theorem \ref{theorem}).

 \vspace{-0.4cm}
\subsection{From STL tasks to assume-guarantee contracts}\label{subsec:stl_to_agc}
%	\bl{Here.}
	
The objective of the paper is to synthesize local controllers $\mathbf{u}_i$, for agents $\Sigma_i$ to achieve the global STL specification $\phi$, where $\phi=\land_{i=1}^N\phi_i$  and $\phi_i$ is the STL task of $\Sigma_i$.
%assigned to the set of agents $\{ \Sigma_i\}$, $i \in I_k \subseteq I$.
Hence, in view of the interconnection between the agents and the distributed nature of the local controllers, one has to make some assumptions on the behaviour of the neighbouring components while synthesizing the local controllers, formalized in terms of assume-guarantee contracts. 
%This property can be formalized in terms of contracts, where the contract should reflect the fact that the objective is to ensure that agent $\Sigma_i$ satisfies ``the guarantee" $\phi_i$ under ``the assumption" that: each of the adversarial neighboring agents $\Sigma_j$, satisfies its local task $\phi_{j}$, $j\in \mathcal{N}_i^a$. 
In this context, and using the concept of  prescribed performance function to cast the STL tasks as presented in Section~\ref{subsec: ppc}, a natural assignment of the local assume-guarantee contract $\C_i \!=\! (A_i^a, G_i)$  for the agents $\Sigma_i$, can be defined formally as follows: 
\begin{enumerate}
\item $A_{i}^a 
%= \{\mathbf{w}_i : E \rightarrow W_i \mid -\gamma_{j_k}(t)+\rho_{j_k}^{max} < \rho^{\psi_{j_k}}(\mathbf{x}_{j_k}(t)) < \rho_{j_k}^{max}, k \in \{1,\dots,|\n_i|\}\}$ \\
= \prod_{j\in \n_i^a} \{ \mathbf{x}_{j}: \mathbb{R}_{\geq 0}  \rightarrow X_{j}  \mid  -\gamma_{j}(t)+\rho_{j}^{max} <  \rho_j^{\psi_{j}}(\mathbf{x}_{\phi_j}(t)) < \rho_{j}^{max}, \forall t \in \mathbb{R}_{\geq 0} \}$;  
\item $G_{i} = \{(\mathbf{x}_i, \mathbf{z}_i): \mathbb{R}_{\geq 0}  \rightarrow X_i \times Z_i   \mid -\gamma_i(t)+\rho_{i}^{max} < \rho_i^{\psi_{i}}({\mathbf{x}}_{\phi_i}(t))  < \rho_{i}^{max}, \forall t \in \mathbb{R}_{\geq 0}\}$;
\end{enumerate}
where 
%$\mathbf{x}_{j}$ denotes the state trajectories of the neighboring agent $\Sigma_j$, $j \!\in\! \n_i$, and 
$-\gamma, \rho^{\psi}, \rho^{max}$ are the functions discussed in Subsection \ref{subsec: ppc} corresponding to their STL tasks. 
%Note that the assumption set $A_{i}^c$ is nothing but a feasibility assumption on the STL task $\phi_i$. 
\bl{Notice that the above definition of the local contracts $\C_i \!=\! (A_i^a, G_i)$ readily satisfies  condition (ii) in Theorem \ref{thm:main_dag} and (iii) in Theorem \ref{thm:main}.}

Once the specification $\phi$ is decomposed into local contracts\footnote{Note that the decomposition of a global STL formula is out of the scope of this paper. In this paper, we use a natural decomposition of the specification, where the assumptions of a component coincide with the guarantees of its neighbours. However, given a global STL for an interconnected system, one can utilize existing methods provided in recent literature, e.g., \cite{charitidou2021signal}, to decompose the global STL task into local ones.} and in view of Theorems~\ref{thm:main_dag} and~\ref{thm:main}, Problem~\ref{pro:1} can be solved by considering local control problems for each agent $\Sigma_i$. These control problems can be solved in a distributed manner and are formally defined as follows:

\begin{problem}
\label{pro:2}
Given an agent $\Sigma_i=(X_i,U_i,W_i, f_i,g_i,h_i)$ associated with an assume-guarantee contract $\C_i \!=\! (A_i^a, G_i)$, where $A_i^a$, and $G_i$ are given by STL formulae by means of \bl{prescribed performance} functions, synthesize a local controller $\mathbf{u}_i :\bar X_k \times \mathbb{R}_{\geq 0} \rightarrow U_i $ such that $\Sigma_i \models_{us} \C_i$.
\end{problem}

 \vspace{-0.25cm}
\section{Distributed Controller Design}\label{sec:localcontroller}

In this section, we first provide a solution to Problem \ref{pro:2} by designing controllers ensuring that  local contracts for agents are uniformly strongly satisfied.
Then, we show that based on our compositionality result proposed in the last section, the global STL task for the network is satisfied by applying the derived local controllers to agents individually. 
First, we utilize the idea of \bl{prescribed performance control} (PPC) \cite{bechlioulis2008robust} in order to enforce the satisfaction of local contracts by prescribing certain transient behavior of the prescribed regions that constrain the closed-loop trajectories of the agents. 

%\textcolor{blue}{We can use any controller synthesis approach, in this paper we use a Lyapunov-based approach, but any other tool can be used (barrieres, LMIs, ...}
 
 \vspace{-0.25cm}
\subsection{Local controller design}\label{Subsec:localcontroller}

As discussed in Subsection \ref{subsec: ppc}, one can enforce STL tasks via PPC-based strategy by prescribing the transient behavior of $\rho_i^{\psi_i}(\mathbf{x}_{\phi_i}(t))$ within the predefined region in  
\begin{align}\label{goalineq1}
    - \gamma_i(t) < \rho_i^{\psi_i}(\mathbf{x}_{\phi_i}(t)) -  \rho_{i}^{max}  < 0,
\end{align}
where  $\psi_i$ is the corresponding non-temporal formula inside the $F, G, FG$ operators as in \eqref{syntax2}.
In order to leverage the idea of prescribed performance control to achieve this, an error term is first defined for each $\rho_i^{\psi_i}(\mathbf{x}_{\phi_i}(t))$ as $e_i(\mathbf{x}_{\phi_i}(t)) = \rho_i^{\psi_i}(\mathbf{x}_{\phi_i}(t))-\rho_{i}^{max}$. \bl{We can then define the modulated error as $\hat e_i(\mathbf{x}_{\phi_i},t)=\frac{e_i(\mathbf{x}_{\phi_i}(t))}{\gamma_i(t)}$ and the corresponding prescribed performance region $\hat {\mathcal{D}_i} := (-1,0)$. Notice that the regions $\hat {\mathcal{D}_i} := (-1,0)$ defined for $\hat e_i(\mathbf{x}_{\phi_i},t)$  are equivalent to the desired prescribed performance bounds for $\rho_i^{\psi_i}(\mathbf{x}_{\phi_i}(t))$ in \eqref{goalineq1}.} 
The transformed error is then defined as 
\begin{align}\label{eq:epsilon}
\epsilon_i({\mathbf{x}}_{\phi_i}, t) :=    T_i(\hat e_i(\mathbf{x}_{\phi_i},t)) = \ln(-\frac{\hat e_i(\mathbf{x}_{\phi_i},t)+1}{\hat e_i(\mathbf{x}_{\phi_i},t)}).
\end{align}
\bl{Let us also define 
\begin{align}\label{eq:J}
\mathcal{J}_i(\hat e_i,t)\!=\! \frac{\partial{T_i(\hat e_i)}}{\partial{\hat e_i}}\frac{1}{\gamma_i(t)} \!=\! -\frac{1}{\gamma_i(t)\hat e_i(1+ \hat  e_i)},
\end{align}
which is the normalized Jacobian of the transformation function. }
%where $T_i: (-1,0) \rightarrow \mathbb{R}$ is the transformation function that is strictly increasing, bijective, and hence admitting an inverse. 
The basic idea of PPC is to derive a control policy such that $ \epsilon_i({\mathbf{x}}_{\phi_i}, t)$ is rendered bounded, which implies the satisfaction of \eqref{goalineq1}.
A more detailed description of the PPC-based strategy is provided in the appendix.

We make the following two assumptions on functions $\rho_i^{\psi_i}$ for formulae $\psi_i$, which are required for the local controller design in the main result of this section.

\begin{assumption}\label{assmprho1}
Each formula within class $\psi$ as in \eqref{syntax1} has the following properties: (i) $\rho_i^{\psi_i}: \mathbb{R}^{n_i} \rightarrow \mathbb{R}$ is concave and (ii) the formula is well-posed in the sense that for all $C \in \mathbb{R}$ there exists $\bar C \geq 0$ such that for all $\mathbf{x}_{\phi_i} \in  \mathbb{R}^{n_{P_i}}$ with $\rho_i^{\psi_i}(\mathbf{x}_{\phi_i}) \geq C$, one has $\Vert \mathbf{x}_{\phi_i} \Vert \leq \bar C < \infty$. 
\end{assumption}

\begin{remark}
\bl{Part (i) of Assumption \ref{assmprho1} imposes that $\rho_i^{\psi_i}$ is concave. Concave functions contain the class of linear functions as well as functions which express, e.g., reachability tasks using predicates such as $\Vert \mathbf{x}_i - [50,20]^\mathsf{T}\Vert \leq 5$ as in Example \ref{eg2}, \blue{for which $\rho_i^{\psi_i}(\mathbf{x}_i) = 5 - \Vert \mathbf{x}_i - [50,20]^\mathsf{T}\Vert$}.
This assumption is required since the controller that will be proposed in  Theorem \ref{theorem} uses the gradient information of $\frac{\partial\rho^{\psi_i}(\mathbf{x}_{\phi_i})}{\partial \mathbf{x}_{i}}$. Hence, local optima or saddle points may lead the system to get stuck at them. 
Part (ii) of Assumption \ref{assmprho1} ensures bounded solutions of $\mathbf{x}_{\phi_i}(t)$ , and thus the well definedness of $\mathbf{x}_{\phi_i}(t)$ for all $t \geq 0$. This assumption is not restrictive in practice since one can combine a new formula $\psi_i^{Asm} := (\Vert \mathbf{x}_{\phi_i}\Vert < \bar C)$ with $\psi_i$ for a sufficiently large $\bar C$ so that $\psi_i^{Asm} \wedge \psi_i $ is well-posed.}
\end{remark}

Define the global maximum of $\rho_i^{\psi_i}(\mathbf{x}_{\phi_i})$ as $\rho_i^{opt}= \sup_{\mathbf{x}_{\phi_i} \in \mathbb{R}^{n_{P_i}}}\rho_i^{\psi_i}(\mathbf{x}_{\phi_i})$. 
\begin{assumption}\label{assmprho2}
(i) The global maximum of $\rho_i^{\psi_i}(\mathbf{x}_{\phi_i})$ is positive, i.e. $\rho_i^{opt} >0$ and (ii) $ \frac{\partial\rho_i^{\psi_i}({\mathbf{x}}_{\phi_i})} {\partial \mathbf{x}_i}$ is a non-zero vector.
\end{assumption}

\begin{remark}
\bl{The global maximum of $\rho_i^{\psi_i}(\mathbf{x}_{\phi_i})$ being positive guarantees that the local STL formula $\psi_i$ is satisfiable, since $\rho_i^{opt} >0$ implies that $\rho_i^{\psi_i}(\mathbf{x}_{\phi_i}, 0) >0$ is possible. Note that $\rho_i^{opt}$ is easy to compute since $\rho_i^{\psi_i}(\mathbf{x}_{\phi_i})$ is concave. 
%If $\rho_i^{opt} \leq 0$ holds, then $\psi_i$ is not satisfiable, and one can resort to \emph{least-violating solutions}, such as the approach proposed in \cite{lindemann2019feedback}.   
The assumption on $\frac{\partial\rho_i^{\psi_i}({\mathbf{x}}_{\phi_i})} {\partial \mathbf{x}_i}$ being a non-zero vector is used to avoid local optima which can cause infeasibility issues. 
\blue{Note that since $\rho_i^{\psi_i}({\mathbf{x}}_{\phi_i}) $ is concave under Assumption \ref{assmprho1}, 
%one can design proper parameters for the prescribed regions such that the local optima are avoided:
one has $\frac{\partial\rho_i^{\psi_i}({\mathbf{x}}_{\phi_i})} {\partial \mathbf{x}_i} = 0$ if and only if $\rho_i^{\psi_i}({\mathbf{x}}_{\phi_i}) = \rho_i^{opt}$, thus (ii) of Assumption \ref{assmprho2} can be guaranteed by choosing $\rho_{i}^{\max}$ such that $0<\rho_{i}^{\max} \!< \! \rho_i^{opt} \!< \!0$ holds as in \eqref{para:rhomax}, under the assumption that $\rho_i^{opt} >0$ as in part (i) of Assumption \ref{assmprho2}.}}
\end{remark}

%Note that the above assumption implies that $\psi_i$ and hence $\phi_i$ is locally feasible. The assumption on $\frac{\partial\rho_i^{\psi_i}({\mathbf{x}}_{\phi_i})} {\partial \mathbf{x}_i}$ being a non-zero vector is used to avoid local optima which can cause infeasibility issues. One can design proper funnel parameters such that the local optima are avoided.

% $\rho_i^{opt} >0$, which leads to the following assumption.

%The following assumptions are required on agents and the task dependency graphs of multi-agent systems in order to design controllers enforcing local contracts. 

%%%%%%%%%

%\textcolor{blue}{Remove the indice i in the next result and its proof}
Now, we are ready to present the main result of this section solving Problem \ref{pro:2} for the local controller design.

\begin{theorem}\label{theorem}
Consider an agent $\Sigma_i$ as in \eqref{eqn:subsys} belonging to cluster $\bar \Sigma_k$  and satisfying Assumption \ref{assmp:lipschitz}. 
Given that $\Sigma_i$ is subject to STL task $\phi_i$, and 
is associated with its local assume-guarantee contract $\C_i \!=\! (A_i^a, G_i)$, where
\begin{enumerate}
% \item $A_{i} 
% %= \{\mathbf{w}_i : E \rightarrow W_i \mid -\gamma_{j_k}(t)+\rho_{j_k}^{max} < \rho^{\psi_{j_k}}(\mathbf{x}_{j_k}(t)) < \rho_{j_k}^{max}, k \in \{1,\dots,|\n_i|\}\}$ \\
% = \prod_{j\in \n_i} \{ \mathbf{x}_{j}: \mathbb{R}_{\geq 0}  \rightarrow X_{j}    \mid  -\gamma_{j}(t)+\rho_{j}^{max} < \rho_j^{\psi_{j}}(\mathbf{x}_{j}(t)) < \rho_{j}^{max}, \forall t \in \mathbb{R}_{\geq 0} \}$,
% \item $G_{i} = \{\mathbf{x}_i : \mathbb{R}_{\geq 0}  \rightarrow X_i \mid -\gamma_i(t)+\rho_{i}^{max} < \rho_i^{\psi_{i}}(\mathbf{x}_i(t)) < \rho_{i}^{max}, \forall t \in \mathbb{R}_{\geq 0} \}$,
%\item $A_{W_i} = \{[\mathbf{x}_{j_1}, \dots, \mathbf{x}_{j_{|\n_i|}} ]| 0 < \rho^{\phi_{j_k}}(\mathbf{x}_{j_k}) < \rho_{{j_k}}^{max}, k \in \{1,\dots,|\n_i|\}\}$ 
% \item $G_{X_i} = \{ \mathbf{x}_{i} | 0 < \rho^{\phi_i}(\mathbf{x}_{i}) < \rho_{i}^{max}\}$,

\item $A_{i}^a 
%= \{\mathbf{w}_i : E \rightarrow W_i \mid -\gamma_{j_k}(t)+\rho_{j_k}^{max} < \rho^{\psi_{j_k}}(\mathbf{x}_{j_k}(t)) < \rho_{j_k}^{max}, k \in \{1,\dots,|\n_i|\}\}$ \\
= \prod_{j\in \n_i^a} \{ \mathbf{x}_{j}: \mathbb{R}_{\geq 0}  \rightarrow X_{j}  \mid  -\gamma_{j}(t)+\rho_{j}^{max} <  \rho_j^{\psi_{j}}(\mathbf{x}_{\phi_j}(t)) < \rho_{j}^{max}, \forall t \in \mathbb{R}_{\geq 0} \}$;  
\item $G_{i} = \{(\mathbf{x}_i, \mathbf{z}_i): \mathbb{R}_{\geq 0}  \rightarrow X_i \times Z_i  \mid -\gamma_i(t)+\rho_{i}^{max} < \rho_i^{\psi_{i}}({\mathbf{x}}_{\phi_i}(t))  < \rho_{i}^{max}, \forall t \in \mathbb{R}_{\geq 0}\}$;
\end{enumerate}
where $\psi_i$ is an atomic formula as in \eqref{syntax1} satisfying Assumptions~\ref{assmprho1} and~\ref{assmprho2}.
%	Suppose that for each agent $\Sigma_i$, the assumption $A_{W_i}$ holds.
Assume $-\gamma_i(0)\!+\!\rho_{i}^{max} \!<\! \rho_i^{\psi_i}({\mathbf{x}}_{\phi_i}(0)) \!<\! \rho_{i}^{max} \!<\! \rho_i^{opt}$, for all $i \in I_k$.
%and $ \frac{\partial\rho_i^{\psi_i}({\mathbf{x}}_{\phi_i})} {\partial \mathbf{x}_i}$ is a non-zero vector.
	%$-\gamma_i^0+\rho_{i}^{max} < \rho_i^\psi(\mathbf{x}_i(0))$ and $-\gamma_i^{\infty} + \rho_{i}^{max} < \rho_i^{opt} < \rho_{i}^{max}$, 
	%\PJ{$\psi_i$ everywhere?} 
If \blue{the task dependency graph  $\mathcal{G}^t = (I, E^t)$ is a directed acyclic graph as in Assumption \ref{assmp:task}}, and all agents $\Sigma_i$ in the cluster apply the controller  
	\begin{align}  \label{controller}	
\hspace{-0.25cm} \mathbf u_i(\bar{\mathbf{x}}_k, t) \! = \! -g_i^\top( \mathbf{x}_i) \!
  \sum_{j \in I_k} ( \frac{\partial\rho_j^{\psi_j}({\mathbf{x}}_{\phi_j} )}{\partial \mathbf{x}_i} \mathcal{J}_j(\hat e_j,t)\epsilon_j({\mathbf{x}}_{\phi_j}, t)), \! \! \! 
	\end{align}
 where \bl{$\mathcal{J}_i(\hat e_i,t), \epsilon_j({\mathbf{x}}_{\phi_j}, t)$ are defined in \eqref{eq:J} and \eqref{eq:epsilon}}, respectively,
then we have $\Sigma_i  \models_{us} \mathcal{C}_i$ for all $i\in I_k$.
%is satisfied for all $t \in \mathbb{R}_{\geq 0} $ for all agents.
%	Then it holds that  $0 < r_i \leq \rho_i^\phi(\mathbf{x}_i,0) \leq \rho_{i}^{max}$, for all agents. 
\end{theorem}
\begin{IEEEproof}
The proof can be found in the appendix. 
\end{IEEEproof}

\blue{As can be seen from \eqref{controller},  the presented control strategy is distributed, but not fully decentralized, as it requires the information from its cooperative neighbors w.r.t. their states and prescribed performance functions. 
From an implementation perspective, such a distributed approach is realistic since communication between neighboring agents is feasible and efficient in many practical applications, and is often necessary to achieve coordinated behavior. 
}

\begin{remark}\label{remark:funneldesign} (Parameter design for the prescribed regions)
Remark that the connection between atomic formulae $\rho_i^{\psi_{i}}(\mathbf{x}_{\phi_i}(t))$ and temporal formulae $ \rho_i^{\phi_i}(\mathbf{x}_{\phi_i},0)$ is made by $\gamma_i$ and $\rho_{i}^{max}$ as in \eqref{goalineq}, which need to be designed as instructed in \cite{larsCDC}. 
Specifically, 
\blue{suppose $\rho_i^{opt} >0$ holds as in part (i) of Assumption \ref{assmprho2}}, and select   
 \vspace{-0.5cm}	

% \begin{align*} \label{funnelpara1}
% t_{i}^{\ast}\in & \begin{cases} a_{i} & \text{if}\ \phi_{i}=G_{[a_{i}, b_{i}]}\psi_{i}\\ 
% 		{[}a_{i}, b_{i}] & \text{if}\ \phi_{i}=F_{[a_{i}, b_{i}]}\psi_{i}\\ 
% 		{[}\underline{a}_{i}+\bar{a}_{i}, \underline{b}_{i} +  \bar{a}_{i}] & \text{if}\ \phi_{i}=F_{[\underline{a}_{i},\underline{b}_{i}]}G_{[\bar{a}_{i}, \bar{b}_{i}]}\psi_{i}\\  \end{cases}\tag{5}\\ 
% 	\rho_{i}^{\max}\in & \left(\max(0, \rho^{\psi_{i}}(\mathbf{x}_{{i}}(0))), \rho_{{i}}^{\text{opt}}\right)\tag{6}\\ 
% 	r_{i}\in & (0,\rho_{i}^{\max})\tag{7}\\ \gamma_{i}^{0}\in&\begin{cases} (\rho_{i}^{\max}-\rho^{\psi_{i}}(\mathbf{x}_{{i}}(0)),\infty)\ & \text{if}\ t_{i}^{\ast} > 0\\ (\rho_{i}^{\max}-\rho^{\psi_{i}}(\mathbf{x}_{{i}}(0)),\rho_{i}^{\max}-r_{i}]\ & \text{else}  \end{cases}\tag{8}\\ \gamma_{i}^{\infty}\in&\left.\left(0, \min(\gamma_{i}^{0}, \rho_{i}^{\max}-r_{i})\right]\right.\tag{9}\\ l_{i}\in&\begin{cases}\mathbb{R}_{\geq 0} & \text{if}- \gamma_{i}^{0}+\rho_{i}^{\max}\geq r_{i}\\   \label{funnelpara2}
% 	\frac{-\ln\left(\frac{r_{i}+\gamma_{i}^{\infty}-\rho_{i}^{\max}}{-\gamma_{i}^{0}-\gamma_{i}^{\infty}}\right)}{t_{i}^{\ast}}& \text{else} \end{cases}\tag{10}
% 	\end{align*}

\begin{align} \label{funnelpara1}
 t_{i}^{\ast}\in & \begin{cases} a_{i} & \text{if}\ \phi_{i}=G_{[a_{i}, b_{i}]}\psi_{i}\\ 
		{[}a_{i}, b_{i}] & \text{if}\ \phi_{i}=F_{[a_{i}, b_{i}]}\psi_{i}\\ 
		{[}\underline{a}_{i}+\bar{a}_{i}, \underline{b}_{i} +  \bar{a}_{i}] \! & \text{if}\ \phi_{i}=F_{[\underline{a}_{i},\underline{b}_{i}]}G_{[\bar{a}_{i}, \bar{b}_{i}]}\psi_{i}  \end{cases} \\ 
  \label{para:rhomax}
	\rho_{i}^{\max}\in & \left(\max(0, \rho_i^{\psi_{i}}(\mathbf{x}_{\phi_i}(0))), \rho_{{i}}^{\text{opt}}\right) \\ 
	r_{i}\in & (0,\rho_{i}^{\max}) \\ 
	\gamma_{i}^{0}\in&\begin{cases} (\rho_{i}^{\max}-\rho_i^{\psi_{i}}(\mathbf{x}_{\phi_i}(0)),\infty)\ & \text{if}\ t_{i}^{\ast} > 0\\ (\rho_{i}^{\max}-\rho_i^{\psi_{i}}(\mathbf{x}_{\phi_i}(0)),\rho_{i}^{\max}-r_{i}]\ & \text{else}  \end{cases} \\ \gamma_{i}^{\infty}\in&\left.\left(0, \min(\gamma_{i}^{0}, \rho_{i}^{\max}-r_{i})\right]\right. \\ l_{i}\in&\begin{cases}\mathbb{R}_{\geq 0} & \text{if}- \gamma_{i}^{0}+\rho_{i}^{\max}\geq r_{i}\\   \label{funnelpara2}
	\frac{-\ln\left(\frac{r_{i}+\gamma_{i}^{\infty}-\rho_{i}^{\max}}{-\gamma_{i}^{0}+\gamma_{i}^{\infty}}\right)}{t_{i}^{\ast}}& \text{else}. \end{cases} 
\end{align} 
\bl{Note the choice of $\gamma_{i}^{0}$ ensures the satisfaction of the assumption on initial conditions $-\gamma_i(0)\!+\!\rho_{i}^{max} \!<\! \rho_i^{\psi_i}({\mathbf{x}}_{\phi_i}(0)) \!<\! \rho_{i}^{max} \!<\! \rho_i^{opt}$ as stated in Theorem \ref{theorem}.}
With $\gamma_i$ and $\rho_{i}^{max}$ chosen properly (as shown above), one can achieve $0 \!<\! r_i \!\leq\! \rho_i^{\phi_i}(\mathbf{x}_{\phi_i},0) \!\leq\! \rho_{i}^{max}$ by prescribing a temporal behavior to $\rho_i^{\psi_{i}}(\mathbf{x}_{\phi_i}(t))$ as in the set of guarantees $G_{i}$  in Theorem \ref{theorem}, i.e., $  -\gamma_i(t)\!+\!\rho_{i}^{max} \!<\! \rho_i^{\psi_{i}}(\mathbf{x}_{\phi_i}(t)) \!< \!\rho_{i}^{max}$ for all $t \!\geq\! 0$.

\end{remark}

% \textcolor{blue}{Maybe we should another result to show completeness of trajectories. of the system.
% \begin{proposition}
% Consider a system $\Sigma$ consisting of interconnected compnents, if the initial condition in the set of guarantees and control law chosen as in the theorem then trajectories of $\Sigma$ are compelte
% \end{proposition}
% }
  
 \vspace{-0.3cm}
\subsection{Global task satisfaction}

% \begin{lemma}\label{lemma1}
% 	(\textcolor{red}{To discuss}) The above assumption is satisfied when the set of guarantees $G_X$ is described as properties in the form of STL formulae/prescribed performance constraints,  $f$ in \eqref{eq:sys} is locally Lipschitz, $\mathbf{x} : \mathbb{R}_{\geq 0}  \rightarrow X$ is left continuous and the set of guarantees $G_X$ is \textcolor{red}{closed?}.
% \end{lemma}
Here, we show that by applying the local controllers to the agents,
the global STL task for the multi-agent system is also  satisfied based on our compositionality result.
%proposed in the last section, the global STL task for the network is satisfied by enforcing the derived local controllers on agents in a decentralized manner 

\begin{corollary}\label{corollary}
Consider a multi-agent system  $\Sigma = \mathcal{I}(\Sigma_1,\dots,\Sigma_N)$ as in Definition \ref{intersys} consisting of $K$ clusters induced by the task dependency graph $\mathcal{G}^t = (I, E^t)$.	 \blue{Suppose Assumption~\ref{assmp:task} holds.}  
Suppose each agent $\Sigma_i$ is associated with a contract $\C_i \!=\! (A_i^a, G_i)$,
 %, and each cluster of agents
and each cluster of agents is associated with a contract $\bar \C_k \!=\! (\bar A_k^a, \bar G_k)$ as in Definition \ref{def:asg_cluster}.
If we apply the controllers as in \eqref{controller} to  agents $\Sigma_i$, then we get $\Sigma \models \C  = (\emptyset,\prod_{k\in \bar I } \bar G_{k})$. This means that the control objective in Problem~\ref{pro:1} is achieved, i.e.,  system $\Sigma$ satisfies signal temporal logic task $\bar\phi$.
\end{corollary}
\begin{IEEEproof}
The proof can be found in the appendix. 
\end{IEEEproof}
\bl{According to the above corollary, we remark that the global task satisfaction holds for multi-agent systems with acyclic task dependency graphs, irrespective of whether the dynamical interconnection graph is acyclic or cyclic. 
}

 \vspace{-0.3cm}
\section{Case Study}\label{sec: case}
%In this section, we demonstrate the effectiveness of our proposed results on two case studies. 

%\subsection{\bl{Room Temperature Regulation}} 

We demonstrate the effectiveness of the proposed results on two case studies: a room temperature regulation and a mobile multi-robot control problem.
 
 \vspace{-0.25cm}
\subsection{Room Temperature Regulation} 
% \begin{figure}[t]
% 	\centering
% 	\includegraphics[width=.16\textwidth]{fig/fig3.pdf}
%  	\caption{A circular building with $N$ rooms.} \label{fig:circularbuilding}
%  \vspace{-0.5cm}	
% \end{figure}

Here, we apply our results to the temperature regulation of a circular building with $N = 1000$ rooms each equipped with a heater. 
 The evolution of the temperature of the interconnected model is described by the differential equation: 
\begin{align}\label{room}
\Sigma:
\dot{\mathbf{T}}(t)=& A\mathbf{T}(t) + \alpha_h T_h \nu(t) + \alpha_e T_e, 
\end{align}
adapted from \cite{girard2015safety}, where $A \!\in\! \mathbb{R}^{N\!\times\! N}$ is a matrix with elements $\{A\}_{ii} \!=\! (- \!2 \alpha\!-\!\alpha_e\!-\!\alpha_h \nu_i)$, $\{A\}_{i,i+1} \!=\! \{A\}_{i+1,i} \! = \{A\}_{1,N} \!=\! \{A\}_{N,1} =\!\alpha$, $\forall i \in \{1, \dots, N-1\}$, and all other elements are identically zero, 
$\mathbf{T}(k)\!=\![\mathbf{T}_1(k);\dots; \mathbf{T}_N(k)]$,   $T_e\!=\![T_{e1};\dots;T_{eN}]$, $\nu(k)\!=\![\nu_1(k);\dots;\nu_N(k)]$, where $\nu_i(k)\!\in\! [0,1]$, $\forall i\!\in\!\{1, \dots, N\}$, represents the ratio of the heater valve being open in room $i$.
Parameters $\alpha\! =\! 0.05$, $\alpha_e\!=\!0.008$, and $\alpha_h\!=\! 0.12$ are heat exchange coefficients, %are the conduction factors, respectively, between the neighboring rooms,  external environment and room $i$, and heater and room $i$. 
$T_{ei}\! =\! -20\,^\circ C$ is the external environment temperature, and  $T_h \!=\!55\,^\circ C$ is the heater temperature. 
Now, by introducing $\Sigma_i$ described by 
\begin{align*}\label{roomi}
\Sigma_i: 
\dot{\mathbf{T}}_i(t)= a\mathbf{T}_i(t) + d \mathbf{w}_i(t)+ \alpha_h T_h \nu_i(t) + \alpha_e T_{ei}, 
\end{align*}
where $a = - \!2 \alpha\!-\!\alpha_e\!-\!\alpha_h \nu_i$, $d = 0.8\alpha$, and $\mathbf{w}_i(t) = [\mathbf{T}_{i-1}(t);\mathbf{T}_{i+1}(t)]$ (with $\mathbf{T}_{0} = \mathbf{T}_{N}$ and $\mathbf{T}_{N+1} = \mathbf{T}_{1}$), one can readily verify that $\Sigma = \mathcal{I}(\Sigma_1,\dots,\Sigma_N)$ as in Definition \ref{intersys}. 
The initial temperatures of these rooms are, respectively, 
$\mathbf{T}_{1}(0)=20\,^\circ C$, $\mathbf{T}_{2}(0)=25.5\,^\circ C$,
$\mathbf{T}_{i}(0)=21\,^\circ C$ if $i \in \{i \text{ is odd } | i \in \{3, \dots, N\} \}$, and $\mathbf{T}_{i}(0)=27\,^\circ C$ if $i \in \{i \text{ is even } | i \in \{4, \dots, N\} \}$.

The room temperatures are subject to the following STL tasks 
$\phi_1$: $F_{[0,5]}G_{[10,20]}(\Vert \mathbf{T}_1 - \mathbf{T}_{999}\Vert  \leq 2)\wedge(\Vert \mathbf{T}_1 - \mathbf{T}_{3}\Vert  \leq 2)\wedge(\Vert \mathbf{T}_1 - 23\Vert  \leq 0.5)$,
$\phi_2$: $F_{[0,5]}G_{[10,20]}(\Vert \mathbf{T}_2 - \mathbf{T}_{1000}\Vert  \leq 2)\wedge(\Vert \mathbf{T}_2 - \mathbf{T}_{4}\Vert  \leq 2)\wedge(\Vert \mathbf{T}_2 - 29\Vert  \leq 0.5)$,
$\phi_i$: $F_{[0,5]}G_{[10,20]}(\Vert \mathbf{T}_i - \mathbf{T}_{i+2}\Vert  \leq 2)$, 
for $i \in \{3,\ldots,N-2\}$,
% for $i \in \{i \text{ is odd } | i \in 3,\ldots,N-2\}$, 
% $\phi_i$: $F_{[0,5]}G_{[10,20]}(\Vert \mathbf{T}_i - \mathbf{T}_{i+2}\Vert  \leq 2)$, 
% for $i \in \{i \text{ is even } | i \in 4,\ldots,N-2\}$, 
and \bl{$\phi_{N-1}$: $F_{[0,25]}(\Vert \mathbf{T}_{N-1} - 23\Vert  \leq 0.5)$, and $\phi_N$: $F_{[0,25]}( \Vert \mathbf{T}_{N} - 29\Vert  \leq 0.5)$}.
Intuitively, the STL tasks require the controllers (heaters) to be synthesized such that the temperature of every other rooms should eventually get close to each other, and the temperature of the rooms reach the specified region ($[22.5, 23.5]$ for rooms $\Sigma_1$  and $\Sigma_{N-1}$, or $[28.5, 29.5]$ for rooms $\Sigma_2$ and $\Sigma_{N}$) in the desired time slots.
In this setting, the circular building consists of two clusters: $I_1 = \{i \text{ is odd } | i \in \{1, \dots, N\} \}$ and $I_2 = \{i \text{ is even } | i \in \{1, \dots, N\} \}$. The cluster interconnection graph is cyclic due to the dynamical couplings between the neighboring rooms.

% \begin{figure}[t]
% 	\centering
% 	\includegraphics[width=.3\textwidth]{fig/eg1/reach_T.pdf}
%  	\caption{Trajectories of the temperature (a) and control input (b) of the closed-loop agents $\Sigma_1$, $\Sigma_2$, $\Sigma_{N-1}$ and $\Sigma_{N}$, where $N=1000$,  
%   under control policy in \eqref{controller}.} \label{fig:room_temperature}
%   \vspace{-0.3cm}	 
% \end{figure}

\begin{figure}[t]
	%\vspace{+ 0.2cm}
	\centering
	\subcaptionbox{Room temperature \label{fig:room_temperature_state}}
	{\includegraphics[width=.241\textwidth]{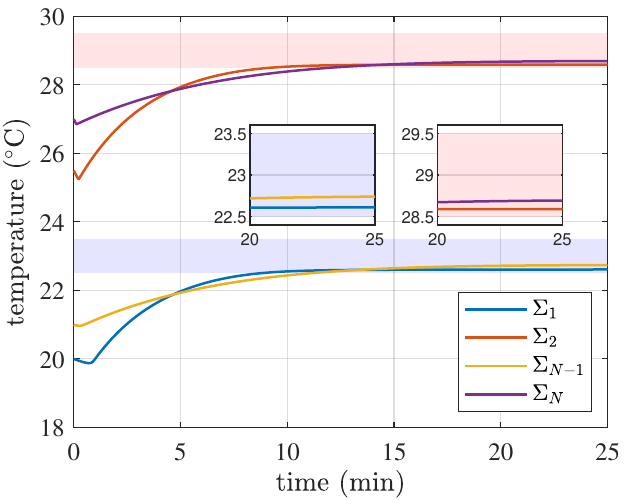}}
	\subcaptionbox{Control inputs \label{fig:temperature_input}}
	{\includegraphics[width=.241\textwidth]{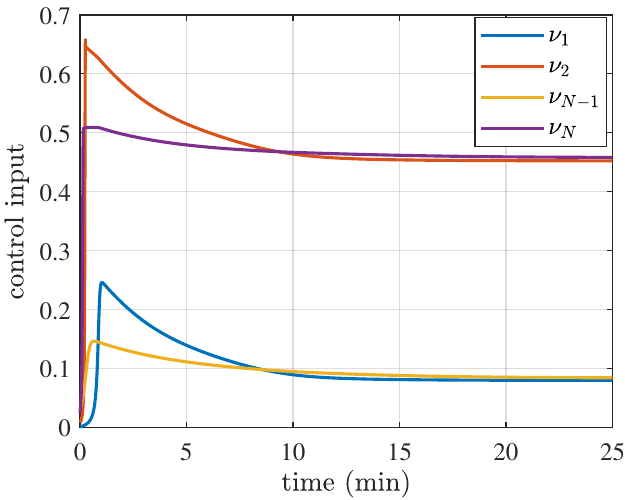}}
	 
\caption{Trajectories of the temperature (a) and \bl{control input (b)} of the closed-loop agents $\Sigma_1$, $\Sigma_2$, $\Sigma_{N-1}$ and $\Sigma_{N}$, where $N=1000$,  
  under control policy in \eqref{controller}.} \label{fig:room_temperature}
  \vspace{-0.7cm}	 
\end{figure}

% \begin{figure}[t]
% 	\centering
% 	\includegraphics[width=.3\textwidth]{fig/eg1/reach_Tdiff.pdf}
%  	\caption{Trajectories of the control iagents $\Sigma_1$ and $\Sigma_{N-1}$, and between $\Sigma_2$ and $\Sigma_{N}$, where $N=1000$,
%   under control policy in \eqref{controller}.} \label{fig:room_temperaturediff}
%  \vspace{-0.3cm}
% \end{figure}

\begin{figure}[t]
	%\vspace{+ 0.2cm}
	\centering
	\subcaptionbox{Prescribed region for $\Sigma_1$ \label{fig:room_rho1}}
	{\includegraphics[width=.241\textwidth]{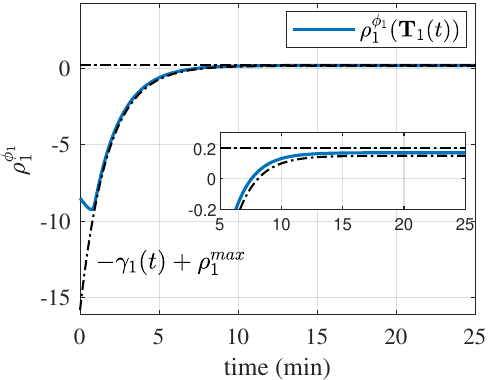}}
	\subcaptionbox{Prescribed region for  $\Sigma_2$ \label{fig:room_rho2}}
	{\includegraphics[width=.241\textwidth]{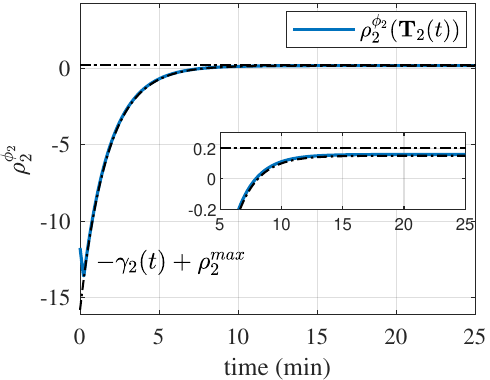}}
	\subcaptionbox{Prescribed region for  $\Sigma_{N-1}$ \label{fig:room_rho3}}
	{\includegraphics[width=.241\textwidth]{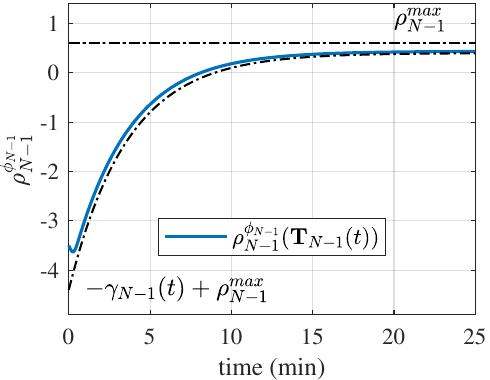}}
	\subcaptionbox{Prescribed region for  $\Sigma_{N}$ \label{fig:room_rho4}}
	{\includegraphics[width=.241\textwidth]{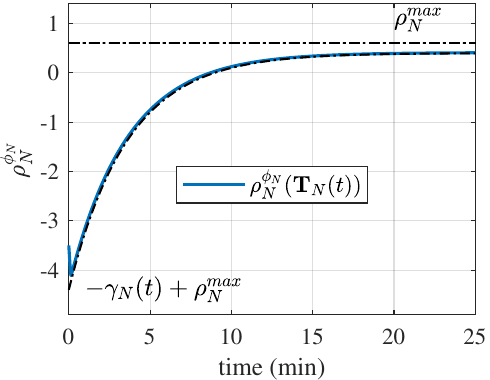}}
 
	\caption{Prescribed regions for the STL tasks for agents $\Sigma_1$, $\Sigma_2$, $\Sigma_{N-1}$ and $\Sigma_{N}$, where $N=1000$. Performance bounds are indicated by dashed lines. Evolution of the prescribed performance of $\rho_i^{\phi_{i}}(\mathbf{T}_i(t))$ are depicted using solid lines.}
	\label{fig:room}
 \vspace{-0.7cm}
\end{figure}

To enforce the STL tasks on this large-scale multi-agent system, we apply the proposed compositional framework by leveraging the results in Theorem \ref{thm:main} and 
applying the proposed PPC-based feedback controllers as in \eqref{controller}.  
Note that Assumptions \ref{assmprho1}-\ref{assmprho2} and \ref{assmp:lipschitz} on the STL formulae and the system dynamics are satisfied, and the task dependency graph also satisfies Assumption \ref{assmp:task}. We can thus apply the feedback controller as in \eqref{controller} on the agents to enforce STL tasks in a distributed manner. 
Numerical implementations were performed using MATLAB on a computer with a processor Intel Core i7 3.6 GHz CPU. 
Note that the computation of local controllers took on average less than 0.1 ms, which is negligible. The computation cost is very cheap since the local controller $\boldsymbol{u}_{i}$ is given by a closed-form expression and computed in a distributed manner. 
The simulation results for agents $\Sigma_1$, $\Sigma_2$, $\Sigma_{N-1}$ and $\Sigma_{N}$ are shown in Figs.~\ref{fig:room_temperature}-\ref{fig:room}. The state and input trajectories of the closed-loop agents are depicted in Fig.~\ref{fig:room_temperature}.  The shaded areas represent the desired temperature regions to be reached by the systems. 
%Fig.~\ref{fig:room_temperaturediff} showcases the temperature difference between rooms $\Sigma_1$  and $\Sigma_{N-1}$, and between rooms $\Sigma_2$ and $\Sigma_{N}$. 
As can be seen from Figs.~\ref{fig:room_temperature}, all these agents satisfy their desired STL tasks. 
In Fig.~\ref{fig:room}, we present the temporal behaviors of $\rho_i^{\psi_{i}}(\mathbf{T}_1(t))$ for the four rooms  $\Sigma_1$, $\Sigma_2$, $\Sigma_{N-1}$ and $\Sigma_{N}$. 
It can be readily seen that the prescribed performances of $\rho_i^{\psi_{i}}(\mathbf{T}_i(t))$ are satisfied, which indicates that the time bounds are also respected.
Remark that the design parameters of the prescribed regions are chosen according to the instructions listed in \eqref{funnelpara1}-\eqref{funnelpara2}, which guarantees the satisfaction of temporal formulae $ \rho_i^{\phi_i}(\mathbf{T}_i,0)$ by prescribing temporal behaviors of atomic formulae $\rho_i^{\psi_{i}}(\mathbf{T}_i(t))$ as in Fig.~\ref{fig:room}. 
We can conclude that all STL tasks are satisfied within the desired time interval. 
Note that the proposed compositional framework allows us to deal with multi-agent systems in a distributed manner, thus rendering the controller synthesis problem of large-scale multi-agent systems tractable. Moreover, the proposed local controller allows us to deal with STL tasks that not only depends on single agents but may also depend on multiple agents. Unlike the methods in \cite{lindemann2019feedback} which require all agents in a cluster to be subject to the same STL task, our designed local controllers can handle different STL tasks in the same cluster. 
Therefore, the methods in \cite{lindemann2019feedback}, cannot be applied to deal with the considered problem in this paper. 

% and thus a comparison is not possible. A comparison with the methods in \cite{lindemann2019robust,lindemann2019feedback} is thus not possible due to high computational complexity. 

 \vspace{-0.4cm}	

\subsection{Mobile Robot Control}

In this subsection, we demonstrate the effectiveness of the proposed results on a network of $N = 5$ mobile robots, where the dynamic of each robot is adapted from \cite{liu2008omni} with induced dynamical couplings.
Each mobile robot has three omni-directional wheels.
%Let $x_{i, j}$ with $j\in\{1,2,3\}$ denote the j-th element of agent $\Sigma_{i}$'s state and let $\boldsymbol{p}_{i}:=[x_{i, 1}; x_{i, 2}]$.[\boldsymbol{p}_{i}; x_{i, 3}]  
The dynamics of each robot $\Sigma_i$, $i\in\{1,2,\ldots,5\}$ can be described by 
\begin{equation*}
\Sigma_i:	\dot{\boldsymbol{x}}_{{i}} \!=\! A_i
\left(B_{i}^{\top}\right)^{-1}\!\!R_{i}\boldsymbol{u}_{i} \!-\!\! f_i(\boldsymbol{x}),
\end{equation*}
where the state variable of each robot is defined as $\boldsymbol{x}_{i}:=[x_{i, 1}; x_{i, 2};x_{i, 3}]$ with $\boldsymbol{p}_{i}:=[x_{i, 1}; x_{i, 2}]$ indicating the robot position and  state $x_{i, 3}$ indicating the robot orientation with respect to the $x_{i, 1}$-axis; $R_{i}: = 0.02$ m is the wheel radius of each robot;  $A_i :=\begin{bmatrix}
		\cos (x_{i, 3}) \! &\! -\sin (x_{i, 3}) \!&\! 0\\
		\sin (x_{i, 3}) \!&\! \cos (x_{i, 3}) \!&\! 0\\
		0 \!&\! 0 \!&\! 1
	\end{bmatrix} $, and
 $B_{i}: = \begin{bmatrix}0 & \cos(\pi/6) & -\cos(\pi/6)\\ -1 & \sin(\pi/6) & -\sin(\pi/6)\\ L_{i} & L_{i} & L_{i}\end{bmatrix}$ describes geometrical constraints with $L_{i}:=0.2$ m being the radius of the robot body; the term 
$f_i(\boldsymbol{x}) = \sum_{j\in \bl{\n_i^a}}k_i 
\frac{[\boldsymbol{p}_{i}  -   \boldsymbol{p}_{j};0]}{\Vert \boldsymbol{p}_{i} - \boldsymbol{p}_{j}\Vert+0.00001}$ is the induced dynamical coupling between agents that is used for the sake of collision avoidance, where
$\boldsymbol{x} = [\boldsymbol{x}_{1};\ldots;\boldsymbol{x}_{N}]$,  $k_i = 0.1$, and $\boldsymbol{p}_{j}$, $j \in \n_i^a$, are the positions of $\Sigma_i$'s adversarial neighboring agents with $\n_2^a = \{4\}$, $\n_3^a = \{4,5\}$, $\n_4^a = \{2,3\}$, $\n_5^a = \{3\}$. 
Each element of the input vector $\boldsymbol{u}_{i}$ corresponds to the angular rate of one wheel.  
The initial states of the robots are, respectively, $\boldsymbol{x}_{1}(0)=[2; 2; -\pi/4]$, $\boldsymbol{x}_{2}(0)=[4; 0; 0]$, $\boldsymbol{x}_{3}(0)=[6; 2; \pi/4]$, $\boldsymbol{x}_{4}(0)=[8; 0; 0]$,
$\boldsymbol{x}_{5}(0)=[10; 2; -\pi/4]$.

\begin{figure}[t]
	\centering
	\includegraphics[width=.22\textwidth]{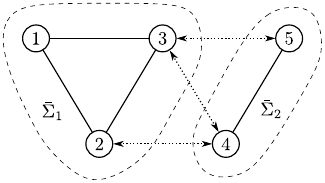}
 	\caption{The multi-robot system with two clusters $\bar \Sigma_1 = \{ \Sigma_1, \Sigma_2, \Sigma_3\}$ and $\bar \Sigma_2 = \{\Sigma_4, \Sigma_5\}$. The solid lines indicate the communication links between agents and the dotted lines indicate the dynamical couplings between clusters.} \label{fig:robotnetwork} 
     
 \vspace{-0.7cm}
\end{figure}

\begin{figure}[t]

	\centering
	\includegraphics[width=.32\textwidth]{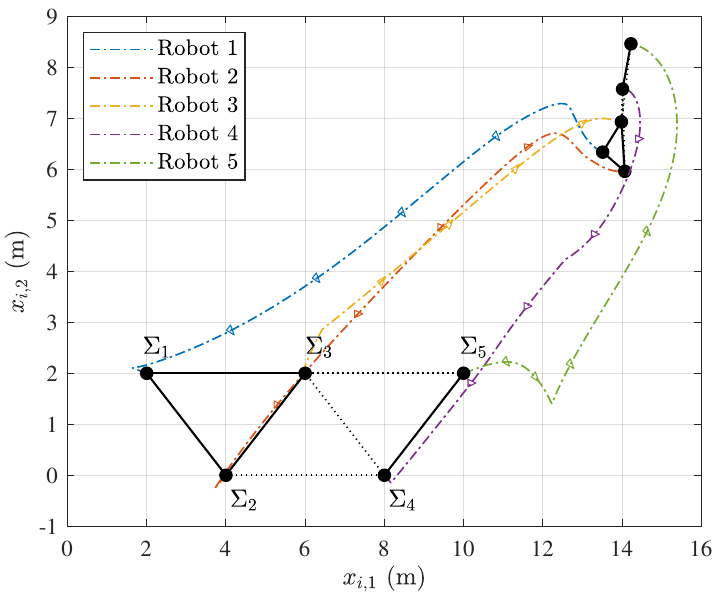}
 	\caption{State trajectories of the closed-loop robot systems on the position plane. Triangles indicate the orientation of robots. 
    %The initial states of the robots are, respectively, $\boldsymbol{x}_{1}(0)=[2; 2; -\pi/4]$, $\boldsymbol{x}_{2}(0)=[4; 0; 0]$, $\boldsymbol{x}_{3}(0)=[6; 2; \pi/4]$, $\boldsymbol{x}_{4}(0)=[8; 0; 0]$,
    %$\boldsymbol{x}_{5}(0)=[10; 2; -\pi/4]$. The goal positions of $\Sigma_3$ and $\Sigma_4$ are, respectively, $[14;7]$ and $[14; 7.5]$.
    } \label{fig:robot_position}
 %	\vspace{-0.5cm}
  \vspace{-0.4cm}
\end{figure}

\begin{figure}[!ht]
	\centering
%	{\includegraphics[width=.45\textwidth]{fig/distance.pdf}}

\subcaptionbox{Relative distances \label{fig:distance_sub}}
	{\includegraphics[width=.24\textwidth]{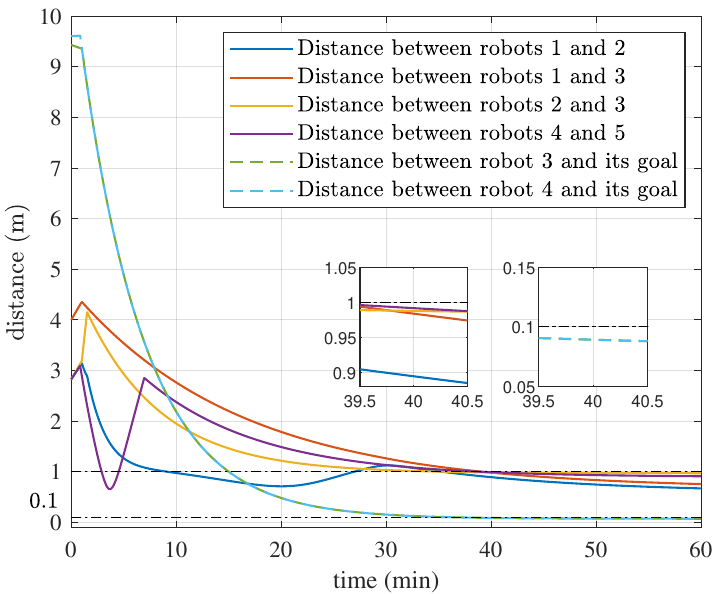}}
 \subcaptionbox{Control input for robot 5 \label{fig:input_5}}
	{\includegraphics[width=.24\textwidth]{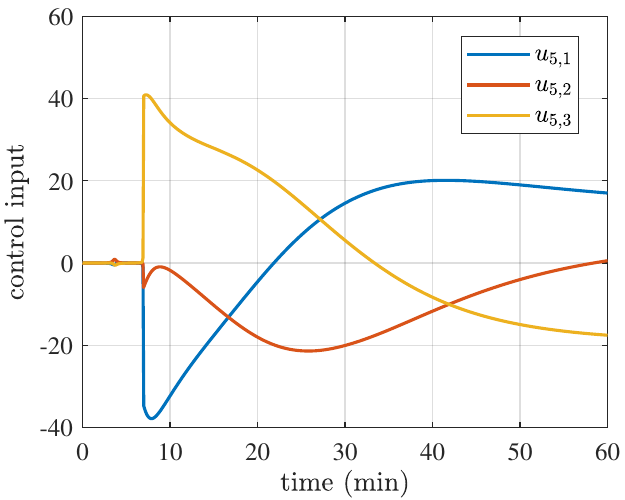}}
 	\caption{
  The evolution of the relative distances between the agents (or between an agent and its goal position) (a) and \bl{control input for robot 5 (b)} as time progresses.   
  \label{fig:distance}} 
   \vspace{-0.4cm}
\end{figure}

\begin{figure}[!ht]
	%\vspace{+ 0.2cm}
	\centering
\subcaptionbox{Prescribed region for $\Sigma_1$ \label{fig:robot_rho1}} {\includegraphics[width=.22\textwidth]{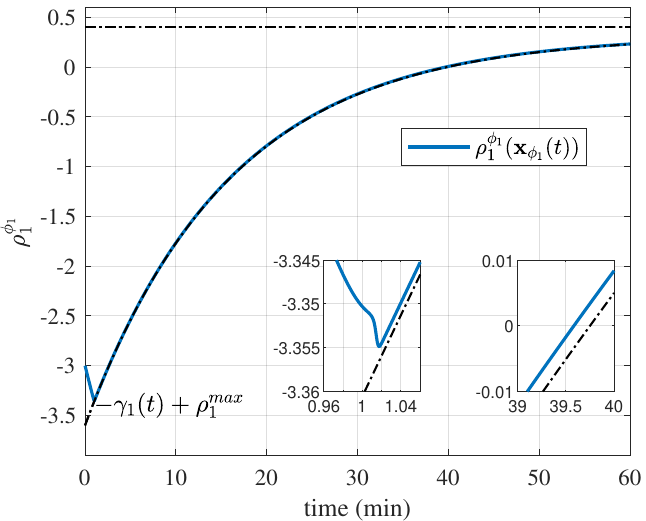}} 
	\subcaptionbox{Prescribed region for $\Sigma_2$  \label{fig:robot_rho2}}
	{\includegraphics[width=.22\textwidth]{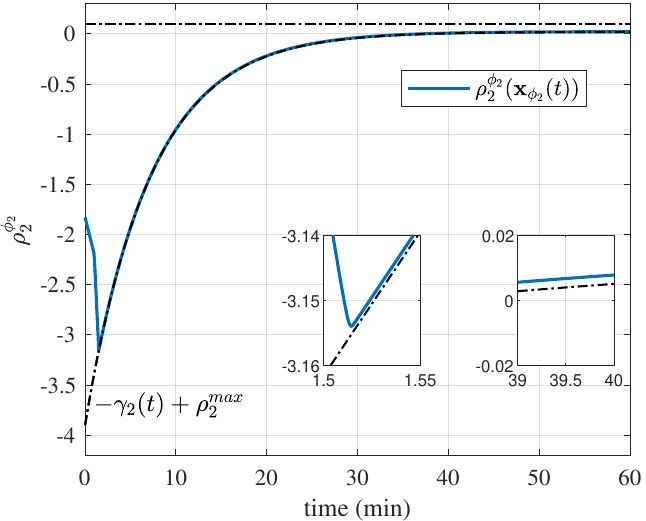}}
\subcaptionbox{Prescribed region for $\Sigma_3$ \label{fig:robot_rho3}}	{\includegraphics[width=.22\textwidth]{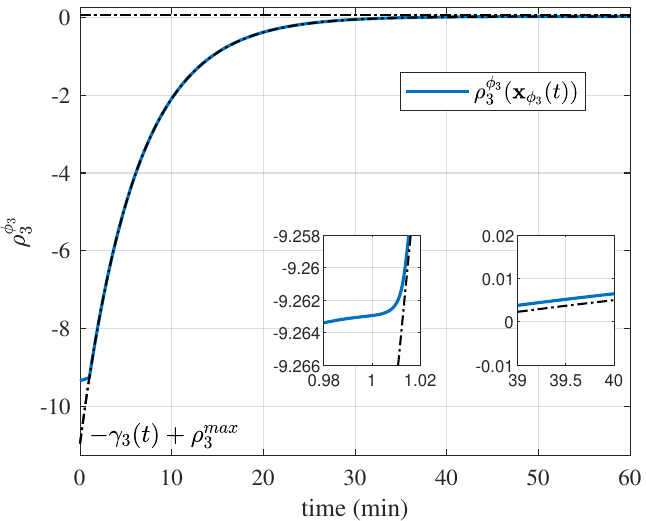}}
%\vspace{+ 0.2cm}
	\subcaptionbox{Prescribed region for $\Sigma_4$  \label{fig:robot_rho4}}
	{\includegraphics[width=.22\textwidth]{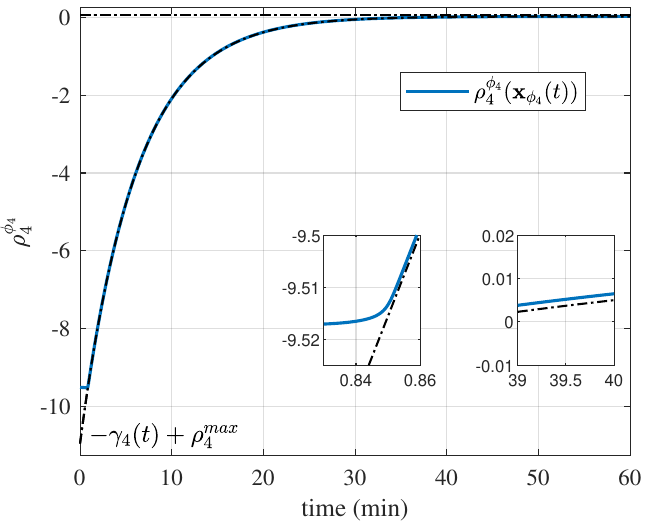}}
 	\subcaptionbox{Prescribed region for $\Sigma_5$ \label{fig:robot_rho5}}
	{\includegraphics[width=.22\textwidth]{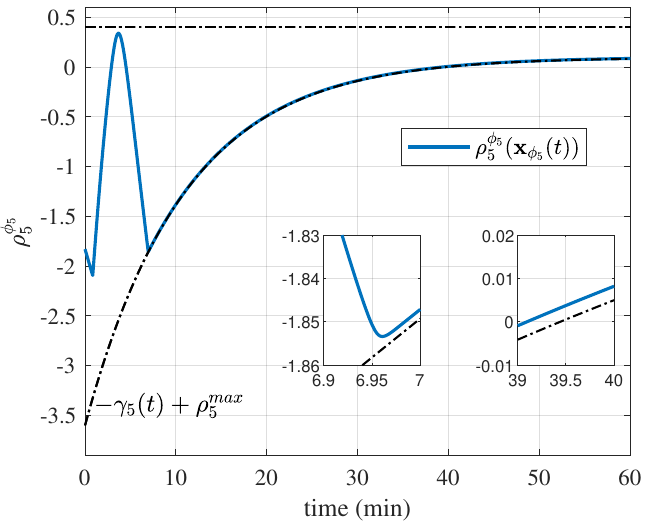}}

	\caption{The prescribed regions for the STL tasks. Performance bounds are indicated by dashed lines. Evolution of $\rho_i^{\phi_{i}}(\boldsymbol{x}_{\phi_i}(t))$ are depicted using solid lines.}
	\label{fig:robot}
 \vspace{-0.3cm}
\end{figure}

The robots are subject to the following STL tasks:
{\small
\begin{align*} 
    \phi_1 &: = F_{[0,40]}G_{[0,20]}((\Vert \boldsymbol{p}_{1}-\boldsymbol{p}_{2}\Vert  \leq 1) \wedge (\Vert \boldsymbol{p}_{1}-\boldsymbol{p}_{3}\Vert \leq 1)\\
    \phi_2 &: = F_{[0,40]}G_{[0,20]}((\Vert \boldsymbol{p}_{2}-\boldsymbol{p}_{3}\Vert  \leq 1)  \\
     & \qquad \qquad \qquad \qquad  \qquad \qquad \wedge (\vert \deg(x_{2,3}) - \deg(x_{3,3}) \vert \leq 7.5) \\
     \phi_3 &:= F_{[0,40]}G_{[0,20]}((\Vert \boldsymbol{p}_{3} -[14;7]\Vert \leq 0.1) \wedge(\vert \deg(x_{3,3}) \vert \leq 7.5) \\
    \phi_4 &:= F_{[0,40]}G_{[0,20]}((\Vert \boldsymbol{p}_{4}-[14;7.5]\Vert \leq 0.1) \wedge(\vert \deg(x_{4,3})\vert \leq 7.5) \\
  \phi_5 &: = F_{[0,40]}G_{[0,20]}((\Vert \boldsymbol{p}_{5} -\boldsymbol{p}_{4}\Vert \leq 1)
\end{align*}
}where $\deg(\cdot)$ converts angle units from radians to degrees.
Intuitively, robots 3 and 4 are assigned to move to their predefined goal points $[14;7]$ and $[14; 7.5]$, respectively, and stay there within the desired time interval, while satisfying the additional requirements on the robots' orientation; robot 2 is required to chase and follow robot 3 within the desired time interval, in the meanwhile keeping a similar orientation as robot 3; robot 2 is required to chase and follow robots 2 and 3 within the desired time interval; and robot 5 is required to chase and follow robot 4. 
Induced by the task dependency graph, we obtain two clusters from the multi-agent system: $\bar \Sigma_1 = \{ \Sigma_1, \Sigma_2, \Sigma_3\}$ and $\bar \Sigma_2 = \{\Sigma_4, \Sigma_5\}$, as depicted in Fig.~\ref{fig:robotnetwork}. 
%Note that in Fig.~\ref{fig:robotnetwork}, the solid lines indicate the communication links between agents, and the dotted lines indicate the dynamical couplings between clusters induced by the interconnection graph. 
The cluster interconnection graph in Fig.~\ref{fig:robotnetwork} is cyclic due to the dynamical couplings between the agents in different clusters.

To enforce the STL tasks on this 5-robot system, we apply the proposed compositional framework by leveraging the results in Theorem \ref{thm:main} and 
the proposed feedback controllers as in \eqref{controller}.  
Note that Assumptions \ref{assmprho1}-\ref{assmprho2} and \ref{assmp:lipschitz} on the STL formulae and the system dynamics are satisfied, and the task dependency graph also satisfies Assumption \ref{assmp:task}. We can thus apply the feedback controller as in \eqref{controller} on the agents to enforce the STL tasks in a distributed manner.  
Numerical implementations were performed using MATLAB on a computer with a processor Intel Core i7 3.6 GHz CPU. Note that the computation of local controllers took on average 0.1 ms, which is negligible since $\boldsymbol{u}_{i}$ is given by a closed-form expression and computed in a distributed manner. 
Simulation results are shown in Figs.~\ref{fig:robot_position}-\ref{fig:robot}. The state trajectories of each robot are depicted as in Fig.~\ref{fig:robot_position} on the position plane. The initial positions and final positions of the agents are represented by solid circles, the solid lines between the circles indicate communication links, and the dotted lines between the circles indicate dynamical couplings between the agents. 
In Fig.~\ref{fig:distance}, we show in subfigure (a) the evolution of the relative distances between the agents (or between an agent and its goal position) and in subfigure (b) the trajectory of control inputs for robot 5 as time progresses.  
As it can be seen from Figs.~\ref{fig:robot_position} and~\ref{fig:distance}, all agents satisfy their desired STL tasks. In particular, agents 3 and 4 finally achieved their tasks to reach and stay around their goal point within a certain distance (0.1 meter) after 40 minutes; agents 1, 2, and 5 achieved their tasks to chase and stay close to their desired agents within a certain distance (1 meter). 
In Fig.~\ref{fig:robot}, we further showcase the temporal behaviors of $\rho_i^{\phi_{i}}(\mathbf{x}_{\phi_i}(t))$ for the robots.
It can be seen that the prescribed performances of   $\rho_i^{\phi_{i}}(\mathbf{x}_{\phi_i}(t))$ are satisfied, which verifies that the time bounds are also respected.
% with respect to the error funnels. 
% evolution of the error funnels with respect to time is presented for each robot 
We can conclude that all STL tasks are satisfied within the desired time interval.

 \vspace{-0.5cm}

\section{Conclusions}
 
\begin{figure}[!t]
	\centering
	\includegraphics[width=.45\textwidth]{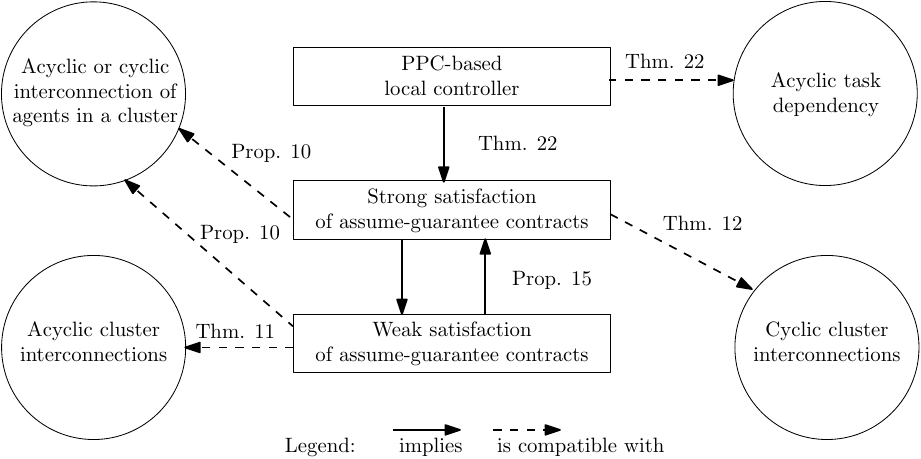}
 	\caption{Summary of the main results.} \label{fig:summary}
 %	\vspace{-0.5cm}
  \vspace{-0.7cm}
\end{figure}
We proposed a compositional approach for the synthesis of signal temporal logic tasks for large-scale multi-agent systems using assume-guarantee contracts. The notions and main results proposed in the paper and their relationships are sketched in Fig. \ref{fig:summary}.
Each agent in the multi-agent system is subject to collaborative tasks in the sense that it does not only depend on the agent itself but may also depend on other agents. 
The STL tasks are first translated to assume-guarantee contracts so that the satisfaction of a contract guarantees the satisfaction of the signal temporal logic task. 
Two concepts of contract satisfaction were introduced to establish our compositionality results, where weak satisfaction was shown to be sufficient to deal with acyclic interconnections, and strong satisfaction was needed for cyclic interconnections. 
We then derived a continuous-time closed-form feedback controller to enforce the uniform strong satisfaction of local contracts in a distributed manner, thus guaranteeing the satisfaction of global STL task for the multi-agent system based on the proposed compositionality result. 
Finally, the theoretical results were validated via two numerical case studies.

% \begin{figure}[t!]
% 	\vspace{+ 0.2cm}
% 	\centering
% 	\subcaptionbox{Funnel for $\Sigma_2$ with task $\psi_2$ \label{fig:robot_rho2}}
% 	{\includegraphics[width=.238\textwidth]{fig/robot_rho2}}
% 	\subcaptionbox{Funnel for $\Sigma_4$ with task $\psi_4$ \label{fig:robot_rho4}}
% 	{\includegraphics[width=.238\textwidth]{fig/robot_rho4}}
	%\quad
	
% 	\bigskip
% 	\subcaptionbox{Funnel for $\Sigma_3$ with task $\psi_3$ \label{fig:robot_rho3}}
% 	{\includegraphics[width=.235\textwidth]{fig/robot_rho3}}
% 	\subcaptionbox{Funnel for $\Sigma_4$ with task $\psi_4$ \label{fig:robot_rho4}}
% 	{\includegraphics[width=.235\textwidth]{fig/robot_rho4}}
	
% 	\bigskip
% 	\subcaptionbox{Funnel for $\Sigma_5$ with task $\psi_5$ \label{fig:robot_rho5}}
% 	{\includegraphics[width=.235\textwidth]{fig/robot_rho5}}
	%\quad
% 	\caption{Funnels for the local STL tasks. Performance bounds are indicated by dashed lines. Evolution of $\rho_i^{\psi_{i}}(\boldsymbol{x}_i(t))$ are depicted using solid lines.}
% 	\label{fig:robot}
% 	\vspace{-0.7cm}
% \end{figure}

\vspace{-0.3cm}
\appendix
%\section{Appendix}

%\counterwithin{definition}{section}
% \renewcommand{\thethm}{\thesection\arabic{thm}}% optional

\begin{definition} (Maximal dependency clusters) \label{def:cluster}
Consider a cluster of agents $\Sigma_i$, $i \!\in\! I_k\!=\! \{{k_1},\!\ldots\!,{k_{|I_k|}}\}$, $k \!\in\! \{1,\dots,K\}$, with each agent as described in \eqref{eqn:subsys}. The product system for the cluster of agents denoted by $\bar \Sigma_k \!=\!\mathcal{I}(\Sigma_{k_1},\!\dots\!,\Sigma_{k_{|I_k|}})$ %and equipped with a directed graph $\G = (I, \E)$, 
is a tuple $\bar \Sigma_k = (\bar X_k,\bar U_k,\bar W_k,\bar f_k,\bar g_k,\bar h_k)$ where 
\begin{enumerate} 
\item $\bar X_k = \prod_{i\in I_k} X_i$, $\bar U_k = \prod_{i\in I_k} U_i$, and $\bar W_k = \prod_{i\in I_k} W_i$ are the state,  external input spaces, and internal input spaces, respectively;
 
\item $\bar f_k $ is the flow drift, $\bar g_k $ is the external input matrix, and
 $\bar h_k $ is the internal input matrix 
defined as:
$\bar f_k (\bar{\mathbf{x}}_k(t)) =  [f_{k_1}(\mathbf{x}_{k_1}(t));
      \dots; f_{k_{|I_k|}}(\mathbf{x}_{k_{|I_k|}}(t))]$, 
    $\bar g_k (\bar{\mathbf{x}}_k(t)) = \diag(g_{k_1}(\mathbf{x}_{k_1}(t)),\dots,g_{k_{|I_k|}}(\mathbf{x}_{k_{|I_k|}}(t))$,
    $ \bar h_k(\bar{\mathbf{w}}_k(t)) = [h_{k_1}(\mathbf{w}_{k_1}(t));\dots;h_{k_{|I_k|}}(\mathbf{w}_{k_{|I_k|}}(t))],$
where  $\bar{\mathbf{x}}_k = [\mathbf{x}_{k_1};\dots;\mathbf{x}_{k_{|I_k|}}] \in \mathbb{R}^{\bar n_{k}} $ with $\bar n_{k} = {n_{k_1}} + \dots + {n_{k_{|I_k|}}}$,  $\bar{\mathbf{w}}_k =[\mathbf{w}_{k_1};\dots;\mathbf{w}_{k_{|I_k|}}] \in \mathbb{R}^{\bar p_{k}}$ with $\bar p_{k} = {p_{k_1}} + \dots + {p_{k_{|I_k|}}}$.

\end{enumerate} 
A trajectory of $\bar \Sigma_k$ is \blue{a} \bl{uniformly continuous} map $(\bar{\mathbf{x}}_k,\bar{\mathbf{w}}_k)\!:\!\mathbb{R}_{\geq 0} \!\rightarrow\!\bar X_k  \!\times\!\bar W_k $ such that 
%there exists an external input trajectory $\mathbf{u}_i\!:\! \mathbb{R}_{\geq 0} \!\rightarrow\! U_i$ such that 
for all $t \!\geq\! 0$  
\begin{equation}
\label{eqn:cluster}
  \dot {\bar{\mathbf{x}}}_k(t)  = \bar f_k(\bar{\mathbf{x}}_k(t))+ \bar g_k(\bar{\mathbf{x}}_k(t)) \bar{\mathbf{u}}_k(t)+\bar h_k(\bar{\mathbf{w}}_k(t)),
\end{equation}
where $\bar{\mathbf{u}}_k\!  =[\mathbf{u}_{k_1};\dots;\mathbf{u}_{k_{|I_k|}}]$. 
\end{definition}

\begin{definition} (Satisfaction of assume-guarantee contracts by the clusters) \label{agc_cluster}
	Consider a cluster of agents
	$\bar \Sigma_k$  an assume-guarantee contract  $\bar \C_k \!=\! (\bar A_k^a, \bar G_k)$ as in Definition \ref{def:asg_cluster}.

We say that $\bar \Sigma_k$  (\emph{weakly}) satisfies $\bar \C_k$, denoted by $\bar \Sigma_k \models \bar \C_k$, if for any trajectory  $(\bar{\mathbf{x}}_k,\bar{\mathbf{w}}_k)  \! : \! \mathbb{R}_{\geq 0}\! \rightarrow\! \bar X_k  \!\times\! \bar W_k$ of $\bar \Sigma_k$, the following holds:
%	for all $t \!\in\! \mathbb{R}_{\geq 0} $ such that  $\mathbf{w}_i(s) \!\in\! A_i^a(s)$ and $\mathbf{z}_i(s) \!\in\! A_i^c(s)$ for all $s \!\in\! [0,t]$, there exists $\delta_i>0$ such that   $(\mathbf{x}_i(s), \mathbf{z}_i(s))\!\in\! G_i(s)$ for all $s \in [0,t+\delta_i]$.
%	for all $t \in \mathbb{R}_{\geq 0} $ such that $\mathbf{z}_i(s) \!\in\! A_i^c(s)$ for all $s \in [0,t]$, 	there exists $\delta_i>0$ such that  $\mathbf{w}_i(s) \!\in\! A_i^a(s)$ for all  
	 for all $t \in \mathbb{R}_{\geq 0} $ such that  $\bar{\mathbf{w}}_k(s) \!\in\! \bar A_k^a(s)$ for all $s \in [0,t]$, we have 
  $ \bar{\mathbf{x}}_k(s)\in \bar G_k(s)$ for all $s \in [0,t]$.
		
%	\bl{To discuss: potential ways to relax the above definition?}
% \bl{To discuss: removing u in the definition of closed-loop systems trajectories?}
	
We say that $\bar \Sigma_k$ \emph{uniformly strongly} satisfies $\bar \C_k$, denoted by $\bar \Sigma_k \models_{us} \bar \C_k$, if for any trajectory $(\bar{\mathbf{x}}_k,\bar{\mathbf{w}}_k)  \! : \! \mathbb{R}_{\geq 0}\! \rightarrow\! \bar X_k  \!\times\! \bar W_k$ of $\bar \Sigma_k$, the following holds: 
	there exists $\bar \delta_k>0$ such that for all $t \in \mathbb{R}_{\geq 0} $, $\bar{\mathbf{w}}_k(s) \!\in\! \bar A_k^a(s)$ for all $s \in [0,t]$, we have 
 %$\bar{\mathbf{x}}_k(s)\in G_i(s)$ 
 $\bar{\mathbf{x}}_k(s)\in \bar G_k(s)$ 
 for all $s \in [0,t+\bar \delta_k]$.
\end{definition}
\vspace{-0.3cm}

\subsection{Supplementaries on prescribed performance control}\label{appendix:funnel}

In order to design feedback controllers to prescribe the transient behavior of $\rho_i^{\psi_i}(\mathbf{x}_{\phi_i}(t))$ within the predefined region:
\begin{align}\label{eq:appendixfunnel}
     - \gamma_i(t) < \rho_i^{\psi_i}(\mathbf{x}_{\phi_i}(t)) -  \rho_{i}^{max}  < 0,
\end{align}
one can translate the prescribed performance functions into notions of errors as follows. 
First, define a one-dimensional error as  
$e_i(\mathbf{x}_{\phi_i}(t)) = \rho_i^{\psi_i}(\mathbf{x}_{\phi_i}(t))-\rho_{i}^{max}$.
Now, by normalizing the error $e_i(\mathbf{x}_{\phi_i}(t))$ with respect to the prescribed performance function $\gamma_i$, we define the modulated error as 
$\hat e_i(\mathbf{x}_{\phi_i},t)=\frac{e_i(\mathbf{x}_{\phi_i}(t))}{\gamma_i(t)}.$
Then, \eqref{eq:appendixfunnel} can be rewritten as 
%$-\gamma_i(t) < e_i(\mathbf{x}_{\phi_i}(t)) < 0$, which in turn leads to 
$-1 < \hat e_i(t) < 0$. We use $\hat {\mathcal{D}_i} := (-1,0)$ 
%\{\hat e_i(t) \mid \hat e_i(t) \in (-1,0)\}$  
to denote the performance region for $\hat e_i(t)$.
Next, the modulated error is transformed through a transformation function  $T_i: (-1,0) \rightarrow \mathbb{R}$ defined as
$T_i(\hat e_i(\mathbf{x}_{\phi_i},t)) = \ln(-\frac{\hat e_i(\mathbf{x}_{\phi_i},t)+1}{\hat e_i(\mathbf{x}_{\phi_i},t)}).$
Note that the transformation function  $T_i: (-1,0) \rightarrow \mathbb{R}$ is a strictly increasing function, bijective and hence admitting an inverse. 
%By applying the transformation function $T$, the inequality
%For the sake of brevity, we omit the argument $t$ from  $\epsilon_i$
By differentiating the transformed error $\epsilon_i := T_i(\hat e_i(\mathbf{x}_{\phi_i},t)) $ w.r.t time, we obtain  
\begin{align}\label{transerror_dynamics}
\dot \epsilon_i  = \mathcal{J}_i(\hat e_i,t)[\dot e_i + \alpha_i(t)e_i],
\end{align}
where  $\mathcal{J}_i(\hat e_i,t)\!=\! \frac{\partial{T_i(\hat e_i)}}{\partial{\hat e_i}}\frac{1}{\gamma_i(t)} \!=\! -\frac{1}{\gamma_i(t)\hat e_i(1+ \hat  e_i)}\!>\!0$, for all $\hat e_i \!\in\! (-1,0)$, is the normalized Jacobian of the transformation function,
and $\alpha_i(t) \!= \!-\frac{\dot \gamma_i(t)}{\gamma_i(t)} \!>\!0$ for all $t \!\in \!\mathbb{R}_{\geq 0} $ is the normalized derivative of the performance function $\gamma_i$. 

It can be seen that, if the transformed error $\epsilon_i$ is bounded for all $t$, then the modulated error $\hat e_i$ is constrained within the performance region $\hat {\mathcal{D}_i}$, which further implies that the error $e_i$ evolves within the prescribed performance bounds as in \eqref{eq:appendixfunnel}. 

\vspace{-0.2cm}
\subsection{Proofs not contained in the main body}

{\bf Proof of Proposition~\ref{proposition1}.} 
We only provide the proof for the case of strong satisfaction of contracts. The case of weak satisfaction can be derived similarly.
Suppose that $\Sigma_i \models_{us} \C_i$ holds for all $i \in I_k$. Consider an arbitrary trajectory $(\bar{\mathbf{x}}_k,\bar{\mathbf{w}}_k)\!:\!\mathbb{R}_{\geq 0} \!\rightarrow\!\bar X_k  \!\times\!\bar W_k $ of the cluster $\bar \Sigma_k$. 
Let us show the existence of $\bar \delta_k>0$ such that
for all $t \in \mathbb{R}_{\geq 0}$ with $\bar{\mathbf{w}}_k(s) \in \bar A_k^a(s)$ for all $s \in [0,t]$, we have $\bar{\mathbf{x}}_k(s) \in \bar G_k(s)$ for all $s \in [0,t+\bar \delta_k]$.

First, given the trajectory $(\bar{\mathbf{x}}_k,\bar{\mathbf{w}}_k)\!:\!\mathbb{R}_{\geq 0} \!\rightarrow\!\bar X_k  \!\times\!\bar W_k $, we have by Definition \ref{def:cluster} that $\bar{\mathbf{x}}_k = [\mathbf{x}_{k_1};\dots;\mathbf{x}_{k_{|I_k|}}]$ and  $\bar{\mathbf{w}}_k =[\mathbf{w}_{k_1};\dots;\mathbf{w}_{k_{|I_k|}}]$, and thus, 
  $(\mathbf{x}_i,\mathbf{w}_i):\mathbb{R}_{\geq 0} \rightarrow X_i\times W_i$ is a trajectory of $\Sigma_i$ for all $i \in I_k$. 
  %, where $\mathbf{w}_i(t)  =[\mathbf{x}_{j_1}(t);\dots;\mathbf{x}_{j_{|\n_i^a|}}(t)]$. 
Now, consider any arbitrary $t \in \mathbb{R}_{\geq 0}$, such that $\bar{\mathbf{w}}_k(s) \in \bar A_k^a(s)$ for all $s \in [0,t]$, it follows that $\mathbf{w}_i(s) \in A_i^a$  for all $s \in [0,t]$ for all $i \in I_k$ since  $\bar A_k^a  = \prod_{i\in I_k}A_i^a$. 
Using the fact that $\Sigma_i \models_{us} \C_i$ for all $i \in I_k$, we have by Definition \ref{asg} that for each $i \in I_k$ there exists $\delta_i$ such that $(\mathbf{x}_i(s), \mathbf{z}_i(s))\in G_i(s)$ for all $s \in [0, t+\delta_i]$,  where $\mathbf{z}_i(t) =[\mathbf{x}_{j_1}(t);\dots;\mathbf{x}_{j_{|I_k|-1}}(t)]$ is the cooperative internal input (the stacked state of its cooperative agents that are involved in the same cluster).		
Let us define $\bar \delta_k = \min_{i \in I_k} \delta_i$.
Then, using the fact that $\bar G_k= \cap_{i \in I_k} G_i$ we get that 
$\bar{\mathbf{x}}_k(s) =  [\mathbf{x}_{k_1}(s);\dots;\mathbf{x}_{k_{|I_k|}}(s)] \in \bar G_k(s)$ for all $s \in [0, t+\bar \delta_k]$, where $G_i(s) \subseteq \bar X_k = X_{k_1} \times \ldots  \times X_i \times \ldots \times \blue{X_{k_{\left|I_k\right|}}}$.	
Therefore, we obtain that $\bar \Sigma_k \models_{us} \bar \C_k$.  
$\hfill\blacksquare$

{\bf Proof of Theorem~\ref{thm:main_dag}.} 	Let $\mathbf{x}\!:\!\mathbb{R}_{\geq 0} \!\rightarrow\! X $ be a trajectory of the multi-agent system $\Sigma$. 
	Then, from the definition of multi-agent systems, we have for all $i \in I$,  $(\mathbf{x}_i,\mathbf{w}_i):\mathbb{R}_{\geq 0} \rightarrow X_i\times W_i$ is a trajectory of $\Sigma_i$, where 
 $\mathbf{x} = [\mathbf{x}_1;\dots;\mathbf{x}_N]$, and
 $\mathbf{w}_i(t)  =[\mathbf{x}_{j_1}(t);\dots;\mathbf{x}_{j_{|\n_i^a|}}(t)]$. 
 %Let $\mathbf{z}_i(t) =[\mathbf{x}_{j_1}(t);\dots;\mathbf{x}_{j_{|I_k|-1}}(t)]$ be the stacked state of its cooperative agents that are involved in the same cluster.
%for any $t \in \mathbb{R}_{\geq 0}$, 
Since the cluster interconnection graph  $\bar{\mathcal{G}}^a=(\bar I,\bar E^a)$  is a directed acyclic graph, there exists at least one initial cluster that does not have adversarial internal inputs. 
Now, consider the initial clusters $\{\bar \Sigma_k\}_{k \in \bar I^{Init}}$.
First, we have by condition (i) that $\Sigma_i \models \C_i$ holds for all $i \in I$, and thus we obtain that  $\bar \Sigma_k \models \bar \C_k$ by Proposition \ref{proposition1}. This implies that for all $t \in \mathbb{R}_{\geq 0} $ with $\bar{\mathbf{w}}_k(s) \in \bar A_k^a(s)$ for all $s \in [0,t]$, we have $\bar{\mathbf{x}}_k(s) \in \bar G_k(s)$ for all $s \in [0, t]$.  Now \blue{consider an arbitrary $t \in \mathbb{R}_{\geq 0}$ with $\bar{\mathbf{w}}_k(s) \in \bar A_k^a(s)$ for all $s \in [0,t]$}.  Since the initial clusters do not have adversarial internal inputs, this implies:
\begin{align}\label{init}
\bar{\mathbf{x}}_k(s) \in \bar G_k(s), \forall s \in [0,t], \forall k \in \bar I^{Init}.
\end{align}
Next, let us prove by contradiction that for all $k\in \bar I$,  $\bar{\mathbf{x}}_k(s) \in \bar G_k(s)$, for all $s \in [0,t]$.
Let us assume that there exists $k \in  \bar I \setminus \bar I^{Init}$, such that  $\bar{\mathbf{x}}_k(s) \notin \bar G_k(s)$, for some $s \in [0,t]$. 
%Then,  there exists $i \in  I_k$, such that  $ (\mathbf{x}_i(s),\mathbf{z}_i(s)) \notin  G_{i}(s)$, for some $s \in [0,t]$.  
Since by condition (i) we have $\Sigma_i \models \C_i$ for all $i \in I$, which implies that $\bar \Sigma_k \models \bar \C_k$ by Proposition \ref{proposition1}, we have that  
$\bar{\mathbf{w}}_k(s) \notin \bar A_k^a(s)$ for some $s \in [0,t]$. 
Since $\bar{\mathbf{w}}_k =[\mathbf{w}_{k_1};\dots;\mathbf{w}_{k_{|I_k|}}]$, 
this means that for some $i\in I_k$, $\mathbf{w}_i(s) \notin A_{i}^a(s)$.
Then using the fact that: $\mathbf{w}_i(s)  =[\mathbf{x}_{j_1}(s);\dots;\mathbf{x}_{j_{|\n_i^a|}}(s)]$ and $\prod_{j\in \n_i^a} \overline{\textbf{Proj}}^j(G_{j}(s) \subseteq A_{i}^a(s)$, we deduce the existence of $j \in \n_i^a$ such that $\mathbf{x}_j(s) \notin \overline{\textbf{Proj}}^j(G_{j}(s))$, which further implies that $\bar{\mathbf{x}}_{k'}(s) \notin \bar G_{k'}(s)$, for some $s \in [0,t]$ where $j \in I_{k'}$. 
\bl{Note that $k' \neq k$, since we assumed that the agents in the same cluster are not dynamically coupled via adversarial internal inputs, i.e., $\n_i^a \cap \n_i^c = \emptyset$.}
\bl{Consider the cluster $I_{k'}$. If $k' \in I^{Init}$, this readily leads to a contradiction with  \eqref{init}; in the case that $k' \in  \bar I \setminus \bar I^{Init}$, by leveraging the same argument as above, we get that there exists $k'' \in  \bar I \setminus \{k, k'\}$ (due to the structure of a DAG and $\n_i^a \cap \n_i^c = \emptyset$) such that $\bar{\mathbf{x}}_{k''}(s) \notin \bar G_{k''}(s)$, for some $s \in [0,t]$.}
By further iterating this argument and the structure of DAG, there exists $l \in  \bar I^{Init}$ such that 
$\bar{\mathbf{x}}_l(s) \notin \bar G_l(s)$, which contradicts \eqref{init}. 
Hence, we have for all $k\in \bar I$,  $\bar{\mathbf{x}}_k(s) \in \bar G_k(s)$, for all $s \in [0,t]$, and thus, $\mathbf{x}(s) = [\bar{\mathbf{x}}_1(s);\ldots;\bar{\mathbf{x}}_K(s)] \in \prod_{k\in \bar I } \bar G_{k}(s) =G(s)$ for all $s \in [0,t]$. Therefore, $\Sigma \models \mathcal{C}$.
$\hfill\blacksquare$

{\bf Proof of Theorem~\ref{thm:main}.} 	
Let $\mathbf{x}\!:\!\mathbb{R}_{\geq 0} \!\rightarrow\! X$ be a trajectory of  the multi-agent system $\Sigma$. 
Then, from the definition of multi-agent systems, we have for all $i \in I$,  $(\mathbf{x}_i,\mathbf{w}_i):\mathbb{R}_{\geq 0} \rightarrow X_i\times W_i$ is a trajectory of $\Sigma_i$, where  $\mathbf{x} = [\mathbf{x}_1;\dots;\mathbf{x}_N]$, and $\mathbf{w}_i(t)  =[\mathbf{x}_{j_1}(t);\dots;\mathbf{x}_{j_{|\n_i^a|}}(t)]$. 
%Let $\mathbf{z}_i(t) =[\mathbf{x}_{j_1}(t);\dots;\mathbf{x}_{j_{|I_k|-1}}(t)]$ be the stacked state of its cooperative agents that are involved in the same cluster.
We prove  $\Sigma \models \mathcal{C}$ by inductively showing the existence of $\delta>0$ such that $\mathbf{x}(s) \in G(s)$ \blue{for all $s \in [0, n\delta]$ for all $n \in \mathbb{N}$}. 
First, we have from (i) that for all $i\in I$, $\mathbf{w}_i(0)\!=\![\mathbf{x}_{j_1}(0);\dots;\mathbf{x}_{j_{|\n_i^a|}}(0)] \in \prod_{j\in \n_i^a} \overline{\textbf{Proj}}^j(G_{j}(0)) \subseteq A_{i}^a(0)$, where the set inclusion follows from (iii). 
Now, consider the clusters $\bar \Sigma_k$, $k \in \bar I$, in the multi-agent system.
We have from (ii) that $\Sigma_i \models_{us} \C_i$ holds for all $i \in I$, and thus $\bar \Sigma_k \models_{us} \bar \C_k$ for all $k \in \bar I$ by 
Proposition \ref{proposition1}. 
Note that by the definition of clusters, we have  $\bar{\mathbf{w}}_k(0)  =[\mathbf{w}_{k_1}(0);\dots;\mathbf{w}_{k_{|I_k|}}(0)] \in \bar A_{k}^a(0)  = \prod_{i\in I_k}A_i^a(0)$, since $\mathbf{w}_i(0)\in A_{i}^a(0)$ holds for all $i \in I$.
Given that $\bar \Sigma_k \models_{us} \bar \C_k$, we thus have the existence of $\bar \delta_k >0$ such that for $\bar{\mathbf{w}}_k(0) \in \bar A_{k}^a(0)$, we have $ \bar{\mathbf{x}}_k(s)  =[\mathbf{x}_{k_1}(s);\dots;\mathbf{x}_{k_{|I_k|}}(s)] \in \bar G_k(s)$ for all $s \in [0,\bar \delta_k]$. 
Let us define $\delta>0$ as $\delta \!:=\! \min_{k \in \bar I} \bar \delta_k$.
Then, it follows that $\mathbf{x}(s) = [\bar{\mathbf{x}}_1(s);\ldots;\bar{\mathbf{x}}_K(s)] \in \prod_{k\in \bar I } \bar G_{k}(s) =G(s)$  for all $s \in [0,\delta ]$.  

Next, let us show that \blue{if $\mathbf{x}(s) \in G(s)$ holds for all $s \in [n\delta, (n+1)\delta)]$, then it implies that $\mathbf{x}(s) \in G(s)$ holds for all $s \in [(n+1)\delta, (n+2)\delta)]$}. 
First, let us assume that $\mathbf{x}(s) \in G(s)$ for all $s \in [n\delta, (n+1)\delta)]$ and show that $\mathbf{x}(s) \in G(s)$ for all $s \!\in\! [(n+1)\delta, (n+2)\delta)]$. 
By $\mathbf{x}(s) \in G(s)$ for all $s \in [n\delta, (n+1)\delta)]$, we obtain that $\bar{\mathbf{x}}_k(s) \in \bar G_k(s)$ for all $s \in [n\delta, (n+1)\delta)]$ for all $k\in \bar I$, 
since $\mathbf{x}(s) = [\bar{\mathbf{x}}_1(s);\ldots;\bar{\mathbf{x}}_K(s)] \in G(s) =\prod_{k\in \bar I } \bar G_{k}(s)$.
This further implies that $(\mathbf{x}_i(s), \mathbf{z}_i(s))\in G_i(s)$ for all $s \in [n\delta, (n+1)\delta)]$ for all $i \in I$, since 
$\bar{\mathbf{x}}_k(s) =  [\mathbf{x}_{k_1}(s);\dots;\mathbf{x}_{k_{|I_k|}}(s)] \in \bar G_k(s) = \cap_{i \in I_k} G_i(s)$, where $G_i(s) \subseteq \bar X_k = X_{k_1} \times \ldots  \times X_i \times \ldots \times \blue{X_{k_{\left|I_k\right|}}}$.	
Then, we obtain that 
for all $i\in I_k$, and for all $s \in [n\delta, (n+1)\delta]$, $\mathbf{w}_i(s)\!=\!$ $[\mathbf{x}_{j_1}(s);\dots;\mathbf{x}_{j_{|\n_i|}}(s)] \in \prod_{j\in \n_i^a}\overline{\textbf{Proj}}^j(G_{j}(s)) \subseteq A_i^a(s)$, where the set inclusion follows from (iii).
This implies that $\bar{\mathbf{w}}_k(s)  =[\mathbf{w}_{k_1}(s);\dots;\mathbf{w}_{k_{|I_k|}}(s)] \in \prod_{i\in I_k}A_i^a(s) = \bar A_{k}^a(s)$ for all $s \in [n\delta, (n+1)\delta]$. 
Again, since $\bar \Sigma_k \models_{us} \bar \C_k$, one gets  for all $k \in \bar I$,   
$ \bar{\mathbf{x}}_k(s)  =[\mathbf{x}_{k_1}(s);\dots;\mathbf{x}_{k_{|I_k|}}(s)] \in \bar G_k(s)$ for all $s \in [(n+1)\delta,(n+1)\delta+\bar \delta_k]$, which further implies that $ \bar{\mathbf{x}}_k(s)  \in \bar G_k(s)$ for all $s \in [(n+1)\delta,(n+2)\delta]$ since $\delta \!:=\! \min_{k \in \bar I} \bar \delta_k$. 
Hence, $\mathbf{x}(s) = [\bar{\mathbf{x}}_1(s);\ldots;\bar{\mathbf{x}}_K(s)] \in \prod_{k\in \bar I } \bar G_{k}(s) =G(s)$ for all $s \in [(n+1)\delta, (n+2)\delta)]$. 
Therefore, by induction, one has that $\mathbf{x}(s) \in G(s)$ \blue{for all $s \in [0, n\delta]$} for all $n \in \mathbb{N}$, and thus for all $s \geq 0$, which concludes that $\Sigma \models \mathcal{C}$.
$\hfill\blacksquare$

{\bf Proof of Proposition~\ref{weaktostrong}.}    Consider $\varepsilon>0$ such that $\Sigma_i \models \C_i^{\varepsilon}$. From uniform continuity of $\mathbf{w}_i:\mathbb{R}_{\geq 0} \rightarrow W_i$, and for $\varepsilon>0$, we have the existence of $\delta_i>0$ such that for all $t \geq 0$, if $\mathbf{w}_i(s) \in A_i^a(s)$, for all $s \in [0,t]$, then  $\mathbf{w}_i(s) \in  \mathcal{B}_{\varepsilon}(A_i^a)(s)$,  for all $s \in [0,t+\delta_i]$. 
 Let us now show the uniform strong satisfaction of contracts. Consider the $\delta_i >0$ defined above, and consider any $t \geq 0$ such that $\mathbf{w}_i(s) \in A_i^a(s)$ for all $s \in [0,t]$.
 First we have by the uniform continuity of $\mathbf{w}_i$, $\mathbf{w}_i(s) \in  \mathcal{B}_{\varepsilon}(A_i^a)(s)$,  for all $s \in [0,t+\delta_i]$. 
Hence, from the weak satisfaction of $\C_i^{\varepsilon}$, one has that  $(\mathbf{x}_i(s), \mathbf{z}_i(s))\in G_i(s)$ for all $s \in [0,t+\delta_i]$, which in turn implies the uniform strong satisfaction of the contract $\mathcal{C}_i$ according to Definition \ref{asg}.  
 %    Moreover, by the weak satisfaction of $\C_i^{\varepsilon}$, we further have 
	% there exists $\delta_i'>0$ such that for all $t \in \mathbb{R}_{\geq 0}$,  such that  $\mathbf{w}_i(s) \!\in\! \mathcal{B}_{\varepsilon}(A_i^a)(s)$ for all $s \in [0,t+\delta_i']$, 	and $\mathbf{z}_i(s) \!\in\! A_i^c(s)$ for all $s \in [0,t]$, we have $(\mathbf{x}_i(s), \mathbf{z}_i(s))\in G_i(s)$ for all $s \in [0,t+\delta_i']$.        
    % Hence, we have from above that $\mathbf{w}_i(s) \in A_i^a(s)$, for all $s \in [0,t+\delta_i+\delta_i']$, which implies that $(\mathbf{x}_i(s), \mathbf{z}_i(s))\in G_i(s)$, for all $s \in [0,t+\delta_i']$.
    Hence, $\Sigma_i \models_{us} \C_i$.
$\hfill\blacksquare$

{\bf Proof of Theorem~\ref{theorem}.}   
We prove the uniform strong satisfaction of the contract using Proposition \ref{weaktostrong}. Let $(\mathbf{x}_i,\mathbf{w}_i) : \mathbb{R}_{\geq 0}  \rightarrow X_i \times W_i$ be a trajectory of $\Sigma_i$. Since $-\gamma_i(0)+\rho_{i}^{max} < \rho_i^{\psi_i}(\mathbf{x}_i(0)) < \rho_{i}^{max}$ holds, we have $\mathbf{x}_i(0) \in G_{i}$. Now, consider  $\varepsilon >0$. Let us prove $\Sigma_i \models \mathcal{C}_i^{\varepsilon}$, where $\mathcal{C}_i^{\varepsilon} = (\mathcal{B}_{\varepsilon}(A_i^a), G_{i})$. 
\blue{Consider any $\mathbf{w}_i = [\mathbf{x}_{j_1} ;\dots;\mathbf{x}_{j_{|\n_i^a|}}]$ with $\mathbf{w}_i(t) \in \mathcal{B}_{\varepsilon}(A_{i}^a)(t)$ for all $t \in \mathbb{R}_{\geq 0}$. By Definition \ref{closeness},   
$\mathbf{w}_i \in \mathcal{B}_{\varepsilon}\left(A_i^a\right)$ implies that $\exists \tilde{\mathbf{w}}_i  = [\tilde{\mathbf{x}}_{j_1};\dots;\tilde{\mathbf{x}}_{j_{|\n_i^a|}}]  \in A_i^a$ such that $\forall t \geq 0$, $\left\|\mathbf{w}_i\left(t\right)-\tilde{\mathbf{w}}_i\left(t\right)\right\| \leq \varepsilon$, which further leads to $\left\|{\mathbf{x}}_{j}\left(t\right)-\tilde{\mathbf{x}}_{j}\left(t\right)\right\| \leq \varepsilon$, $\forall t \geq 0$, $\forall j \in \mathcal{N}_i^a$.
}
Let us recall that the STL robustness functions  $\rho_j^{\psi_j}$ are continuously differentiable in  $\mathbf{x}$, thus 
there exists a real constant $D \in \mathbb{R}_{\geq 0}$ such that  for any ${\mathbf{x}}_{j}\left(t\right)$ and \blue{$\tilde{\mathbf{x}}_{j}\left(t\right)$}, it holds that 
$\left|\rho_j^{\psi_j}\left({\mathbf{x}}_{j}\left(t\right)\right)-\rho_j^{\psi_j}\left(\tilde{\mathbf{x}}_{j}\left(t\right)\right)\right| \leq 
D \left\|{\mathbf{x}}_{j}\left(t\right)-\tilde{\mathbf{x}}_{j}\left(t\right)\right\| \leq D\varepsilon$. 
Also note that by the definition of $A_i^a$, $\tilde{\mathbf{w}}_i(t) \in A_i^a$ gives us for all $j \in \mathcal{N}_i^a$, $ -\gamma_{j}(t)+\rho_{j}^{max} <  \rho_j^{\psi_{j}}(\tilde{\mathbf{x}}_{j}(t)) < \rho_{j}^{max}, \forall t \geq 0$.
As a consequence, we get $ -\gamma_{j}(t)+\rho_{j}^{max} -D\varepsilon <  \rho_j^{\psi_j}\left({\mathbf{x}}_{j}\left(t\right)\right) < \rho_{j}^{max} + D\varepsilon, \forall t \geq 0$.
%Note that by definition of $A_{i}^a$ we have: for all $j \in  \n_i^a$, $-\gamma_{j}(t)+\rho_{j}^{max}-\varepsilon < \rho_j^{\psi_{j}}(\mathbf{x}_{j}(t)) < \rho_{j}^{max}+  \varepsilon $ holds for all $t \in \mathbb{R}_{\geq 0}$. 

Next, we show that ${\mathbf{x}_{\phi_i}}(t) \in G_{i}$. \bl{The underlying idea of proving this is based on showing that the transformed error $\epsilon_i$ is bounded. This is proved by showing that the prescribed performance region $\hat{\mathcal{D}}_i = (-1,0)$ for $\hat e_i$ is forward invariant.}  
% \begin{itemize}
% \item for all $t \in \mathbb{R}_{\geq 0} $, if $\mathbf{w}_{|[0,t]} \in A_W$, then $\mathbf{x}_{|[0,t]} \in G_X$.
% \end{itemize}
Consider the cluster of agents $\bar \Sigma_k$ with $i \in I_k$ where each agent is subject to STL task $\phi_i$. 
\bl{Let us define the stack vectors $\epsilon = [\epsilon_1;\ldots;\epsilon_{|I_k|}]$, $\hat{e} = [\hat{e}_1;\ldots;\hat{e}_{|I_k|}]$, 
and the prescribed performance region $\hat{\mathcal{D}} := \hat{\mathcal{D}}_{1} \times \ldots \times \hat{\mathcal{D}}_{|I_k|}$ for $\hat{e}$, where $\hat {\mathcal{D}_i} := (-1,0)$.}  
%Consider a potential function $V : \mathbb{R}^{|I_k|} \rightarrow \mathbb{R}_{\geq 0}$ defined as $V(\epsilon) = \frac{1}{2}\epsilon^{\top}\epsilon$. 
\bl{Consider a potential function 
 $V : \hat{\mathcal{D}} \rightarrow \mathbb{R}_{\geq 0}$ defined as $V(\hat e) = \frac{1}{2}\epsilon(\hat e)^{\top}\epsilon(\hat e)$ with $\epsilon_i (\hat e_i)$ defined as in  \eqref{eq:epsilon}.}
By differentiating $V$ with respect to time, we obtain $\dot V = \epsilon^{\top} \dot \epsilon$, where each $\dot \epsilon_i$ can be obtained by $ \dot \epsilon_i \stackrel{\eqref{transerror_dynamics}}=  
 \mathcal{J}_i(\hat e_i,t)[\dot e_i \!+\! \alpha_i(t)e_i] \blue{=} \mathcal{J}_i(\hat e_i,t)[ 
\blue{\frac{\partial{\rho_i^{\psi_i}(\mathbf{x}_{\phi_i})}}{\partial \bar{\mathbf{x}}_k}^\top \dot {\bar{\mathbf{x}}}_k(t)
 }
 -\! \dot \gamma_i(t)\hat e_i]$.  
%&- \epsilon_i\mathcal{J}_i(\hat e_i,t)\dot \gamma_i(t)\hat e_i,
Thus, we get
\begin{align} \label{dotV0}
    \dot V = \epsilon^{\top} \dot \epsilon=\epsilon^{\top}\mathcal{J}(\Gamma \dot {\bar{\mathbf{x}}}_k - \mathbf{p}),
\end{align}
where $\mathcal{J} \in \mathbb{R}^{|I_k|\times |I_k|} $ is a diagonal matrix with diagonal entries 
$\mathcal{J}_i(\hat e_i)\!=\! -\frac{1}{\gamma_i(t)\hat e_i(1+ \hat  e_i)}$, $i \in I_k$, $\Gamma \in \mathbb{R}^{|I_k|\times \bar n_{k} } $ is a matrix with row vectors $ \frac{\partial{\rho_i^{\psi_i}(\mathbf{x}_{\phi_i})}}{\partial \bar{\mathbf{x}}_k}^\top$, and $\mathbf{p} \in  \mathbb{R}^{|I_k|} $ is a column vector with entries  $\dot \gamma_i(t)\hat e_i$.
By inserting the dynamics of $\dot {\bar{\mathbf{x}}}_k $ \eqref{eqn:cluster} into \eqref{dotV0}, we obtain that 
\begin{align}  \notag
\!\!\!\!    \dot V =&\epsilon^{\top}\mathcal{J}(\Gamma (\bar f_k(\bar{\mathbf{x}}_k)+ \bar g_k(\bar{\mathbf{x}}_k) \bar{\mathbf{u}}_k+\bar h_k(\bar{\mathbf{w}}_k))- \mathbf{p})\\  \label{dotV1} 
    =&\epsilon^{\top}\mathcal{J}(\Gamma (\bar f_k(\bar{\mathbf{x}}_k)+ \bar h_k(\bar{\mathbf{w}}_k))- \mathbf{p}) + \epsilon^{\top}\mathcal{J}\Gamma\bar g_k(\bar{\mathbf{x}}_k) \bar{\mathbf{u}}_k, 
\end{align}
where 
$\bar{\mathbf{u}}_k\!  =[\mathbf{u}_{k_1};\dots;\mathbf{u}_{k_{|I_k|}}]$, with the local control law 
$\mathbf u_i= -g_i^\top( \mathbf{x}_i)
  \sum_{j \in I_k} ( \frac{\partial\rho_j^{\psi_j}({\mathbf{x}}_{\phi_j} )}{\partial \mathbf{x}_i} \mathcal{J}_j(\hat e_j,t)\epsilon_j({\mathbf{x}}_{\phi_j}, t))$
as in \eqref{controller}, $i \in I_k= \{{k_1},\ldots,{k_{|I_k|}}\}$.
Then, we get
\begin{small}
\begin{align} \label{term1}
 \!\!\!   \dot V =\epsilon^{\top}\mathcal{J}(\Gamma (\bar f_k(\bar{\mathbf{x}}_k)+ \bar h_k(\bar{\mathbf{w}}_k)- \Theta\mathbf{d} )- \mathbf{p} ) 
    -\epsilon^{\top}\mathcal{J}\Gamma \Theta \Lambda \mathcal{J} \epsilon, 
\end{align}
\end{small}where $\Theta := \diag(g_{k_i}(\mathbf{x}_{k_i})g_{k_i}^{\top}(\mathbf{x}_{k_i})) \in  \mathbb{R}^{\bar n_{k}\times \bar n_{k}} $, $\mathbf{d} \in  \mathbb{R}^{\bar n_{k}} $ is a column vector with entries $h_i(d_i)$, 
and  
$\Lambda \in \mathbb{R}^{\bar n_{k}\times |I_k|} $ is a matrix with column vectors $ \frac{\partial{\rho_i^{\psi_i}(\mathbf{x}_{\phi_i})}}{\partial \bar{\mathbf{x}}_k} = [\frac{\partial{\rho_i^{\psi_i}(\mathbf{x}_{\phi_i})}}{\partial {\mathbf{x}}_{i_1}}; \ldots; \frac{\partial{\rho_i^{\psi_i}(\mathbf{x}_{\phi_i})}}{\partial {\mathbf{x}}_{i_{P_i}}}]$. 
Recall that $\Gamma \in \mathbb{R}^{|I_k|\times \bar n_{k} } $ is a matrix with row vectors $ \frac{\partial{\rho_i^{\psi_i}(\mathbf{x}_{\phi_i})}}{\partial \bar{\mathbf{x}}_k}^\top $,
we get that $\Gamma = \Lambda ^\top$. Then by \eqref{term1}, and thus $-\epsilon^{\top}\mathcal{J}\Gamma \Theta \Lambda \mathcal{J} \epsilon    = -\epsilon^{\top}\mathcal{J}\Lambda ^\top\Theta \Lambda \mathcal{J} \epsilon$. 
Note that by Assumption \ref{assmp:task}, 
we get that $\Lambda$ is a block matrix that can be converted to an upper triangular form. Moreover, the diagonal entries of the block matrix $\Lambda$ are non-zero since $ \frac{\partial\rho_i^{\psi_i}({\mathbf{x}}_{\phi_i})}{\partial \mathbf{x}_i}$ are non-zero. Therefore, we can obtain that the rank of matrix  $\Lambda$ is $\text{rank}(\Lambda) = |I_k|$.
Note that according to Assumption \ref{assmp:lipschitz}, $g_i(\mathbf{x}_i)g_i(\mathbf{x}_i)^\top$ is positive definite.
Therefore, the matrix $\Lambda ^\top\Theta \Lambda$ is positive definite. 
Then, we can obtain that $\Lambda ^\top\Theta \Lambda \geq \alpha_L I_{|I_k|}$ for some $\alpha_L >0$.  
Recall that 
$\mathcal{J} \in \mathbb{R}^{|I_k|\times |I_k|} $ is a diagonal matrix with positive diagonal entries 
$\mathcal{J}_i(\hat e_i)\!=\! -\frac{1}{\gamma_i(t)\hat e_i(1+ \hat  e_i)}$, thus, 
$\mathcal{J}^2 \geq \alpha_J I_{|I_k|}$, where $\alpha_J = \min_{i \in I_k}\{\frac{1}{\sup_{t \in \mathbb{R}_{\geq 0}  }\gamma_i(t)^2}\min_{\hat e_i \in \hat{\mathcal{D}_i}} (\frac{1}{\hat e_i(1+ \hat  e_i)})^2\} >0$.
Hence, we obtain that 
$ -\epsilon^{\top}\mathcal{J}\Lambda ^\top\Theta \Lambda \mathcal{J} \epsilon \leq -\alpha_L\alpha_J\epsilon^{\top}\epsilon \leq -2\alpha_L\alpha_JV$.
Hence, we get the chain of inequality
\begin{small}
\begin{align} \notag
  \dot V =&\epsilon^{\top}\mathcal{J}(\Gamma (\bar f_k(\bar{\mathbf{x}}_k)+ \bar h_k(\bar{\mathbf{w}}_k)- \Theta\mathbf{d} )- \mathbf{p} ) 
     -\epsilon^{\top}\mathcal{J}\Lambda ^\top\Theta \Lambda \mathcal{J} \epsilon , \\\notag
    \leq&  \epsilon^{\top}\mathcal{J}(\Gamma (\bar f_k(\bar{\mathbf{x}}_k)+ \bar h_k(\bar{\mathbf{w}}_k)- \Theta\mathbf{d} )- \mathbf{p} ) \\ \notag
   & -\epsilon^{\top}\mathcal{J}(\alpha_L-\xi) \mathcal{J} \epsilon - \xi\Vert  \epsilon^{\top}\mathcal{J} \Vert^2 \\ \label{dotV2}
     \leq& -\kappa V + \eta(t)  
\end{align}
  \end{small}for some $\xi$ satisfying $0 < \xi < \alpha_L $, where $\kappa=2(\alpha_L-\xi)\alpha_J$,
%$\eta(t) = \Vert \epsilon\Vert \Vert \mathcal{J}\Vert (\Vert \Gamma\Vert  (\Vert \bar f_k(\bar{\mathbf{x}}_k)\Vert + \Vert \bar h_k(\bar{\mathbf{w}}_k)\Vert + \Vert \Theta\mathbf{d} \Vert )+ \Vert \mathbf{p} \Vert )$.
$\eta(t) = \frac{1}{4\xi}(\Vert \Gamma\Vert  (\Vert \bar f_k(\bar{\mathbf{x}}_k)\Vert + \Vert \bar h_k(\bar{\mathbf{w}}_k)\Vert + \Vert \Theta\mathbf{d} \Vert )+ \Vert \mathbf{p} \Vert )^2$.

Now, we proceed with finding an upper bound $\bar \eta$ of $\eta(t)$ \bl{for all $\hat e \in  \hat{\mathcal{D}}$ and all times $t \in \mathbb{R}_{\geq 0}$.}
\bl{Let us recall the definition of the modulated error  $\hat e_i(\mathbf{x}_{\phi_i},t) := \frac{\rho_i^{\psi_i}(\mathbf{x}_{\phi_i})-\rho_{i}^{max}}{\gamma_i(t)}$ and its corresponding prescribed performance region $\hat{\mathcal{D}_i} := (-1,0)$ for each $i \in I_k$.
Next, we define $\mathcal{X}_i(t) := \{\mathbf{x}_{\phi_i} \in \mathbb{R}^{n_{P_i}}| -1< \hat e_i(\mathbf{x}_{\phi_i},t) = \frac{\rho_i^{\psi_i}(\mathbf{x}_{\phi_i})-\rho_{i}^{max}}{\gamma_i(t)} <0 \}$ as the set of states $\mathbf{x}_{\phi_i}$ such that $\hat e_i(\mathbf{x}_{\phi_i},t) \in \hat{\mathcal{D}_i}$ at time $t \in \mathbb{R}_{\geq 0}$. 
One can also observe that $\mathcal{X}_i(t)$ has the property that for $t_1<t_2$, $\mathcal{X}_i(t_2) \subseteq  \mathcal{X}_i(t_1)$ holds since $\gamma_i(t)$ is non-increasing in $t$.
Thus, $\mathcal{X}_i(0)$ collects all states $\mathbf{x}_{\phi_i}$ such that $\hat e_i(\mathbf{x}_{\phi_i},t) \in \hat{\mathcal{D}_i}$ at all times $t \in \mathbb{R}_{\geq 0}$. 
Note also that $\mathcal{X}_i(0)$ is bounded due to condition (ii) of Assumption \ref{assmprho1} and $\gamma_i$ is bounded by definition, for all $i \in I_k$. 
}
%Since $\mathbf{x}_{\phi_i} \in \mathcal{X}_i(0)$ is bounded, we obtain that  $\mathbf{x}_{\phi_i}$ for all $i \in I_{\phi_i}$ are evolving in a bounded region. 
Thus, by the continuity of functions $f_i$, $g_i$ and $h_i$, 
%on $cl(\mathcal{X}_i(0))$, 
it holds that  $\Vert f_i(\mathbf{x}_i)\Vert$ and $\Vert g_i(\mathbf{x}_i)g_i(\mathbf{x}_i)^\top h_i(d_i(t))\Vert$
are upper bounded, where 
$d_i(t)=[\gamma_{j_1}(t)\mathbf{1}_{n_{j_1}};\dots;\gamma_{j_{|\n_i|}}(t)\mathbf{1}_{n_{|\n_i|}}]$.
Thus, $\Vert \bar f_k(\bar{\mathbf{x}}_k)\Vert$ and  $\Vert \Theta\mathbf{d} \Vert $ are bounded. 
Note that $\mathbf{p} $ is a column vector with entries  $\dot \gamma_i(t)\hat e_i$. Since $| \dot \gamma_i(t)\hat e_i| \leq |\dot \gamma_i(0)|$ and $|\dot \gamma_i(0)|$ is bounded by definition, $ \Vert \mathbf{p}  \Vert $ is bounded as well.
\bl{Moreover, recall from the beginning of the proof that by the definition of $A_{i}^a$ in the assume-guarante contract, we have $\mathbf{w}_i(t) = [\mathbf{x}_{j_1}(t);\dots;\mathbf{x}_{j_{|\n_i^a|}}(t)] \in \mathcal{B}_{\varepsilon}(A_{i}^a)$, which implies $ -\gamma_{j}(t)+\rho_{j}^{max} -D\varepsilon <  \rho_j^{\psi_j}\left({\mathbf{x}}_{j}\left(t\right)\right) < \rho_{j}^{max} + D\varepsilon, \forall t \geq 0$.
By further combining this with Assumption \ref{assmprho1}, we get that $\mathbf{w}_i(t)$ is bounded, and thus  
it holds that $\Vert h_i(\mathbf{w}_i) \Vert$ and $\Vert \bar h_k(\bar{\mathbf{w}}_k)\Vert$ are also upper bounded.}
Additionally, note that $\frac{\partial{\rho_i^{\psi_i}(\mathbf{x}_{\phi_i})}}{\partial \mathbf{x}_{\phi_i}} = 0$ if and only if  $\rho_i^{\psi_i}(\mathbf{x}_{\phi_i}) = \rho_i^{opt}$ since  $\rho_i^\psi(\mathbf{x}_{\phi_i})$ is concave under Assumption \ref{assmprho1}. 
However, since  
%$-\gamma_i(0)+\rho_{i}^{max} < \rho_i^\psi(\mathbf{x}_i(0)) < \rho_{i}^{max}$ and 
$ \rho_i^{\psi_i}(\mathbf{x}_{\phi_i}(0)) < \rho_{i}^{max} < \rho_i^{opt}$, and 
for all states $\mathbf{x}_{\phi_i} \in \mathcal{X}_i(0)$, $\rho_i^{\psi_i}(\mathbf{x}_{\phi_i}) < \rho_{i}^{max}$ holds,
then, we have for all states $\mathbf{x}_{\phi_i} \in \mathcal{X}_i(0)$, $\frac{\partial{\rho_i^{\psi_i}(\mathbf{x}_{\phi_i})}}{\partial \mathbf{x}_{\phi_i}} \neq 0_{n_i}$, and $\Vert \frac{\partial{\rho_i^{\psi_i}(\mathbf{x}_{\phi_i})}}{\partial \mathbf{x}_{\phi_i}} \Vert^2 \geq k_{\rho} >0$ holds for a positive constant $k_{\rho}$. 
Since $\Gamma $ is a matrix with row vectors $ \frac{\partial{\rho_i^{\psi_i}(\mathbf{x}_{\phi_i})}}{\partial \bar{\mathbf{x}}_k}^\top$, we get that $\Vert \Gamma  \Vert$ is bounded. 
% Let us denote by $k_f \in  \mathbb{R}_{\geq 0} $, $k_h \in  \mathbb{R}_{\geq 0} $, and $k_g \in  \mathbb{R}_{\geq 0} $ the upper bounds satisfying $\max_{\mathbf{x}_i \in \mathcal{X}_i(0)} \Vert f_i(\mathbf{x}_i)\Vert \leq k_f$, $\max_{\mathbf{w}_i \in \mathcal{B}_{\varepsilon}(A_{W_i})}$ $\Vert h_i(\mathbf{w}_i)\Vert \leq k_h$, and $\max_{\mathbf{x}_i \in \mathcal{X}_i(0)} \Vert g_i(\mathbf{x}_i)g_i(\mathbf{x}_i)^\top h_i(d_i(t))\Vert \leq k_g$, respectively. 
Consequently, we can define an upper bound $\bar \eta$ of $\eta(t)$, for all $\mathbf{x}_{\phi_i} \in \mathcal{X}_i(0)$ for all $i \in I_k$. 
%$\forall \mathbf{x}_i \in \mathcal{X}_i(0)$, $\forall t \in \mathbb{R}_{\geq 0} $, as $\bar \eta =  \frac{k_f^2+k_h^2+k_g^2}{\xi} + \frac{|\dot \gamma_i(0)|^2}{2\xi k_{\rho}}$, where $|\dot \gamma_i(0)|$ is bounded by definition.

Next, we show that $\hat{\mathcal{D}}$ is forward invariant. 
To do this, we first introduce a function $\mathcal{S}(\hat{e}) = 1- e^{-V(\hat{e})}$ for which $0 < \mathcal{S}(\hat{e}) <1$, $\forall \hat{e}_i \in \hat {\mathcal{D}_i}$ , $\forall i \in I_k$, and $\mathcal{S}(\hat{e}) \rightarrow 1$ as $\hat{e} \rightarrow \partial \hat {\mathcal{D}}$. By differentiating $\mathcal{S}(\hat{e})$ we get 
\begin{align}\label{dotmV}
\dot {\mathcal{S}}(t) = \dot V(\hat{e})\big( 1-\mathcal{S}(\hat{e})\big) .
\end{align}
By substituting \eqref{dotV2} and inserting $V(\hat{e}) = - \ln{(1-\mathcal{S}(\hat{e}))}$ in \eqref{dotmV}, we get
\begin{align}\label{dotmVb}
\hspace{-0.1cm}
\dot {\mathcal{S}}(t) \leq -\kappa \big(1-\mathcal{S}(\hat{e}) \big)\big(\ln{( e^{-\frac{\eta(t)}{\kappa}})}- \ln{(1-\mathcal{S}(\hat{e}))}\big).
\end{align}
Note that by definition, we have $\kappa>0$ and $1-\mathcal{S}(\hat{e})> 0$. 
Now define the region $\Omega_{\hat e} = \{\hat e \in \hat{\mathcal{D}} | \mathcal{S}(\hat e) \leq 1- e^{-\frac{\bar \eta}{\kappa}}\}$. 
%It can be readily seen that  $\dot {\mathcal{S}}(t) < 0$ holds for all $\hat e_i \notin \Omega_{\hat e_i}$. 
%Next, we show that $\hat{\mathcal{D}_i}$ is a set of attraction.
Since $-\gamma_i(0)+\rho_{i}^{max} < \rho_i^{\psi_i}(\mathbf{x}_{\phi_i}(0)) < \rho_{i}^{max}$ holds $\forall i \in I_k$, then we obtain that $\hat e_i(\mathbf{x}_{\phi_i}(0)) \in \hat{\mathcal{D}_i} = (-1,0)$ $\forall i \in I_k$,
%which means that the initial error are defined within the prescribed performance bounds.
and consequently, $\mathcal{S}(\hat{e}(0)) <1$ holds. Let us define $c = \mathcal{S}(\hat{e}(0))$ and the set $\Omega_c = \{ \hat e \in \hat{\mathcal{D}} | \mathcal{S}(\hat e) \leq c \}$. Now, consider the case when $c < 1- e^{-\frac{\bar \eta}{\kappa}}$. In this case, $\Omega_c \subset \Omega_{\hat e}$, and by \eqref{dotmVb}, $\dot {\mathcal{S}}(t) \leq 0$ for all  $\hat e  \in \partial{\Omega_{\hat e}}$, therefore, $\hat e(t) \in \Omega_{\hat e}$, $\forall t \in \mathbb{R}_{\geq 0} $. Next, consider the other case when $c \geq 1- e^{-\frac{\bar \eta}{\kappa}}$. In this case $\Omega_{\hat e} \subseteq \Omega_c$, and by \eqref{dotmVb}, $\dot {\mathcal{S}}(t) < 0$ for all  $\hat e \in \Omega_c \setminus \Omega_{\hat e}$, hence, $\mathcal{S}(\hat e) \rightarrow \Omega_{\hat e}$. Thus, starting from any point within the set $\Omega_c$, $\mathcal{S}(\hat e(t))$ remains less than $1$. Consequently, the modulated error $\hat e$ always evolves within a closed strict subset of  $\hat{\mathcal{D}}$  (that is, set $\Omega_{\hat e}$ in the case that $c < 1- e^{-\frac{\bar \eta}{\kappa}}$,  or set $\Omega_c$ in the case that $c \geq 1- e^{-\frac{\bar \eta}{\kappa}}$), which implies that $\hat e$ is not approaching the boundary $\partial \hat {\mathcal{D}}$. It follows that  $\epsilon$ is bounded, and thus, $\epsilon_i$ is bounded for all $i \in I_k$.
Thus, we can conclude that  $\rho_i^{\psi_i}(\mathbf{x}_{\phi_i}(t))$ evolves within the predefined region \eqref{goalineq}, i.e., ${\mathbf{x}_{\phi_i}}(t) \in G_{i}$ for all $t \in \mathbb{R}_{\geq 0}$. 
%the guarantee $G_{X_i}$ for agent $\Sigma_i$ is satisfied. 
Therefore, we have $\Sigma_i \models \mathcal{C}_i^{\varepsilon}$. By Proposition \ref{weaktostrong}, it implies that $\Sigma_i \models_{us} \mathcal{C}_i$.
$\hfill\blacksquare$

Part of the proof above was inspired by \cite[Thm. 1]{karayiannidis2012multi}, where similar Lyapunov arguments were used in the context of PPC-based control of multi-agent systems. However, the results there deal with consensus control of multi-agent systems only and cannot handle temporal logic properties.

{\bf Proof of Corollary~\ref{corollary}.}  
	From Theorem \ref{theorem}, one can verify that the closed-loop agents under controller \eqref{controller} satisfy: 
	for all $i \in I$, $\Sigma_i \models_{us} \C_i$, and  for all $i \in I$, $\prod_{j\in \n_i^a} \overline{\textbf{Proj}}^j(G_{j}) \subseteq A_{i}^a$.
	Moreover, for all $i \in I$ and for any trajectory $(\mathbf{x}_i,\mathbf{w}_i):\mathbb{R}_{\geq 0} \rightarrow X_i\times W_i$ of $\Sigma_i$, the choice of parameters of the prescribed regions as in \eqref{funnelpara1}--\eqref{funnelpara2} ensures that $\mathbf{x}_i(0) \in \overline{\textbf{Proj}}^i(G_i)(0)$.
Hence, all conditions required in Theorem \ref{thm:main} are satisfied.
\bl{Note that since $\Sigma_i \models_{us} \mathcal{C}_i$ implies $\Sigma_i \models \mathcal{C}_i$, the conditions required in Theorem \ref{thm:main_dag} hold as well. 
Thus, irrespective of whether the dynamical interconnection graph is acyclic or cyclic, 
we can conclude that $\Sigma \models \C  = (\emptyset,\prod_{k\in \bar I } \bar G_{k})$ as a consequence of Theorems~\ref{thm:main} and~\ref{thm:main_dag}. } 
%, and thus, we conclude that $\Sigma \models \C  = (\emptyset,\prod_{k\in \bar I } \bar G_{k})$ as a consequence of Theorem \ref{thm:main}. 
Therefore, the multi-agent system $\Sigma$ satisfies the STL task  $\bar\phi=\land_{i=1}^N\phi_i$.
$\hfill\blacksquare$

%\vspace{-0.25cm}
%\newpage
 
%\section*{References}
%\vspace{-0.5cm}
\bibliography{biblio}

\begin{thebibliography}{10}
\providecommand{\url}[1]{#1}
\csname url@samestyle\endcsname
\providecommand{\newblock}{\relax}
\providecommand{\bibinfo}[2]{#2}
\providecommand{\BIBentrySTDinterwordspacing}{\spaceskip=0pt\relax}
\providecommand{\BIBentryALTinterwordstretchfactor}{4}
\providecommand{\BIBentryALTinterwordspacing}{\spaceskip=\fontdimen2\font plus
\BIBentryALTinterwordstretchfactor\fontdimen3\font minus \fontdimen4\font\relax}
\providecommand{\BIBforeignlanguage}[2]{{%
\expandafter\ifx\csname l@#1\endcsname\relax
\typeout{** WARNING: IEEEtran.bst: No hyphenation pattern has been}%
\typeout{** loaded for the language `#1'. Using the pattern for}%
\typeout{** the default language instead.}%
\else
\language=\csname l@#1\endcsname
\fi
#2}}
\providecommand{\BIBdecl}{\relax}
\BIBdecl

\bibitem{tanner2003stable}
H.~G. Tanner, A.~Jadbabaie, and G.~J. Pappas, ``Stable flocking of mobile agents, part i: Fixed topology,'' in \emph{42nd Conf. Decis. Control}, vol.~2.\hskip 1em plus 0.5em minus 0.4em\relax IEEE, 2003, pp. 2010--2015.

\bibitem{ren2005consensus}
W.~Ren and R.~W. Beard, ``Consensus seeking in multiagent systems under dynamically changing interaction topologies,'' \emph{IEEE Trans. Autom. Control}, vol.~50, no.~5, pp. 655--661, 2005.

\bibitem{olfati2007consensus}
R.~Olfati-Saber, J.~A. Fax, and R.~M. Murray, ``Consensus and cooperation in networked multi-agent systems,'' \emph{Proceedings of the IEEE}, vol.~95, no.~1, pp. 215--233, 2007.

\bibitem{cortes2004coverage}
J.~Cortes, S.~Martinez, T.~Karatas, and F.~Bullo, ``Coverage control for mobile sensing networks,'' \emph{IEEE Transactions on robotics and Automation}, vol.~20, no.~2, pp. 243--255, 2004.

\bibitem{mesbahi2010graph}
M.~Mesbahi and M.~Egerstedt, \emph{Graph theoretic methods in multiagent networks}.\hskip 1em plus 0.5em minus 0.4em\relax Princeton University Press, 2010.

\bibitem{belta2017formal}
C.~Belta, B.~Yordanov, and E.~A. Gol, \emph{Formal methods for discrete-time dynamical systems}.\hskip 1em plus 0.5em minus 0.4em\relax Springer, 2017, vol.~15.

\bibitem{tabuada2009verification}
P.~Tabuada, \emph{Verification and control of hybrid systems: a symbolic approach}.\hskip 1em plus 0.5em minus 0.4em\relax Springer Science \& Business Media, 2009.

\bibitem{kloetzer2008fully}
M.~Kloetzer and C.~Belta, ``A fully automated framework for control of linear systems from temporal logic specifications,'' \emph{IEEE Trans. Autom. Control}, vol.~53, no.~1, pp. 287--297, 2008.

\bibitem{fainekos2009temporal}
G.~E. Fainekos, A.~Girard, H.~Kress-Gazit, and G.~J. Pappas, ``Temporal logic motion planning for dynamic robots,'' \emph{Automatica}, vol.~45, no.~2, pp. 343--352, 2009.

\bibitem{loizou2004automatic}
S.~G. Loizou and K.~J. Kyriakopoulos, ``Automatic synthesis of multi-agent motion tasks based on ltl specifications,'' in \emph{43rd Conf. Decis. Control}, vol.~1.\hskip 1em plus 0.5em minus 0.4em\relax IEEE, 2004, pp. 153--158.

\bibitem{guo2015multi}
M.~Guo and D.~V. Dimarogonas, ``Multi-agent plan reconfiguration under local ltl specifications,'' \emph{Int. J. Robot Res.}, vol.~34, no.~2, pp. 218--235, 2015.

\bibitem{liu2017distributed}
Z.~Liu, B.~Wu, J.~Dai, and H.~Lin, ``Distributed communication-aware motion planning for multi-agent systems from stl and spatel specifications,'' in \emph{56th Conf. Decis. Control}, 2017, pp. 4452--4457.

\bibitem{jiang1994small}
Z.~P. Jiang, A.~R. Teel, and L.~Praly, ``Small-gain theorem for iss systems and applications,'' \emph{Mathematics of Control, Signals and Systems}, vol.~7, pp. 95--120, 1994.

\bibitem{7857702}
M.~Zamani and M.~Arcak, ``Compositional abstraction for networks of control systems: A dissipativity approach,'' \emph{IEEE Trans. Control Netw. Syst.}, vol.~5, no.~3, pp. 1003--1015, 2018.

\bibitem{kim2017small}
E.~S. Kim, M.~Arcak, and S.~A. Seshia, ``A small gain theorem for parametric assume-guarantee contracts,'' in \emph{20th Int. Conf. Hybrid Syst., Comput. Control}, 2017, pp. 207--216.

\bibitem{jagtap2020compositional}
P.~Jagtap, A.~Swikir, and M.~Zamani, ``Compositional construction of control barrier functions for interconnected control systems,'' in \emph{23rd Int. Conf. Hybrid Syst., Comput. Control}, 2020, pp. 1--11.

\bibitem{liu2021symbolic}
S.~Liu, N.~Noroozi, and M.~Zamani, ``Symbolic models for infinite networks of control systems: A compositional approach,'' \emph{Nonlinear Analysis: Hybrid Systems}, vol.~43, p. 101097, 2021.

\bibitem{saoud2021assume}
A.~Saoud, A.~Girard, and L.~Fribourg, ``Assume-guarantee contracts for continuous-time systems,'' \emph{Automatica}, vol. 134, p. 109910, 2021.

\bibitem{sharf2021assume}
M.~Sharf, B.~Besselink, A.~Molin, Q.~Zhao, and K.~H. Johansson, ``Assume/guarantee contracts for dynamical systems: Theory and computational tools,'' \emph{IFAC-PapersOnLine}, vol.~54, no.~5, pp. 25--30, 2021.

\bibitem{saoud2020contract}
A.~Saoud, A.~Girard, and L.~Fribourg, ``Contract-based design of symbolic controllers for safety in distributed multiperiodic sampled-data systems,'' \emph{IEEE Trans. Autom. Control}, vol.~66, no.~3, pp. 1055--1070, 2020.

\bibitem{ghasemi2020compositional}
K.~Ghasemi, S.~Sadraddini, and C.~Belta, ``Compositional synthesis via a convex parameterization of assume-guarantee contracts,'' in \emph{23rd Int. Conf. Hybrid Syst., Comput. Control}, 2020, pp. 1--10.

\bibitem{barttac}
B.~M. Shali, A.~van~der Schaft, and B.~Besselink, ``Composition of behavioural assume-guarantee contracts,'' \emph{IEEE Trans. Autom. Control}, pp. 1--16, 2022.

\bibitem{benveniste2018contracts}
A.~Benveniste, B.~Caillaud, D.~Nickovic, R.~Passerone, J.-B. Raclet, P.~Reinkemeier, A.~Sangiovanni-Vincentelli, W.~Damm, T.~A. Henzinger, K.~G. Larsen \emph{et~al.}, ``Contracts for system design,'' 2018.

\bibitem{maler2004monitoring}
O.~Maler and D.~Nickovic, ``Monitoring temporal properties of continuous signals,'' in \emph{{F}{O}{R}{M}{A}{T}{S} - {F}{T}{R}{T}{F}{T}}, 2004, pp. 152--166.

\bibitem{donze2010robust}
A.~Donz{\'e} and O.~Maler, ``Robust satisfaction of temporal logic over real-valued signals,'' in \emph{Int. Conf. {F}{O}{R}{M}{A}{T}{S} Syst.}, 2010, pp. 92--106.

\bibitem{raman2014model}
V.~Raman, A.~Donz{\'e}, M.~Maasoumy, R.~M. Murray, A.~Sangiovanni-Vincentelli, and S.~A. Seshia, ``Model predictive control with signal temporal logic specifications,'' in \emph{53rd Conf. Decis. Control}.\hskip 1em plus 0.5em minus 0.4em\relax IEEE, 2014, pp. 81--87.

\bibitem{larsCDC}
L.~Lindemann, C.~K. Verginis, and D.~V. Dimarogonas, ``Prescribed performance control for signal temporal logic specifications,'' in \emph{56th Conf. Decis. Control}, 2017, pp. 2997--3002.

\bibitem{lindemann2019feedback}
L.~Lindemann and D.~V. Dimarogonas, ``Feedback control strategies for multi-agent systems under a fragment of signal temporal logic tasks,'' \emph{Automatica}, vol. 106, pp. 284--293, 2019.

\bibitem{lindemann2020barrier}
------, ``Barrier function based collaborative control of multiple robots under signal temporal logic tasks,'' \emph{IEEE Control Netw. Syst.}, vol.~7, no.~4, pp. 1916--1928, 2020.

\bibitem{ye2024decentralized}
H.~Ye, C.~Wen, and Y.~Song, ``Decentralized and distributed control of large-scale interconnected multi-agent systems in prescribed time,'' \emph{IEEE Transactions on Automatic Control}, 2024.

\bibitem{nuzzo2015compositional}
P.~Nuzzo, ``Compositional design of cyber-physical systems using contracts,'' Ph.D. dissertation, UC Berkeley, 2015.

\bibitem{saoud2019compositional}
A.~Saoud, ``Compositional and efficient controller synthesis for cyber-physical systems,'' Ph.D. dissertation, Universit{\'e} Paris-Saclay, 2019.

\bibitem{liu2022compositional}
S.~Liu, A.~Saoud, P.~Jagtap, D.~V. Dimarogonas, and M.~Zamani, ``Compositional synthesis of signal temporal logic tasks via assume-guarantee contracts,'' in \emph{61st Conf. Decis. Control}, 2022, pp. 2184--2189.

\bibitem{fainekos2009robustness}
G.~E. Fainekos and G.~J. Pappas, ``Robustness of temporal logic specifications for continuous-time signals,'' \emph{Theor. Comput. Sci.}, vol. 410, no.~42, pp. 4262--4291, 2009.

\bibitem{aksaray2016q}
D.~Aksaray, A.~Jones, Z.~Kong, M.~Schwager, and C.~Belta, ``Q-learning for robust satisfaction of signal temporal logic specifications,'' in \emph{55th Conf. Decis. Control}.\hskip 1em plus 0.5em minus 0.4em\relax IEEE, 2016, pp. 6565--6570.

\bibitem{lindemann2020efficient}
L.~Lindemann and D.~V. Dimarogonas, ``Efficient automata-based planning and control under spatio-temporal logic specifications,'' in \emph{Amer. Control Conf.}\hskip 1em plus 0.5em minus 0.4em\relax IEEE, 2020, pp. 4707--4714.

\bibitem{sastry2013nonlinear}
S.~Sastry, \emph{Nonlinear systems: analysis, stability, and control}.\hskip 1em plus 0.5em minus 0.4em\relax Springer Science \& Business Media, 2013, vol.~10.

\bibitem{isidori2013nonlinear}
A.~Isidori, \emph{Nonlinear control systems II}.\hskip 1em plus 0.5em minus 0.4em\relax Springer.

\bibitem{marchesini2024communication}
G.~Marchesini, S.~Liu, L.~Lindemann, and D.~V. Dimarogonas, ``Communication-constrained stl task decomposition through convex optimization,'' in \emph{Amer. Control Conf.}, 2024.

\bibitem{gross2005graph}
J.~L. Gross and J.~Yellen, \emph{Graph theory and its applications}.\hskip 1em plus 0.5em minus 0.4em\relax CRC press, 2005.

\bibitem{guo2013reconfiguration}
M.~Guo and D.~V. Dimarogonas, ``Reconfiguration in motion planning of single-and multi-agent systems under infeasible local ltl specifications,'' in \emph{52nd Conf. Decis. Control}.\hskip 1em plus 0.5em minus 0.4em\relax IEEE, 2013, pp. 2758--2763.

\bibitem{bechlioulis2008robust}
C.~P. Bechlioulis and G.~A. Rovithakis, ``Robust adaptive control of feedback linearizable mimo nonlinear systems with prescribed performance,'' \emph{IEEE Trans. Autom. Control}, vol.~53, no.~9, pp. 2090--2099, 2008.

\bibitem{charitidou2021signal}
M.~Charitidou and D.~V. Dimarogonas, ``Signal temporal logic task decomposition via convex optimization,'' \emph{IEEE Control Syst. Lett.}, 2021.

\bibitem{girard2015safety}
A.~Girard, G.~G{\"o}ssler, and S.~Mouelhi, ``Safety controller synthesis for incrementally stable switched systems using multiscale symbolic models,'' \emph{IEEE Trans. Autom. Control}, vol.~61, no.~6, pp. 1537--1549, 2015.

\bibitem{liu2008omni}
Y.~Liu, J.~J. Zhu, R.~L. Williams~II, and J.~Wu, ``Omni-directional mobile robot controller based on trajectory linearization,'' \emph{Robot. Auton. Syst.}, vol.~56, no.~5, pp. 461--479, 2008.

\bibitem{karayiannidis2012multi}
Y.~Karayiannidis, D.~V. Dimarogonas, and D.~Kragic, ``Multi-agent average consensus control with prescribed performance guarantees,'' in \emph{51st Conf. Decis. Control}.\hskip 1em plus 0.5em minus 0.4em\relax IEEE, 2012, pp. 2219--2225.

\end{thebibliography}
\bibliographystyle{IEEEtran}

\end{document}